\documentclass{article}

\newlength{\abstractwidth}
\usepackage{cancel}
\usepackage{hyperref}
\usepackage{color}
\usepackage{graphicx}
\usepackage{tikz}
\usepackage{mathtools}
\usepackage{bbm}

\usepackage{float}

\usepackage{cancel}
\usepackage{hyperref}
\newcommand{\doiurl}[2]{{\hypersetup{urlcolor=darkred}\href{http://dx.doi.org/#2}{#1}\hypersetup{urlcolor=blue}}}

\usepackage{amsmath}
\usepackage{amssymb}
\usepackage{latexsym}

 \usepackage{setspace}
 
 \usepackage{caption}
\usepackage{subcaption}

\numberwithin{equation}{section}

\newcommand{\abs}[1]{\left\lvert #1 \right\rvert}
\renewcommand{\thefootnote}{\fnsymbol{footnote}}
\renewcommand{\thanks}[1]{\footnote{#1}}
\newcommand{\starttext}{
\setcounter{footnote}{0}
\renewcommand{\thefootnote}{\arabic{footnote}}}
\newcommand{\bea}{\begin{eqnarray}}
\newcommand{\eea}{\end{eqnarray}}
\newcommand{\be}{\begin{eqnarray}}
\newcommand{\ee}{\end{eqnarray}}

\def\ie{\begin{equation}\begin{aligned}}
\def\fe{\end{aligned}\end{equation}}
\def\half{{\scriptstyle \frac 12}}

\def\sevenh{{\scriptstyle \frac 72}}
\def\threeh{{\scriptstyle \frac 32}}
\def\fiveh{{\scriptstyle \frac 52}}

\def\nineh{{\scriptstyle \frac 92}}

\makeatletter
\newcommand{\setword}[2]{%
  \phantomsection
  #1\def\@currentlabel{\unexpanded{#1}}\label{#2}%
}
\makeatother

\def\ie{\begin{equation}\begin{aligned}}
\def\fe{\end{aligned}\end{equation}}
\def\cF{{\cal A}}

\def\cD{{\cal D}}
\def\cE{{\cal E}}
\def\cF{{\cal F}}

\def\cN{{\cal N}}

\def\cS{{\cal S}}

\def\cV{{\cal V}}

\def\bt{{b.t.}}

\def\ZZ{{\mathbb Z}}

\def\NN{{\mathbb N}}

\def\nn{\nonumber}

\def\Im{{\rm Im \,}}

\def\p{\partial}

\def\stau{\tau}

\def\ttau{\rho}

\def\Z{{\mathbb Z}}

\begin{document}

\begin{flushright}
{\small QMUL-PH-25-01}
\end{flushright}

\starttext

\setcounter{footnote}{0}

\vskip 0.3in

\begin{center}

\centerline{\large \bf
Modular Features of Superstring Scattering Amplitudes:}
\vskip 0.1in
\centerline{\large \bf Generalised Eisenstein Series and Theta Lifts}

\vskip 0.2in

\vskip 0.2in
{Daniele Dorigoni$^{1,2}$, Michael B. Green$^{3}$, Congkao Wen$^{4}$} 
\vskip 0.15in

\vskip 0.1in

{\small $^{1}$ Centre for Particle Theory \& Department of Mathematical Sciences, 
}\\
\small{Durham University, Lower Mountjoy, Stockton Road, Durham DH1 3LE, UK}
\vskip 0.1in

{\small $^{2}$  Max-Planck-Institut f\"ur Gravitationsphysik (Albert-Einstein-Institut),}\\
\small{
am M\"uhlenberg 1, Potsdam, 14476, Germany}
\vskip 0.1in

{ \small $^{3}$ Department of Applied Mathematics and Theoretical Physics }\\
{\small  Wilberforce Road, Cambridge CB3 0WA, UK}

\vskip 0.1in

{\small  $^{4}$ Centre for Theoretical Physics, Department of Physics and Astronomy,  }\\ 
{\small Queen Mary University of London,  London, E1 4NS, UK}

\vskip 0.5in

\begin{abstract}
\vskip 0.1in

In previous papers it has been shown that the coefficients of terms in the large-$N$ expansion of a certain integrated four-point correlator of superconformal primary operators in $\mathcal{N}=4$ supersymmetric Yang--Mills theory are rational  sums of  real-analytic Eisenstein series and ``generalised Eisenstein series''.  The latter are novel modular functions first encountered in the context of graviton amplitudes in type IIB superstring theory.  Similar modular functions, known as two-loop modular graph functions, are also encountered in the low-energy expansion of the integrand of genus-one closed superstring  amplitudes.  In this paper we further develop the  mathematical structure of such generalised Eisenstein series emphasising, in particular, the occurrence of $L$-values of holomorphic cusp forms in their Fourier mode  decomposition. We show that both the coefficients in the large-$N$ expansion of this integrated correlator and two-loop modular graph functions admit a unifying description in terms of four-dimensional lattice sums generated by theta lifts of certain local Maass functions, which generalise the structure of real-analytic Eisenstein series. Through the theta lift representation, we demonstrate that elements belonging to these two families of non-holomorphic modular functions can be expressed as rational linear combinations of generalised Eisenstein series for which all the $L$-values of holomorphic cusp forms precisely cancel.  
 \end{abstract}  
                     
\end{center}
\nopagebreak 
\newpage

\tableofcontents

\newpage

\section{Overview and results}
\label{sec:overview}

\subsection{Introduction}
\label{sec:intro}

Two interesting and related classes of modular functions have arisen from the study of distinct aspects of closed  superstring scattering amplitudes.   Elements in these families can both be expressed as rational linear combinations of \textit{Generalised Eisenstein Series} (GESs), which are defined to be modular invariant functions satisfying the inhomogeneous Laplace eigenvalue equation
\bea
\label{eq:geneisen}
\left( \Delta_\tau -s(s-1) \right) \cE(s;s_1,s_2;\tau) =  E(s_1;\tau)\, E(s_2; \tau)\,,
\eea
where $s\in \NN$ and $\tau\coloneqq \tau_1+i\tau_2 \in \mathfrak{H}$ is a complex modulus, which takes values in a upper-half complex plane $\mathfrak{H}\coloneqq\{\tau\in \mathbb{C}\,\vert\,{\rm Im}(\tau)>0\}$ and $\Delta_\tau$ is the hyperbolic laplacian, $\Delta_\tau\coloneqq \tau_2^2 (\partial_{\tau_1}^2+ \partial_{\tau_2}^2)$.  The modular invariant function $E(s;\tau)$ appearing in the source term in \eqref{eq:geneisen} is a non-holomorphic Eisenstein series, which satisfies 
\ie
\left( \Delta_\tau -s(s-1) \right)  E(s;\tau)=0 \, . 
\fe

The first class of GESs that we consider has $s_1,s_2 \in \NN$  and it appears to be the relevant vector space of non-holomorphic modular invariant functions  to describe \textit{two-loop Modular Graph Functions} (MGFs) {\cite{Green:1999pv,DHoker:2015wxz, DHoker:2016mwo}.}  Interest in MGFs originated from the study of  the low energy expansion of the integrand for  genus-one contributions to four-graviton scattering in either of the type II superstring theories \cite{Green:1999pv, Green:2008uj}. The name MGF originates from the fact that these functions are the values of  connected Feynman graphs for a free scalar field theory living on a two-torus with complex structure $\tau$. Modular invariance (${\rm SL}(2,\Z)$-invariance) is a direct consequence of diffeomorphism invariance on the genus-one world-sheet torus.  The mathematical structure of these functions generalises in an obvious manner to modular graph {\it forms} whose study has  led to an impressive symbiosis with novel ideas in algebraic geometry \cite{Brown:mmv,Brown:I,Brown:II}. The construction of MGFs may be extended to higher loops but in this paper we will focus on the two-loop cases since these are related to GESs. With a slight abuse of terminology, in what follows the acronym MGFs specifically refers to two-loop modular graph functions.

Two-loop MGFs are defined by the following lattice sums, 
\begin{equation}\label{eq:Cabc}
C_{a,b,c}(\tau) \coloneqq  \sum_{\substack{ p_1,p_2,p_3 \in\Lambda' \\ p_1 +p_2 +p_3=0}}  \frac{(\tau_2/\pi)^{a+b+c}}{|p_1|^{2a}|p_2|^{2b}|p_3|^{2c}}\,,
\end{equation} 
where $a,b,c\in \mathbb{N}$ and the sum is over three complex lattice momenta $p_i = {m}_i + {n}_i \tau \in  \Lambda' = \mathbb{Z} +\tau\mathbb{Z}\setminus\{0\}$, with $i=1,2,3$, subject to the conservation condition $p_1+p_2+p_3=0$.
Notably, specific linear combinations \cite{DHoker:2015gmr} of  MGFs satisfy inhomogeneous Laplace eigenvalue equations with source terms that are bilinear in non-holomorphic  Eisenstein series  of the form \eqref{eq:geneisen} with $s_1,s_2\in \mathbb{N}$. However, for values of $s\geq 6$ modular-invariant solutions of  \eqref{eq:geneisen} with integer indices cannot all be expressed as superpositions of MGFs. The construction of GESs requires additional contributions proportional to $L$-functions of  holomorphic cusp forms \cite{Dorigoni:2021jfr,Dorigoni:2021ngn} with modular weight $2s$. Conjecturally, MGFs select only rational combinations of this family of GES for which the $L$-value contributions precisely cancel. 
               
A second class of GESs consists of  solutions to \eqref{eq:geneisen} with sources given by non-holomorphic Eisenstein series with half-integer indices, i.e. with $s_1,s_2\in \NN+1/2$.
{An element in this family of GESs arises in the low-energy expansion of four-graviton scattering amplitudes in type IIB superstring theory}  \cite{Green:1999pu,Green:2005ba}. 
In this case ${\rm SL}(2,\Z)$ covariance is known as S-duality and can be motivated by interpreting superstring theory in terms of eleven-dimensional M-theory compactified on a two-torus.  The complex structure of the torus is here identified with the complex axion-dilaton field,  $\tau=\tau_1+i\tau_2$, where $\tau_2=1/g_s$ and $g_s$ is the string coupling constant.   
For lack of better terminology, we refer to this second class of non-holomorphic modular invariant functions arising in the context scattering amplitudes in type IIB superstring theory as \textit{S-Dual Modular Functions} (SMFs).

SMFs also appear in the study of correlation functions in $\cN=4$ supersymmetric  Yang--Mills (SYM) in four dimensions. This four-dimensional quantum field theory exhibits Montonen--Olive duality~\cite{Montonen:1977sn}, in which ${\rm SL}(2,\Z)$ acts on the complex Yang--Mills coupling constant $\tau= \theta/2\pi + 4\pi i /g_{_{\rm YM}}^2$. This fits  beautifully with the AdS/CFT conjecture \cite{Maldacena:1997re}, where type IIB superstring theory in an $AdS_5\times S^5$ background is identified with $\cN=4$ SYM with $SU(N)$ gauge group.  S-duality in the dual superstring theory here is interpreted as the holographic image of  Montonen--Olive duality  of $\cN=4$ SYM.  {It was shown in  \cite{Binder:2019jwn, Chester:2020dja}, that certain integrated correlators  of four superconformal primary operators in the stress tensor multiplet are {computable using supersymmetric localisation for all $N$ \cite{Pestun:2007rz},} and their properties depend sensitively on the choice of integration measures (see e.g. the review \cite{Dorigoni:2022iem}). In the simplest example the large-$N$ expansion of the correlator is a series of  half-integer powers of $1/N$ and the coefficients are rational linear combinations of non-holomorphic Eisenstein series with half-integer indices  \cite{Chester:2019jas}. The second example of integrated correlators has a large-$N$ expansion  with  both half-integer and integer powers of $1/N$ \cite{Chester:2020vyz,Alday:2023pet}.  The half-integer powers again have coefficients that are rational linear combinations of non-holomorphic Eisenstein series with half-integer indices, while the coefficients of integer powers of $1/N$ are given by finite rational sums of SMFs, which are of relevance to the observations in this paper.}

SMFs satisfy inhomogeneous Laplace eigenvalue equations of the form~\eqref{eq:geneisen} with source terms that are bilinear in non-holomorphic Eisenstein series, just as was the case with MGFs, but here the Eisenstein series in the sources have half-integer indices, $s_1,s_2\in \NN+1/2$. 
Once again, for $s\ge 6$ not every GES solution to~\eqref{eq:geneisen} with half-integer indices can be expressed as a superposition of SMFs. As before, the construction of this second family of GESs involves $L$-functions of holomorphic cusp forms~\cite{Fedosova:2023cab} with modular weight $2s$.   SMFs are conjecturally equal to special linear combinations  of GESs, where the $L$-values cancel as  was the case for MGFs.

We will demonstrate that one of the main results of our paper establishes that both MGFs and SMFs emerge as specific instances of a broader class of non-holomorphic modular-invariant functions. These functions can be systematically constructed via a particular theta lift applied to a family of local modular Maass functions. This construction automatically leads to rational linear combinations of GES solutions to \eqref{eq:geneisen}, where, as previously noted, the contributions from holomorphic cusp forms are entirely cancelled.

\subsection{Results}
\label{sec:Results}

 As discussed in the introduction, both MGFs and SMFs can be expressed as specific rational linear combinations of GESs. We will see that these two families of non-holomorphic modular invariant functions are special cases of the theta-lifts defined by
\begin{equation}
\mathcal{E}^w_{i,j}(\tau)  = \int_{( \mathbb{R}^+)^3 } B^w_{i,j}(t)\,  \Gamma_{2,2}'(\tau;t)\,  {\rm d}^3 t          \,,
 \label{eq:thetlift0}
\end{equation}  
where $t$ represents three cartesian coordinates $t_1,t_2,t_3$, while the indefinite theta-series $\Gamma_{2,2}'(\tau;t)$ is expressed as a four-dimensional lattice sum, {as defined in \eqref{eq:G22tred}}.  The  lifted functions  $B^w_{i,j}(t)$ are related to local Maass functions that will be defined in detail later in the paper, and the indices $i,j,w\in \NN$. 

 We will show that the function $\mathcal{E}^w_{i,j}(\tau)$ is  a specific sum of modular invariant solutions to the inhomogeneous Laplace equation \eqref{eq:geneisen}.
 More concretely, we will prove that every $\cE_{i,j}^w(\tau)$  defined in \eqref{eq:thetlift0} can be expressed as a rational linear combination of GESs and non-holomorphic Eisenstein series, 
\ie
 \cE_{i,j}^w(\tau) =  \sum_{r=-\frac{(i+j-1)}{2}}^{\frac{(i+j-1)}{2}} b_{i,j}(w,r) \,\cE(s;\tfrac{w}{2}+r  ,\tfrac{w}{2}-r ;\tau)  + d_{i,j}(w) \, E(w;\tau)\,,
\label{eq:Eijsolved0}
\fe
where the eigenvalue $s=3i+j+1$ and  the explicit expression of the general coefficients $b_{i,j}(w,r)$ is given by \eqref{eq:bijGen}. 
All MGFs are obtained by considering $\cE_{i,j}^w(\tau)$ when $s=3i+j+1$ and $w$ have the same parity with $1 \leq s \leq w-2$ and $w\geq3$. Conversely, SMFs are generated by $\cE_{i,j}^w(\tau)$ when $s$ and $w$ have opposite parity with $s\geq w+1$ and $w\geq 3$.\footnote{{It is worth noting that many of our conclusions do apply to a wider range of weights $w$, for example SMFs with $s< w$.  However, we will focus on modular functions with parameters satisfying these specific ranges since they are most relevant for superstring amplitudes and the integrated correlator.}}

We then analyse in more detail the properties of an individual GES, $ \cE(s;s_1,s_2;\tau)$ of the type that appears in the rational linear combination \eqref{eq:Eijsolved0}.
Given the parity arguments mentioned above, we retrieve the claim made in the introduction that GESs relevant for MGFs have integer indices $s_1,s_2\in\mathbb{N}$ with $s_1,s_2\geq 2$, while for SMFs they have half-integral indices $s_1,s_2\in \mathbb{N}+1/2$. 

In \cite{Dorigoni:2021jfr, Dorigoni:2021ngn} it was argued that the modular invariant solution to \eqref{eq:geneisen} with $s_1,s_2\in\mathbb{N}$ and $s_1,s_2\geq 2$ and eigenvalue $s$ with the same parity as $s_1+s_2$ must take the following form,  
\begin{equation}\label{eq:PartplusHomoN}
 \cE(s;s_1,s_2;\tau)=  \cE_{{\rm p}}(s;s_1,s_2;\tau) + \sum_{\Delta\in \cS_{2s}} \lambda_{\Delta}(s;s_1,s_2) H_{\Delta}(\tau)\,,
\end{equation}
where $\cE_{{\rm p}}$ is a specific particular solution to \eqref{eq:geneisen} constructed explicitly in \cite{Dorigoni:2021jfr, Dorigoni:2021ngn}  {and subject to the boundary condition that the coefficient ot $\tau^{2s}$  in the  solution of the  homogeneous equation correspondng to  \eqref{eq:geneisen} is zero.  Our main interest here is the second term in \eqref{eq:PartplusHomoN}, which is a solution to the homogeneous equation that is exponentially suppressed as $\tau_2\to \infty$. }Here $H_{\Delta}(\tau)$ denotes a particular Eichler integral \eqref{eq:Hdelta} of the Hecke-normalised cusp form $\Delta\in\mathcal{S}_{2s}$ with $\mathcal{S}_{2s}$ the vector space of holomorphic cusp forms of weight $2s$.  The coefficients $\lambda_{\Delta}(s;s_1,s_2)$ were determined in a multitude of cases  in \cite{Dorigoni:2021jfr, Dorigoni:2021ngn},  demonstrating that these contain the completed $L$-functions, $\Lambda(\Delta, t)$, of the corresponding holomorphic cusp form. 

For the case $s_1,s_2\in\mathbb{N}+1/2$ and eigenvalue $s\geq s_1+s_2+1$ of opposite parity to $w=s_1+s_2$, we use the method presented in \cite{Fedosova:2022zrb} to construct a particular solution to \eqref{eq:geneisen} and show that whenever ${\rm dim}\,\mathcal{S}_{2s} \neq 0$ this does not correspond to the GES solution. In such cases, a suitable multiple of the solution of the homogeneous equation must be added to the particular solution to ensure that the complete expression is the modular invariant solution $ \cE(s;s_1,s_2;\tau)$. 
We show that again in this case, the GES solution to \eqref{eq:geneisen} with $s_1,s_2\in\mathbb{N}+1/2$ and eigenvalue $s\geq s_1+s_2+1$ of opposite parity to $w=s_1+s_2$ must take the same form   \eqref{eq:PartplusHomoN}. Using the results of \cite{Fedosova:2023cab} concerning certain convolutions of divisor sigma functions, we find an analytic expression for the coefficients  of the solution of the homogeneous equation, which is given as 
\begin{align}
&\nn\lambda_{\Delta}(s;s_1,s_2) =\\
&    \frac{  \pi  (-1)^{\frac{s+s_1-s_2+2}{2}} {2^{5-4s}}  \Gamma \left(2s-1\right) \Gamma\left(\frac{s_1+s_2-s}{2}\right)  } {\Gamma (s_1) \Gamma (s_2) \Gamma\left(\frac{s_1+s_2+s}{2}\right)  \Gamma \left(\tfrac{s+s_1-s_2+1}{2} \right) \Gamma \left(\tfrac{s-s_1+s_2+1}{2} \right) }   \frac{ \Lambda (\Delta, s+s_1-s_2)\Lambda (\Delta, s+s_1+s_2-1)}{\langle \Delta, \Delta\rangle}\,,\label{eq:LambdaRes}
\end{align}
where $\langle \Delta, \Delta\rangle$ denotes the Petersson norm of the Hecke eigenform $\Delta\in \mathcal{S}_{2s}$.

Given the analytic dependence of the coefficient of the solution of the homogeneous equation \eqref{eq:LambdaRes} on the indices $s_1,s_2$, we have verified that if we specialise \eqref{eq:LambdaRes} to the case of GESs with $s_1,s_2\in\mathbb{N}$ and $s_1,s_2\geq 2$ and eigenvalue $s$ with the same parity as $s_1+s_2$ we reproduce all the particular cases presented in \cite{Dorigoni:2021jfr, Dorigoni:2021ngn}\footnote{In these references, the  coefficient of the solution of the homogeneous equation  $\lambda_{\Delta}(s;s_1,s_2)$  was denoted by $a^+_{\Delta_{2s},s_1,s_2}$. }.
This leads to the following conjecture:
\vskip 0.2cm
{\bf \setword{Conjecture 1}{Conj1}.} 
\vskip 0.1 cm
\textit{The generalised Eisenstein series with $s_1,s_2\in\mathbb{N}$ and $s_1,s_2\geq 2$ and eigenvalue $s$ with the same parity as $s_1+s_2$ must take the form \eqref{eq:PartplusHomoN}, where the particular solution has been constructed in \cite{Dorigoni:2021jfr, Dorigoni:2021ngn} and the coefficient of the solution of the homogeneous equation  is given by \eqref{eq:LambdaRes}.}
\vskip 0.2cm

This result is quite striking, since the particular solutions constructed in \cite{Dorigoni:2021jfr, Dorigoni:2021ngn} and  \cite{Fedosova:2022zrb} for the two different cases of GESs have in principle nothing to do with one another. Yet in both cases, the ``modular form'' content of the particular solution (which we stress is not modular invariant on its own) captures everything \textit{but} the very same Eichler integral of holomorphic cusp forms.

{Importantly,  for the case where the GESs have half-integer indices, $s_1,s_2\in \NN+1/2$, a theorem on convolution identities for divisor sums proved by Fedosova, Klinger-Logan and Radchenko \cite{Fedosova:2023cab}  enables us to prove \ref{Conj1} in Section \ref{sec:Lvalues}. 
For GESs with integer indices, $s_1,s_2\in \NN^{\geq 2}$ \ref{Conj1} remains unproven, albeit reproducing all particular results presented in \cite{Dorigoni:2021jfr, Dorigoni:2021ngn}. }

Having clarified the form of the solution given in \eqref{eq:PartplusHomoN} for the families of GESs relevant to both MGFs and SMFs, we will revisit the general rational linear combination of GESs \eqref{eq:Eijsolved0}, which is produced by the theta lift \eqref{eq:thetlift0}.
The explicit linear combination \eqref{eq:Eijsolved0} together with the solution for the coefficients of the homogeneous equation \eqref{eq:LambdaRes} lead to a further  conjecture:

\vskip 0.2cm
{\bf \setword{Conjecture 2}{Conj2}.} 
\vskip 0.1 cm
\textit{
The modular invariant function ${\cal E}_{i,j}^w(\tau)$ defined in \eqref{eq:thetlift0} is a rational linear combination of generalised Eisenstein series that {has the form  of a four-dimensional lattice sum and for which the  $L$-values of holomorphic cusp forms cancel.}}
\vskip 0.2 cm

{Consequently,  while there is no known general lattice-sum representation  for a single GES, the modular invariant functions ${\cal E}_{i,j}^w(\tau)$ do provide  a four-dimensional lattice sum representation of ``special'' rational linear combinations of GESs. Perhaps the key defining property of these special linear combinations of GESs is the absence of contributions proportional to $L$-values of holomorphic cusp forms, despite the presence of such terms in each individual GES that appears in the linear combination.}

For MGFs that are given by rational linear combinations of GESs with $s_1,s_2\in\mathbb{N}$ and $s_1,s_2\geq 2$ and eigenvalue $s$ with the same parity as $w=s_1+s_2$, this is indeed a conjecture since we do not prove \eqref{eq:LambdaRes}  for  the  coefficient of the  solution of the homogeneous  equation. 
However, for SMFs \ref{Conj2} can be proven since these are expressed in terms of GESs with $s_1,s_2\in\mathbb{N}+1/2$ and eigenvalue $s\geq s_1+s_2+1$ of opposite parity to $w=s_1+s_2$ for which the proof of \eqref{eq:LambdaRes} follows from \cite{Fedosova:2023cab}.

In either case, assuming the general validity of \eqref{eq:LambdaRes}, we will show that the $L$-function contribution to $ \cE_{i,j}^w(\tau)$ coming from the holomorphic cusp form $\Delta \in \mathcal{S}_{2s}$ with $s=3i+j+1$ is proportional to 
\begin{equation}
\mathcal{K}_{i,j}(\Delta) =  \int_0^{i \infty} \Delta(\tau) \, [\tau^2 (1-\tau)^2]^i (\tau^2 -\tau+1)^j {\rm d} \tau\,.\label{eq:cuspijIntro}
\end{equation}
The polynomials in $\tau$ appearing as part of the integrand in the above equation play a fundamental role in the construction of the local Maass functions, $B^w_{i,j}(t)$, whose theta lift \eqref{eq:thetlift0} we will analyse.
We will show that $\mathcal{K}_{i,j}(\Delta)=0$ for all $i,j\in \mathbb{N}$ and for every holomorphic cusp form $\Delta \in \mathcal{S}_{2s}$ with $s=3i+j+1$, thus providing strong evidences that the modular functions $\cE_{i,j}^w(\tau)$ as defined in \eqref{eq:thetlift0}, or equivalently the particular rational linear combinations of GESs as in \eqref{eq:Eijsolved0}, are indeed free of holomorphic cusp forms. 

{Furthermore, in Section \ref{sec:Absence} we summarise a counting argument, first noticed for the MGF case in \cite{Dorigoni:2021jfr}, which shows that \textit{any} rational linear combination of GESs relevant for the discussion of MGFs and SMFs and for which the cusp form contribution cancels out must be given by a linear combination with rational coefficients of theta lifts ${\cal E}_{i,j}^w(\tau)$.}

Remarkably,  precisely these particular linear combinations of GESs appear to be the relevant ones for the study of important physical observables in superstring theory and its holographic dual CFT.  This is true for MGFs for perturbative genus-one contributions to superstring amplitudes, as well as for SMFs arising in the low-energy expansion of superstring amplitudes and the large-$N$ expansion of integrated correlators in $\mathcal{N}=4$ SYM.

\subsection{Outline}
\label{sec:Outline}

The paper is organised as follows.
In section \ref{sec:Theta} we introduce the main characters of our story and define an infinite class of modular invariant non-holomorphic functions, constructed as theta lifts
of certain modular local Maass forms. The defining property of these Maass forms is that they are constructed via a raising procedure from particular negative modular weight holomorphic polynomials, known as modular local polynomials. In appendix \ref{sec:nonhol} we present some basic properties of Maass functions and non-holomorphic Eisenstein series, while in appendix \ref{app:ModPol} we discuss more technical details regarding modular local polynomials.
We show in section \ref{sec:LapEq} that the theta lift thus defined satisfies an inhomogeneous Laplace eigenvalue equation with source terms that are bilinear in non-holomorphic Eisenstein series. Hence, we derive a general expression for the theta lift as a rational linear combination of generalised Eisenstein series and standard non-holomorphic Eisenstein series.
We discuss various properties of the generalised Eisenstein series in section \ref{sec:GenEis}. In particular, we revisit two methods for constructing a particular solution to the differential equation \eqref{eq:geneisen} and show how to determine the homogeneous part so as to obtain a modular invariant solution. 
The solution of the homogenous  equation takes the form of an Eichler integral of a special holomorphic cusp form multiplied by the product of two of its completed $L$-values that we determine analytically. Some fundamental properties of holomorphic cusp forms and their $L$-values are reviewed  in appendix \ref{app:Cusps}.
We conclude section \ref{sec:GenEis} by showing that the sum of the holomorphic cusp form solutions to the homogeneous equation that contribute to $\cE^w_{i,j}(\tau)$, i.e. to the sum of  generalised Eisenstein series obtained from the theta lifts,  cancels out completely.  
Finally, in section \ref{sec:Discussion} we present some interesting open problems that we believe our analysis will help in solving.

\section{Theta lift of local Maass functions}
\label{sec:Theta}

In this section we will define a particular example of a non-holomorphic indefinite theta-series that 
is a  four-dimensional lattice sum, which will enter the subsequent analysis.  This theta-series is 
 related to the genus-one partition function for a closed string moving on a target space compactified on $T^2$.
We will use this theta-series to define a class of modular invariant functions via theta lift of certain local Maass functions.  This theta lift provides an elegant and unified approach to both MGFs and SMFs and clarifies some  intriguing properties shared by these two families of modular functions.

\subsection{Theta lift: definitions and properties}
\label{sec:thetadef}


The simplest examples of MGFs and SMFs are the non-holomorphic Eisenstein series, $E(s;\tau)$, with $s\in \NN^+$ and $s\in \NN+{1\over 2}$, respectively.  Some properties of these modular functions, which enter into the source terms in  \eqref{eq:geneisen}, are reviewed in appendix~\ref{sec:nonhol}. 

However, the space of modular invariant functions needed to discuss higher loop MGFs and higher order coefficients in the low-energy effective action of type IIB scattering amplitudes in becomes significantly more complicated and goes beyond the standard world of non-holomorphic Eisenstein series or Maass functions.
Importantly, both  MGFs and the SMFs  considered here are closely  related, albeit for very different physical reasons, to a particular non-holomorphic modular function of three complex variables that can be expressed in terms of an indefinite theta-series.

The particular indefinite theta series  of relevance arises quite naturally in string theory  when considering a { closed-string partition function for a genus-one world-sheet with modular parameter $\tau$, embedded in  a $T^2$-compactified target space.}
The moduli space of the target {space} is the symmetric space: 
{
\begin{equation}
[{\rm SO}(2;\mathbb{R})\times {\rm SO}(2;\mathbb{R})] \backslash {\rm SO}(2,2;\mathbb{R}) / {\rm SO}(2,2;\mathbb{Z})\,,
\end{equation}
which describes the product of two hyperbolic planes, with each parameterised by a complex coordinate.  These are  the complex structure denoted as $\rho=\rho_1+i\rho_2$,  and the Kahler structure denoted by $\nu = B+i V$, where $V$ is the volume of the torus and $B$ is the value of an antisymmetric tensor on the torus.}

We now define a lattice sum  $\Gamma_{2,2}\notag : \mathfrak{H}^3 \to \mathbb{R}$, which is 
a real-analytic function of the three upper-half plane variables $\tau,\rho,\nu \, {\in \mathfrak{H}} \coloneqq \{z\in \mathbb{C}\,,\Im\,z>0\}$,  
\begin{align}
\Gamma_{2,2}(\tau; \rho,\nu) &\coloneqq  \sum_{\underline{m},\underline{n}\in \mathbb{Z}^2} V \exp\Big( -\frac{\pi}{\tau_2} (\underline{m}+\underline{n}\tau) G(\rho,V) (\underline{m}+\underline{n}\bar{\tau})+2\pi i B \,\mbox{det}\,M\Big)\,,
\label{eq:Gamma22}
\end{align}
where we have introduced the two $2\times2$ matrices
\begin{align}
G(\rho,V) \coloneqq  \frac{V}{\rho_2}  \left(\begin{matrix} 1 & \rho_1 \\ \rho_1 & |\rho|^2\end{matrix}\right)\,,\qquad
M \coloneqq \left(\begin{matrix} m_1 & n_1 \\ m_2 & n_2\end{matrix}\right)=\left(\underline{m} \,\underline{n} \right)\,.
\end{align}
{with $m_1, m_2, n_1, n_2 \in \mathbb{Z}$.}
This function has the important symmetry properties,
\begin{align}
\Gamma_{2,2}(\tau; \rho,\nu) &= \Gamma_{2,2}(\rho; \tau,\nu)\,,\\
\Gamma_{2,2}(-\frac{1}{\tau}; \rho,\nu) & = \Gamma_{2,2}(\tau; \rho,\nu) \,, \qquad \Gamma_{2,2}(\tau+1; \rho,\nu)  = \Gamma_{2,2}(\tau; \rho,\nu) \, ,  \\
\Gamma_{2,2}(\tau; \rho, -\frac{1}{\nu}) & \label{eq:G22nuS}= \Gamma_{2,2}(\tau; \rho, \nu) \,, \qquad  \Gamma_{2,2}(\tau; \rho, \nu+1)= \Gamma_{2,2}(\tau; \rho, \nu) \, .
\end{align}
{The first identity can be derived from \eqref{eq:Gamma22} by exchanging the summation variables $n_1\leftrightarrow m_2$},
while the $S$-transformation for the $\nu$ variable $\nu \to -1/\nu$ can be derived upon performing a quadruple Poisson resummation over $m_1,m_2,n_1,n_2$.

The Laplace operator on this symmetric space reduces to
\begin{equation}
\Delta_{SO(2,2)} = \Delta_\rho+\Delta_\nu = \rho_2^2 (\partial_{\rho_1}^2+\partial_{\rho_2}^2) + V^2 (\partial_B^2 +\partial_V^2)\,,
\end{equation}
and the lattice-sum $\Gamma_{2,2}$ satisfies
\begin{equation}
\Delta_{SO(2,2)} \Gamma_{2,2}(\tau;\rho,\nu) = 2 \Delta_\tau  \Gamma_{2,2}(\tau;\rho,\nu)  =2 \tau_2^2 (\partial_{\tau_1}^2+\partial_{\tau_2}^2) \Gamma_{2,2}(\tau;\rho,\nu)\,.
\end{equation}
We note also that 
\begin{equation}
 \Delta_\tau  \Gamma_{2,2}(\tau;\rho,\nu) =  \Delta_\rho  \Gamma_{2,2}(\tau;\rho,\nu)= \Delta_\nu  \Gamma_{2,2}(\tau;\rho,\nu)\,.\label{eq:LapVar}
\end{equation}

In string theory, one usually integrates over $\tau$, the modular parameter of the genus-one worldsheet, for example, in obtaining a genus-one correction to a closed-string scattering amplitude for external  states moving on a toroidally compactified target space.
However, in the present discussion this is not the case, instead of integrating over $\tau$ we are interested in considering a particular theta lift that involves integration over $\rho$ and $V={\mbox{Im}}\,\nu$ after having set $B=0$ in \eqref{eq:Gamma22}.
More explicitly,  we will consider
\begin{equation}
{\cal E}_{i,j}^w(\tau)  = \int_{ \mathbb{R}^+ \times \mathcal{F}_\rho } V^{{ w}} A_{i,j}(\rho)  \Gamma_{2,2}(\tau;\rho,i V)\, 2 \frac{{\rm d}V}{V^2} \frac{{\rm d}^2 \rho}{\rho_2^2}\,,\label{eq:lift}
\end{equation}
where, as will appear natural later, the domain of integration $\mathcal{F}_\rho$ is the fundamental domain with respect to the congruence subgroup\footnote{The congruence subgroup $\Gamma_0(2)$ is defined as $\Gamma_0(2)\coloneqq\{\left(\begin{smallmatrix} a& b \\ c & d\end{smallmatrix}\right)\in {\rm SL}(2,\mathbb{Z})\,\vert\, c \equiv 0\,\rm{mod}\,2\}$.} $\Gamma_0(2)$, i.e. $\mathcal{F}_\rho = \Gamma_0(2) \backslash  \mathfrak{H} $.  The parameter $w$ is a non-negative integer, $w\in \mathbb{N}$.
We will shortly provide some string theory arguments that justify the importance of the integral representation \eqref{eq:lift} in the discussion of MGFs as well as in the context of higher-derivative corrections to the low-energy expansion of type IIB string theory.

s{We emphasise that \eqref{eq:lift} is quite an unusual type of theta-lift in that the integrand does not seem to be a modular function in the $\nu$ variable. It is possible to think of the $V$ integral as being originally defined over the fundamental domain $\mathcal{F}_\nu ={\rm SL}(2,\mathbb{Z}) \backslash \mathfrak{H}$ where the integrand $V^w$ is replaced by a modular invariant function $ F(\nu)$. We choose this function to be defined by the Poincar\'e series $F(\nu) = \sum_{\gamma \in \Gamma_\infty \backslash {\rm SL}(2,\mathbb{Z})} \varphi(\gamma\cdot \nu)$ where $\Gamma_\infty =\{ \pm \left(\begin{smallmatrix} 1 & n \\ 0& n \end{smallmatrix}\right)\,\vert\, n \in \mathbb{Z}\} $ is the Borel stabiliser of the cusp and where we consider the distribution-valued Poincar\'e seed $\varphi(\nu) \coloneqq V^w \delta(B)$. By performing the standard unfolding trick for $F(\nu)$ we change the domain of integration from $\mathcal{F}_\nu ={\rm SL}(2,\mathbb{Z}) \backslash \mathfrak{H}$ to the strip $\Gamma_\infty \backslash \mathfrak{H}$ while simultaneously replacing $F(\nu)$ by its Poincar\'e seed $\varphi(\nu)$. The integral over ${\rm Re}(\nu) = B$ can then be easily performed thanks to the delta function, thus showing that \eqref{eq:lift} is indeed a theta-lift of a distributional automorphic form\footnote{We thank Kathrin Bringmann for related discussions.}. With a slight abuse of standard terminology, we still refer to \eqref{eq:lift} as a theta-lift.}

 In the rest of this paper we consider the theta lift \eqref{eq:lift} where the lifted elements, $A_{i,j}(\rho)$, belong to a special class of Maass functions.
The construction of the relevant family of functions $A_{i,j}(\rho)$ was given in unpublished notes by Don Zagier \cite{Zagier2008} that are reviewed and slightly expanded in section 5.4 of   \cite{DHoker:2018mys}, and will be summarised later in section \ref{sec:ModLoc}. The functions $A_{i,j}(\rho)$ are local Maass functions that 
 transform as modular functions under $\Gamma_0(2)$ apart from on an `exceptional set'  of points forming the boundaries of  $\cF_{\rho}$ (hence the denomination `local'). These functions are constructed from special holomorphic polynomials with negative modular weight via a Maass raising operation so as to become non-holomorphic  modular invariant functions.
 As a result $A_{i,j}(\rho)$ is a Laurent series in $\rho_2$ with coefficients that are polynomial in $(\rho_1^2-\rho_1)$ and satisfies the Laplace equation
\bea
(\Delta_\rho - s(s-1))A_{i,j}(\rho)=0\,,
\label{eq:lapA}
\eea
with $s=3i+j+1$.   

Before moving on to discuss the construction of the functions $A_{i,j}(\rho)$ and the properties of their theta lifts \eqref{eq:lift} in more detail in section~\ref{sec:ModLoc} we present a different parametrisation for the indefinite theta series $\Gamma_{2,2}$, which is related to the string-theory origins of  MGFs \cite{Green:1999pv} and SMFs \cite{Green:1999pu,Green:2005ba}.  One property we immediately note is that ({temporarily neglecting issues about regularisation}) the result of performing a {{\it formal}} quadruple Poisson resummation of the integers in  the theta-series \eqref{eq:Gamma22} leads to $\Gamma_{2,2}(\tau;\rho, i V) = \Gamma_{2,2}(\tau;\rho , i V^{-1}) $ so that by changing variables $V\to 1/V$ in \eqref{eq:lift} we find that ${\cal E}_{i,j} ^w(\tau) $ satisfies the functional relation
\begin{equation}
{\cal E}_{i,j}^{w}(\tau)   = {\cal E}_{i,j}^{2-w}(\tau)  \label{eq:FuncRelGen}\,,
\end{equation}
which is very reminiscent of the functional equation \eqref{eq:functional} for the non-holomorphic Eisenstein series. We now discuss an alternative parameterisation of $\Gamma_{2,2}$, which will prove to be very useful. 

\subsection*{An alternative parameterisation of $\Gamma_{2,2}$}

\begin{figure}[t!]
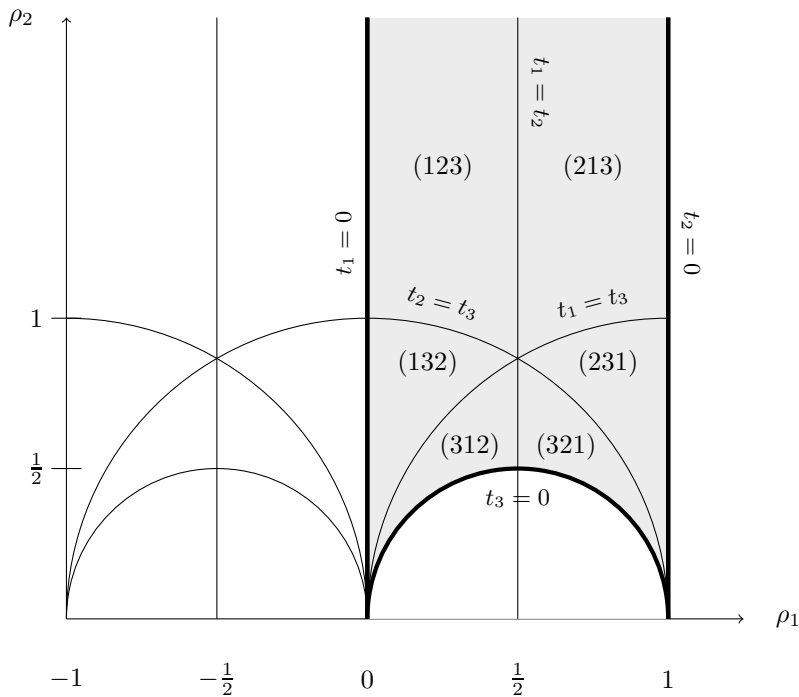

\begin{center}
\tikzpicture[scale=2]
\scope[xshift=-3cm,yshift=0.0]
\draw[white,fill=gray!15] (0,0) rectangle (2,4.0);
\draw (-2,0) -- (2,0) ;
\draw  [ultra thick] (0,0) -- (0,4.) ;
\draw   [ultra thick](2,0) -- (2,4) ;
\draw [->] (2,0) -- (2.5,0);
\draw (2.8, 0.0) node{$\ttau_1$};
\draw [->] (-2,0) -- (-2,4.0) ;
\draw  (-2.3, 4.0) node{$\ttau_2$};
\draw [fill=white,ultra thick] (2,0) arc(0:180:1.0) ;
\draw  (2,0) arc(0:90:2.0) ;
\draw (0,0) arc(180:90:2.0) ;
\draw (1,0) -- (1,4) ;
\draw (-2.1,1) -- (-1.9,1) ;
\draw (-2.1,2) -- (-1.9,2) ;
\draw  [ultra thick] (0,0) -- (0,4.) ;
\draw   [ultra thick](2,0) -- (2,4) ;
\draw (0,0) arc(0:180:1.0) ;
\draw  (0,0) arc(00:90:2.0) ;
\draw (-2,0) arc(180:90:2.0) ;
\draw (-1,0) -- (-1,4) ;
%
\draw (0, -0.4) node{$0$};
\draw (1, -0.4) node{$\frac{1}{2}$};
\draw (2, -0.4) node{$1$};
\draw (-1, -0.4) node{$-\frac{1}{2}$};
\draw (-2, -0.4) node{$-1$};

\draw (-2.2, 1.0) node{$\frac{1}{2}$};
\draw (-2.2, 2.0) node{$1$};

\draw (0.5, 3) node{{$(123)$}};
\draw (1.5, 3) node{{$(213)$}};
\draw (0.4, 1.7) node{{$(132)$}};
\draw (1.6, 1.7) node{{$(231)$}};
\draw (0.68, 1.15) node{{$(312)$}};
\draw (1.32, 1.15) node{{$(321)$}};
\draw (-0.15, 2.5) node[rotate=90]{\small $t_1=0$};
\draw (1.15, 3.5) node[rotate=-90]{\small $t_1=t_2$};
\draw (2.15, 2.5) node[rotate=-90]{\small $t_2=0$};
\draw (0.5, 2.1) node[rotate=-15]{\small $t_2=t_3$};
\draw (1.5, 2.1) node[rotate=15]{\small $t_1=t_3$};
\draw (1, 0.8) node{\small $t_3=0$};
\endscope

\endtikzpicture
\end{center} 
\caption{In the $\rho$-plane, the domain $t_i>0$ is mapped to the region shaded in grey, isomorphic to a fundamental domain of $\Gamma_0(2)$.  Here the labels $(ijk)$ in each sub-domain denote the ordering $t_i<t_j<t_k$.}
\label{fig:fundomain}
\end{figure}

This reparameterization maps the real integration parameters $\rho_1$, $\rho_2$ and $V$ in \eqref{eq:lift} into the three Schwinger parameters, $t_1,t_2,t_3$ of a two-loop vacuum Feynman diagram  defined in the range $0\le t_i\le \infty$ and given by the following equations\footnote{The notation $f(t)$ denotes a function of the three $t_i$ variables with $i=1,2,3$.} 
\begin{align}
\ttau_1(t) = \frac{t_1}{t_1+t_2} \,,\qquad \ttau_2(t) =\frac{V(t)}{t_1+t_2}\,,  \qquad   V(t)=  \sqrt{t_1 t_2+t_1t_3 +t_2t_3}\, ,
\label{eq:rhodef}
\end{align}
where the measure in terms of the new variables is given by
 \ie
 2  {\rm d}V {V^2} \frac{{\rm d}^2 \rho}{\rho_2^2} = {\rm d} t_1  {\rm d} t_2  {\rm d} t_3 \, . 
 \fe
 The domain of integration $0\le t_i\le \infty$  translates into the domain 
 in \eqref{eq:lift} that is given by the real positive axis $0\le V \le  \infty$ and
\bea
 0\le \rho_1 \le 1\,, \ \ \qquad (\rho_1-\half)^2 + \rho_2^2 \ge \frac{1}{4}\,, 
\label{eq:domain} 
 \eea
 which is precisely the fundamental domain of $\Gamma_0(2)$ illustrated by the grey domain in figure \ref{fig:fundomain} (which is reproduced from \cite{Green:1999pu}).

We will now express \eqref{eq:Gamma22} in these new variables. With a slight abuse of notation, we denote  $\Gamma_{2,2}(\tau;t) \coloneqq \Gamma_{2,2}(\tau; \rho(t),iV(t))$, and after a simple redefinition of summation variables we arrive at
\begin{equation}\label{eq:G22t}
 \Gamma_{2,2}(\tau;t)\coloneqq \Gamma_{2,2}(\tau; \rho(t),iV(t)) = \sqrt{t_1 t_2 +t_1t_3+t_2t_3} \sum_{\substack{ p_1,p_2,p_3 \in\Lambda \\ p_1 +p_2 +p_3=0}}  \exp\Big( -\frac{\pi}{\tau_2} \sum_{j=1}^3 t_j |p_j|^2 \Big)\,,
\end{equation} 
where the sum is over three complex lattice momenta $p_i = {m}_i + {n}_i \tau \in  \Lambda = \mathbb{Z} +\tau\mathbb{Z}$, with $i=1,2,3$, subject to the conservation condition $p_1+p_2+p_3=0$.

 The integrand for the theta lift \eqref{eq:lift} can be re-expressed in terms of the $t_i$ variables in the form 
\bea
\cE^w_{i,j}(\tau) =   \int_{(\mathbb{R}^+)^3}\, V(t)^{w-4} A_{i,j}(t) \, \Gamma_{2,2}(\tau;t)\,{\rm d}^3 t\,.
\label{eq:intlatt}
\eea
 The functions $A_{i,j}(t)$ must be symmetric functions of $t_1$, $t_2$, $t_3$ and the integrand in \eqref{eq:intlatt} is invariant under the symmetric group  $\mathfrak{S}_3$, a property that is not immediately apparent in \eqref{eq:lift}.
The manifest invariance of the expression \eqref{eq:intlatt} under permutations of $t_1$, $t_2$ and $t_3$ is exhibited by \eqref{eq:lift}  in terms of the $\rho$ variable, which is invariant  under the six elements of transformation:
\begin{align}
\label{eq:symm}
\ttau  \to \ttau, \qquad  \ttau  \to 1-\ttau^{-1}, \qquad   \ttau  \to (1-\ttau)^{-1}, \qquad
\ttau \to 1- \bar \ttau,\qquad   \ttau \to  \bar \ttau^{-1}, \qquad  \ttau  \to  (1-\bar \ttau^{-1} )^{-1}\,.
\end{align} 
The first three of these correspond to transpositions of pairs of $t_i$ whereas the last three correspond to three-cycles with $t_1$, $t_2$, $t_3$ in different orders.

 The Laplace identities \eqref{eq:LapVar}  imply the relation
\begin{equation}
\Delta_\tau  \Gamma_{2,2}(\tau;t ) = \Delta_t  \Gamma_{2,2}(\tau;t)\,,
\end{equation}
where the laplacian in the $t_i$ variables is given by
\begin{equation} \label{eq:deltat}
\Delta_t [ F(t)] = -2\sum_{i=1}^3 \partial_i[t_i F(t)] + \sum_{i,j=1}^3 \partial_i \partial_j \{ [t_i t_j+(2\delta_{ij}-1) V^2(t)] F(t)\}\, .
\end{equation}
Importantly, we see that $\Delta_t [ V(t)^\alpha  F(t)] = V(t)^\alpha \Delta_t [F(t)]$ for all $\alpha \in \mathbb{R}$, with $V(t) = \sqrt{t_1t_2 +t_1t_3 +  t_2 t_3}$.

We will now discuss the particular  functions $A_{i,j}(\rho)$  that are invariant under the group of symmetries \eqref{eq:symm}, which will play a central role in the theta lift  \eqref{eq:lift}.

\subsection{Modular local polynomials and local Maass functions}
\label{sec:ModLoc}

We present here the systematic construction of a basis of $\Gamma_0(2)$-invariant functions  $A_{i,j}(\rho)$, given by Don Zagier  in unpublished notes that are expanded on in section {5.4} of \cite{DHoker:2018mys}. 
The function  $A_{1,0}(\ttau)$ originally arose in the construction of the modular invariant coefficient of the higher-derivative interaction $d^6R^4$ in the four-graviton amplitude of type IIB superstring theory  \cite{Green:2005ba,Green:2008bf}.
This coefficient function is  $\cE(4; \threeh,\threeh;\stau)$ in the notation of \eqref{eq:geneisen} \footnote{In \cite{Green:2005ba} the function $A_{1,0}(\ttau)$ was called  $A(\tau)$ and the coefficient function was called $\cE_{\threeh,\threeh}(\Omega,\bar \Omega)$. See also \cite{Green:2008bf} for alternative early notation.}. 

The functions $A_{i,j}(\rho)$ here considered  are particular examples of  `modular local Maass forms’, constructed from special modular forms known as modular local polynomials \cite{Bringmann,BringmannMaas}.
\vspace{0.2cm}

\textbf{Definition} \cite{Bringmann}: For an ${\rm SL}(2, \mathbb{Z})$-invariant nowhere dense set $E$, a function $\mathcal{P}:\mathfrak{H}\to \mathbb{C}$ is called a weight-$k$  modular local polynomial with exceptional set $E$ if (with $k$ even and non-positive) $\mathcal{P}$ is such that
\begin{itemize}
\item[(i)]For every $\gamma=( \begin{smallmatrix} a & b \\ c & d \end{smallmatrix}) \in  {\rm SL}(2, \mathbb{Z})$ we have $\mathcal{P}(\rho)\vert_k \gamma = \mathcal{P}(\rho)$, where the usual $\vert_k \gamma$ action is defined as 
\begin{equation}
\mathcal{P}(\rho)\big\vert_k \gamma = (c\rho+d)^{-k} \,\mathcal{P}\left(\frac{a\rho+b}{c\rho+d}\right)\,;
\end{equation}
\item[(ii)] On each connected component $\mathcal{C} \subset \mathfrak{H}\setminus E$ we have $\mathcal{P}(\rho) = { P}_\mathcal{C}(\rho)$ where ${ P}_\mathcal{C}$ is a polynomial in $\rho$.
\end{itemize}
Furthermore, as explained in \cite{Bringmann} we must impose some regularity condition so that the limiting values exist when $\rho \to \rho^*\in E$. Importantly, while the modular local polynomials are finite on the exceptional set $E$ their derivatives are discontinuous.

In \cite{BringmannMaas} modular local polynomials were introduced  for which the exceptional sets are
\begin{equation}
E_D \coloneqq \bigcup_{Q\in \mathcal{Q}_D} S_Q\,,\label{eq:ExSet}
\end{equation}
where $\mathcal{Q}_D$ denotes the set of binary quadratic forms with discriminant $D$, i.e. $Q=[a,b,c] \in \mathcal{Q}_D$ is such that $Q(X,Y) = [a,b,c](X,Y) \coloneqq aX^2 +b XY + cY^2$ with $a,b,c\in \mathbb{Z}$ and discriminant $D = b^2-4ac$.
The sets $S_Q$ are defined by
\begin{equation}
S_Q \coloneqq  \left\lbrace \rho \in \mathfrak{H}\,\Big\vert\,a|\rho|^2 + b\,{ \mbox Re}(\rho)+c =0\right\rbrace\,.
\end{equation}
For the case of interest the discriminant we want to consider is $D=1$. As we show in appendix \ref{app:ModPol}, the connected component $\mathcal{C}_{\rho^*} \subset \mathfrak{H}\setminus E_1$ which contains $\rho^* \coloneqq e^{\frac{\pi i }{3}}$ coincides with a fundamental domain of $\Gamma_0(2)$, i.e. 
\begin{equation}
 \mathcal{C}_{\rho^*}  = \Gamma_0(2) \backslash \mathfrak{H} \coloneqq \left\lbrace \rho \in \mathfrak{H}\,\Big\vert\, 0<{\mbox Re}(\rho) <1 \,,\,  |\rho-\frac{1}{2}|^2 >  \frac{1}{4}\right\rbrace\,.
\end{equation}
We can then think of these modular local polynomials as the modular analogues of the periodic seesaw function. Once we define the polynomial on one of the connected components, say e.g. $\mathcal{C}_{\rho^*}$, we extend it to a function on the whole upper-half plane $\mathfrak{H}$ by simply patching together all of its images under $\Gamma_0(2)$.

The function $A_{i,j}(\rho)$ in the integrand of the theta lift \eqref{eq:lift} is modular invariant, so it has zero holomorphic and anti-holomorphic weight with respect to all $\gamma \in \Gamma_0(2)$.  It can be constructed by  successive applications of the Maass raising operator $D_k$ to a modular polynomial  $\mathcal{P}(\rho)$ of even negative weight, where $D_k \coloneqq \p_\ttau + k/(\ttau-\bar \ttau)$  maps a modular form of weight $(k,0)$ to a modular form of weight $(k+2,0)$.
 More explicitly,  starting with a modular form of weight $-2n$  we can raise it to a real-analytic modular function by acting with the iterated Maass derivative operator  
\be
\label{eq:Dn}
D_{-2n}^{(n)} \coloneqq {(-2i)^n n! \over (2n)!} D_{-2} \circ D_{-4} \circ \cdots \circ D_{-2n+2} \circ D_{-2n}\,,
\ee
which can also be rewritten as
\be
D_{-2n}^{(n)} = {(-2i)^n n! \over (2n)!} \label{eq:Dn2}
\sum_{m=0}^n \begin{pmatrix} n \cr m \end{pmatrix}\frac{(-n-m)_m}{(\ttau -\bar \ttau )^m}\, \frac{\partial^{n-m}}{\partial \ttau ^{n-m}}\,,
\ee
where $(x)_m\coloneqq x(x+1)\dots (x+m-1)$ is the Pochhammer symbol.
We note  that the Maass operator $D_k$ satisfies the property $\Delta_{k+2} \cdot D_k - D_{k} \Delta_k = -k D_k$, where $\Delta_k\coloneqq 4 D_{k-2} \, \ttau_2^2 \p_{\bar \ttau}$ is the Laplacian acting on weight $k$ modular forms.
As a consequence, given that a modular local polynomial, $\mathcal{P}(\rho)$, of weight $k=-2n$ is a holomorphic function and therefore satisfies $\Delta_{-2n}\mathcal{P}(\rho) = 0$. It follows that $A(\rho) \coloneqq D_{-2n}^{(n)}  \mathcal{P}(\rho)$ is a modular function satisfying the Laplace eigenvalue equation
\begin{equation}
\left[ \Delta_\rho - n(n+1) \right] A(\rho) = 0\,,\label{eq:LapGen}
\end{equation}
where $\Delta_\rho = \Delta_0 =\rho_2^2(\partial_{\rho_1}^2+\partial_{\rho_2}^2)$.

As proved in \cite{Zagier2008, DHoker:2018mys}, the subspace of the modular local polynomials with exceptional set $E_1$ and weight $k$ that have been lifted as described above to be modular invariant functions with respect to $\Gamma_0(2)$ as well as being invariant with respect to the group of automorphism of $\mathfrak{H}$ generated by the involutions $\rho \to 1- \bar\rho$ and $\rho \to 1/\bar\rho$, is spanned by the basis:
\begin{equation}
A_{i,j}(\rho) \coloneqq D_{-2n}^{(n)} [ u^i v^j] \,,\label{eq:AijDef}
\end{equation}
where $i,j\in \mathbb{N}$ such that $k=-2n = -2(3i+j)$, and
\bea
u\coloneqq \ttau^2(1- \ttau)^2\, ,  \qquad v\coloneqq \ttau^2-\ttau+1 \,.
\label{eq:uvdef}
\eea

From \eqref{eq:Dn2} and \eqref{eq:LapGen}, it is easy to show that the $A_{i,j}(\ttau)$  are  Laurent polynomials in $\ttau_2$ with coefficients that are polynomial in $\ttau_1$, satisfying the Laplace equation
\bea
\label{DeltaA}
\left[\Delta_\rho   - s(s-1) \right]  A_{i,j}  (\ttau)=0\label{eq:lap}\,,
\eea
for all $\rho \in \Gamma_0(2)\backslash \mathfrak{H}$,  where $s=3i+j+1$.
For example, we have
\begin{align} \label{eq:Aij-eg}
A_{0,1}(\ttau)&\nn = \ttau_2+ \frac{\ttau_1(\ttau_1-1)+1}{\ttau_2}\,,\\
A_{1,0}(\ttau) &= \frac{\ttau_2 }{5} +\frac{[1+6\ttau_1(\ttau_1-1)]}{5 \ttau_2} + \frac{\ttau_1^2(\ttau_1-1)^2}{\ttau_2^{3}}\,,\\
A_{1,1}(\ttau) &\nn = \frac{\ttau_2^2}{7}+ \frac{3[4+15\ttau_1(\ttau_1-1)]}{35 }+ \frac{  3 \ttau_1 (\ttau_1-1) [ 5 \ttau_1 (\ttau _1-1) +3 ]+1}{7\ttau_2^2} +\frac{\ttau_1^2(\ttau_1-1)^2 [\ttau_1(\ttau_1-1) +1]}{\ttau_2^4}\,.
\end{align}
The particular theta lifts we want to consider are $\Gamma_{2,2}$ lifts of the modular-invariant local Maass forms $A_{i,j}(\rho)$, which are defined by  \eqref{eq:lift}.   \
As will be clarified in the next section, the integral in \eqref{eq:lift} does not quite converge if one uses the definition \eqref{eq:Gamma22} for $\Gamma_{2,2}$ as a complete lattice sum over $\mathbb{Z}^4$.  To obtain a convergent expression it is necessary to exclude the points $(m_1,n_1)=0$, $(m_2,n_2)=0$ and $(m_1+m_2,n_1+n_2)=0$ from the lattice sum.  In terms of the $t_i$ variables the expression for the theta lift \eqref{eq:intlatt} becomes 
\begin{equation}
\mathcal{E}^w_{i,j}(\tau)  \coloneqq \int_{( \mathbb{R}^+)^3 } [V(t)]^{w-4} A_{i,j}(\rho(t))  \,\Gamma_{2,2}'(\tau;t)\,  {\rm d}^3 t\,, \label{eq:Ewij2}
\end{equation}
where using \eqref{eq:G22t} we have defined
\begin{equation}\label{eq:G22tred}
 \Gamma_{2,2}'(\tau;t) \coloneqq \sqrt{t_1 t_2 +t_1t_3+t_2t_3} \sum_{\substack{ p_1,p_2,p_3 \in\Lambda' \\ p_1+p_2+p_3=0}}   \exp\Big( -\frac{\pi}{\tau_2} \sum_{j=1}^3 t_j |p_j|^2 \Big)\,,
\end{equation} 
and the sum runs over three lattice momenta $p_i = {m}_i + {n}_i \tau \in  \Lambda' = \mathbb{Z} +\tau\mathbb{Z}\setminus\{0\}$, with $i=1,2,3$, again subject to the conservation condition $p_1+p_2+p_3=0$.  The symbol $ \Lambda'$ indicates that we exclude the point $p=0$ from the lattice $p\in  \Lambda = \mathbb{Z} +\tau\mathbb{Z}$, which is a requirement for the convergence of the integral \eqref{eq:Ewij2}.
The objects of interest can then be rewritten in terms of the lattice sum integral
\begin{equation}
\mathcal{E}^w_{i,j}(\tau)  = \sum_{\substack{ p_1,p_2,p_3 \in\Lambda' \\ p_1+p_2 +p_3=0}} \int_{( \mathbb{R}^+)^3 } B^w_{i,j}(t)   \exp\Big( -\frac{\pi}{\tau_2} \sum_{j=1}^3 t_j |p_j|^2 \Big)\,  {\rm d}^3 t\,, \label{eq:Ewijt}
\end{equation}
where we have   defined the auxiliary functions 
\begin{equation}
B^w_{i,j}(t) \coloneqq  [V(t)]^{w-3} A_{i,j}(\rho(t)) \,.\label{eq:BwijDef}
\end{equation}
For example when expressed in terms of the $t_i$ variables, the $A_{i,j}(\rho)$ given in \eqref{eq:Aij-eg} become 
\begin{align} \label{eq:examplesAij}
 A_{0,1} (\rho(t)) &= \sigma_2^{-\half}  \sigma_1  \,  , \cr
  A_{1,0} (\rho(t)) &= \sigma_2^{-\threeh} \left(  \frac{1}{5} \sigma_1 \sigma_2-  \sigma_3 \right) \, , \\
 A_{1,1} (\rho(t)) &= \sigma_2^{-2} \left(  \frac{1}{7} \sigma_1^2 \sigma_2+\frac{2}{35}\sigma_2^2 - \sigma_1 \sigma_3 \right) \,, \nonumber
\label{eq:aijsome}
\end{align} 
where we have expressed the results  in terms of  the basis for symmetric polynomials in three variables $\sigma_1$, $\sigma_2$, and $\sigma_3$, defined as
\begin{equation}
\sigma_1 \coloneqq t_1+t_2+t_3\,,\qquad \sigma_2 \coloneqq V(t)^2 = t_1 t_2 +t_1 t_3 +t_2 t_3\,,\qquad \sigma_3\coloneqq t_1 t_2 t_3\, . \label{eq:symVar}
\end{equation} 

It is important to realise that whereas the reflection identity \eqref{eq:FuncRelGen} can be obtained by performing a quadruple Poisson resummation this procedure cannot be carried out  for the restricted lattice sum \eqref{eq:G22tred}.  However,
after correcting for the absence of the $(m_1,n_1)=(m_2,n_2)=(m_1+m_2,n_1+n_2)=(0,0)$ terms in the sums  the identity \eqref{eq:FuncRelGen} still holds modulo the addition of terms proportional to non-holomorphic Eisenstein series, namely we find
 \begin{equation}\label{eq:FuncEijw}
 \mathcal{E}^w_{i,j}(\tau)  = \mathcal{E}^{2-w}_{i,j}(\tau) +\Big( {\mbox{non-holomorphic Eisenstein series}}\Big)\,.
 \end{equation}
 
 We can now better clarify why the theta lift considered here is a suitable framework for discussing both  MGFs and SMFs. 
 Firstly, it follows directlly from \eqref{eq:Ewijt}  that  $\mathcal{E}^w_{i,j}(\tau)$ is closely related to MGFs \eqref{eq:Cabc} introduced in \cite{Green:1999pv}. If in equation \eqref{eq:Ewijt}  we replace the integrand $ B^w_{i,j}(t)$ by the simple monomial $t_1^{a-1} t_2^{b-1} t_3^{c-1} $ with $a,b,c\in \mathbb{N}$ we  deduce
 \begin{equation}
 \label{eq:cabcint}
\sum_{\substack{ p_1,p_2,p_3 \in\Lambda' \\ p_1+p_2 +p_3=0}} \int_{( \mathbb{R}^+)^3 } t_1^{a-1} t_2^{b-1} t_3^{c-1} \exp\Big( -\frac{\pi}{\tau_2} \sum_{j=1}^3 t_j |p_j|^2 \Big)\,  {\rm d}^3 t = \Gamma(a)\Gamma(b)\Gamma(c) \, C_{a,b,c}(\tau)\,.
\end{equation}
Here the parameters $t_i$ correspond  to the Schwinger parameters for the two-loop graph associated with the MGF, $C_{a,b,c}(\tau)$.

Similarly in the context of SMFs the integral representation \eqref{eq:Ewijt}  originates by considering the four-graviton two-loop scattering amplitude in eleven-dimensional supergravity compactified on a two-torus \cite{Green:2005ba}, where the $t_i$ variables are related to the Schwinger parameters for the two-loop Feynman diagrams. While in \cite{Green:2005ba} a prominent role was played by \eqref{eq:BwijDef} specialised to the case of the local Maass function $A_{1,0}(\rho)$, in \cite{Alday:2023pet} it was shown that at large-$N$ the expansion coefficient of a certain integrated correlator in $\mathcal{N}=4$ SYM can be represented as a rational linear combination of theta lifts \eqref{eq:Ewijt}  involving more general $A_{i,j}(\rho)$.
The precise connection between the theta lift \eqref{eq:Ewijt} and both MGFs and SMFs will be derived by analysing the Laplace equation satisfied by $\mathcal{E}^w_{i,j}(\tau)$.

In the next section we will study the real-analytic modular invariant functions $\mathcal{E}^w_{i,j}(\tau)$ of the variable $\tau \in \mathfrak{H}$ and show that they satisfy an inhomogeneous Laplace eigenvalue equation with sources given by bilinears in non-holomorphic Eisenstein series. 
By varying the integers $w,i,j$, we will show that  $\mathcal{E}^w_{i,j}(\tau)$ corresponds to either MGFs that appear in genus-one superstring amplitudes or to the SMFs that appear  in the low energy effective action of the scattering amplitudes in type IIB superstring theory and in the context of the large-$N$ expansion of integrated correlators in $\mathcal{N}=4$ SYM.

\section{{Inhomogeneous Laplace Equation}}
\label{sec:LapEq}

In this section we analyse the properties of the theta-lifted modular local Maass functions defined in \eqref{eq:Ewij2} by studying their behaviour under the action of the hyperbolic Laplace operator $\Delta_\tau$.
 
From their definition \eqref{eq:AijDef}, it follows that the functions $A_{i,j}(\ttau)$ are Laurent polynomials in $\ttau_2$ with coefficients that are polynomial in $\ttau_1$. We recall that  $A_{i,j}(\ttau)$ satisfies the  homogeneous Laplace eigenvalues equation \eqref{eq:lap} with respect to the Laplace-Beltrami operator $\Delta_\ttau = \ttau_2^2(\partial_{\ttau_1}^2 +\partial_{\ttau_2}^2)$
\begin{equation}
\label{DeltaOm}
\left[\Delta_\ttau   - s(s-1) \right]  A_{i,j}  (\ttau)=0\, ,
\end{equation}
where $\rho$ is inside the domain $\ttau \in \mathcal{C}_{\rho^*}= \Gamma_0(2) \backslash \mathfrak{H}$ and $s=3i+j+1$.

When translated into $t$-variables, the differential equation \eqref{DeltaOm} implies that the function $B_{i,j}^w(t)$ defined in \eqref{eq:BwijDef} satisfies another 
homogeneous Laplace eigenvalue equation
 \begin{equation}
\label{DeltaT}
\left[\Delta_t   - s(s-1) \right]  B^w_{i,j}  (t)=0\,,
\end{equation}
where again $s=3i+j+1$ and the laplacian $\Delta_t$ is defined in \eqref{eq:deltat}. 
Furthermore it is straightforward to show that
\begin{equation}\label{eq:DeltaTauOm}
 \Delta_\stau \exp\Big( - \frac{\pi}{\tau_2}\sum_{i=1}^3 t_i |p_i|^2  \Big) = \Delta_t \exp\Big( - \frac{\pi}{\tau_2}\sum_{i=1}^3 t_i |p_i|^2  \Big)\,,
\end{equation}
so by  acting with $\Delta_\stau$ on \eqref{eq:Ewij2} and combining \eqref{eq:DeltaTauOm} and  \eqref{DeltaT} we deduce
\begin{equation}
\big[ \Delta_\stau - s(s-1) \big] \cE^w_{i,j}(\stau) = \rm{\bt} \,,
\end{equation}
where `$\rm{\bt}$' denotes the boundary terms that arise from integrating the Laplacian $\Delta_t$ by parts.

We will shortly analyse these boundary terms and show that they produce the source terms for the inhomogeneous Laplace eigenvalue equation, 
\begin{align}
\left[\Delta_\stau -s(s-1) \right]  \cE_{i,j}^{w}(\stau) = \sum_{r=-\frac{(i+j-1)}{2}}^{\frac{(i+j-1)}{2}} b_{i,j}(w,r) E\left(\frac{w}{2}+r;\stau \right) E \left(\frac{w}{2}-r;\stau \right) +{\tilde d}_{i,j}(w) E(w;\stau)\,,\label{eq:LapcEijw}
\end{align}
for particular values of the coefficients $b_{i,j}(w,r)$ and  $\tilde{d}_{i,j}(w)$. 
Note that the source terms  in this equation have total ``transcendental weight''   $w=s_1+s_2$.
We also note that when the weight $w$ and the eigenvalue $s=3i+j+1$ have the same parity the source terms contain bilinears in non-holomorphic Eisenstein series with integer indices.  This is relevant to the discussion of MGFs for which $s\leq w-2$ so that the indices of the non-holomorphic Eisenstein series on the right hand-side of \eqref{eq:LapcEijw} are always integers greater or equal than two.
Conversely, when the $w$ and $s$ have opposite parity the source terms contain bilinears in non-holomorphic Eisenstein series with half-integer indices. This is the case that is relevant to the discussion of SMFs, in which the eigenvalues are such that $s\geq w+1$. Hence the indices of the non-holomorphic Eisenstein series in the bilinear term in \eqref{eq:LapcEijw} may have different signs.  
Nonetheless, it is always possible to use the functional equation 
\begin{equation}
\Gamma(s) E(s;\tau) = \Gamma(1-s)E(1-s;\tau)\,,
\end{equation}
to rewrite \eqref{eq:LapcEijw} so that all Eisenstein series in the source term appear with positive indices, although in that case the functional equation does not respect  uniform transcendentality in weight.

It is important to notice that for all the cases we are interested in the eigenvalue $s\neq w$ hence it is always possible to use \eqref{eq:EisenLap} and invert the laplacian\footnote{\label{footnote}We can check from the lattice sum representation \eqref{eq:Ewijt} that for the range of eigenvalues relevant for the SMFs the coefficient of the solution of the homogeneous equation  $E(s;\tau)$ vanishes. Similarly, for the case relevant for discussing MGFs we can directly evaluate \eqref{eq:Ewijt} to find a vanishing coefficient for the solution of the homogeneous equation  $E(s;\tau)$, apart from the case $s=1$ and $w$ odd where this solution reduces to a simple rational multiple of $\zeta(w)$ which can be computed  \cite{DHoker:2017zhq}. This implies that there is no issue in inverting the Laplace operator in \eqref{eq:LapcEijw}.} over the single Eisenstein series appearing as source to obtain 
\begin{equation} \label{eq:cijw}
\cE_{i,j}^{w}(\stau) = \sum_{r=-\frac{(i+j-1)}{2}}^{\frac{(i+j-1)}{2}} b_{i,j}(w,r) \, \cE\! \left(s; \frac{w}{2}+r,\frac{w}{2}-r;\stau \right) +d_{i,j}(w) E(w;\stau)\,,
\end{equation}
where $d_{i,j}(w) \coloneqq {\tilde d}_{i,j}(w)/[w(w-1)-s(s-1)]$.
We now describe how to compute the coefficients {$b_{i,j}(w,r)$} and $d_{i,j}(w)$ starting from the definition \eqref{eq:Ewij2} for the theta lift of local Maass functions.

\subsection{Boundary terms}
\label{sec:bc}

Applying the laplacian $\Delta_\stau$  to  \eqref{eq:Ewij2}  gives
\begin{align}  
\Delta_\tau  \cE^w_{i,j}(\stau) =  \sum_{\substack{ p_1,p_2,p_3 \in\Lambda' \\ p_1+p_2 +p_3=0}} \int_0^\infty   [V(t)]^{w-3}  \, A_{i,j}(\ttau(t))\, \Delta_t \exp\Big( - \frac{\pi}{\tau_2}\sum_{i=1}^3 t_i |p_i|^2  \Big)\, {\rm d}^3 t\,,
\label{eq:defEwij2}
\end{align}
where  we have used  \eqref{eq:DeltaTauOm}.  We integrate \eqref{eq:defEwij2} by parts and note that 
\begin{equation}
\Delta_t [V(t)^\alpha F(t)] = V(t)^\alpha \Delta_t [ F(t)] \, , \label{eq:deltatV}
\end{equation}
for all $\alpha$, hence by using the definition \eqref{eq:BwijDef} and the relation 
 \eqref{DeltaA} it follows that
\begin{equation}
\Delta_t\Big[ B^{w}_{i,j}(t)\Big] =  [V(t)]^{w-3} \Big[\Delta_\ttau A_{i,j}(\ttau)\Big]_{\ttau=\ttau(t)} = s(s-1) B^{w}_{i,j}(t)\, , 
\end{equation}
with $s=3i+j+1$. From this we we see that
\begin{align}
\Delta_\stau \cE^w_{i,j}(\stau) &\nn=  \sum_{\substack{ p_1,p_2,p_3 \in\Lambda' \\ p_1+p_2 +p_3=0}}\int_0^\infty B^{w}_{i,j}(t)\, \Delta_t\Big[ \exp\Big( - \frac{\pi}{\tau_2}\sum_{i=1}^3 t_i |p_i|^2  \Big) \Big] {\rm d}^3 t\\
&\nn=  \sum_{\substack{ p_1,p_2,p_3 \in\Lambda' \\ p_1+p_2 +p_3=0}} \int_0^\infty  \Delta_t\big[B^{w}_{i,j}(t)\big]\,  \exp\Big( - \frac{\pi}{\tau_2}\sum_{i=1}^3 t_i |p_i|^2  \Big)  {\rm d}^3 t+{\rm{\bt}}\\
& = s(s-1)  \cE^w_{i,j}(\stau) +{\rm{\bt}}\, ,\label{eq:LapIn}
\end{align}
where `${\rm{\bt}}$' represents  boundary terms. 

The  boundary terms  in \eqref{eq:LapIn} are easily collected by integrating $\Delta_t$ by parts using the definition \eqref{eq:deltat}.
We start by noting that since all the momenta $p_1,p_2,p_3=-p_1-p_2$ are non-vanishing we only need to worry about boundary contributions coming from $t_i\to0$ since the integral is exponentially suppressed along any direction $t_i\to \infty$.
Using the fact that the integrand is invariant under permutations of $(t_1,t_2,t_3)$ we arrive at the expression
\begin{align}
{\rm{\bt}} = &\nn \sum_{\substack{ p_1,p_2,p_3 \in\Lambda' \\ p_1+p_2 +p_3=0}} 3 \int_0^\infty  \,\Big[t_1t_2(\partial_3 -\partial_1-\partial_2) B^{w}_{i,j}(t_1,t_2,t_3)\Big]_{t_3=0} \exp\Big(\! -\frac{ \pi t_1 |p_1|^2}{\tau_2}-\frac{ \pi t_2 |p_2|^2}{\tau_2}\Big){\rm d}^2t\\
&+\sum_{p\in \Lambda' } \frac{6\pi |p|^2}{\tau_2} \int_0^\infty   \Big[t_1t_2B^{w}_{i,j}(t_1,t_2,t_3)\Big]_{t_3=0} \exp\Big(-\frac{\pi |p|^2 (t_1+t_2)}{\tau_2}\Big){\rm d}^2t\,,\label{eq:BT}
\end{align}
where we have defined $p=m+n \tau\in \Lambda'$ with $(m,n) \in \mathbb{Z}^2 \setminus \{ (0,0)\}$.
We now show that  with the definition of $B^w_{i,j}(t)$ given earlier  that the first term produces a bilinear in Eisenstein series, while the second term produces an expression  linear in Eisenstein series.

Although we do not have a closed formula for the complete boundary terms associated with a general $B^w_{i,j}(t)$, we can make some general observations.
It follows from  \eqref{eq:AijDef} that when $A_{i,j}(\rho)$ is rewritten in terms of  $t_i$ via~\eqref{eq:rhodef} it can be expressed in terms of the basis  $\sigma_1,\sigma_2,\sigma_3$  for symmetric polynomials in three variables defined in \eqref{eq:symVar}. Using these variables, $A_{i,j}(\rho)$ takes the following general form 
\begin{equation}\label{eq:Aijexp}
A_{i,j}(\ttau(t)) = \sigma_2^{-\frac{s-1}{2}}\!\!\! \sum_{\substack{  \alpha,\beta,\gamma \geq 0\\ \alpha+2\beta+3\gamma = s-1 }}\!\! c(\alpha,\beta,\gamma)\,  \sigma_1^\alpha  \sigma_2^{\beta}  \sigma_3^\gamma\,,
\end{equation}  
 where the coefficients $c(\alpha,\beta,\gamma)\in \mathbb{Q}$ implicitly depend on $i,j$ (and  $s=3i+j+1$). 
We note that when $s$ is even the overall factor is an half-integer power of $\sigma_2$ while for $s$ odd the power becomes integer.  The condition on the sum that {constrains} the powers to satisfy $\alpha+2\beta+3\gamma=s-1$ comes from the fact that the $A_{i,j}(\rho)$ are homogeneous functions of $t_i$ since $\ttau_1$ and $\ttau_2$ defined in \eqref{eq:rhodef} are.
From  \eqref {eq:rhodef} we also deduce that the transformation $\ttau\to \bar\ttau$ is equivalent to $t_i\to -t_i$, or equivalently 
\ie
\sigma_1\to -\sigma_1\, , \qquad \sigma_2\to \sigma_2\, , \qquad \sigma_3 \to -\sigma_3 \, , 
\fe 
hence using \eqref{eq:Dn2} and \eqref{eq:AijDef} we see that 
\begin{align}
&\nn A_{i,j}(\bar\ttau)  = (-1)^{i+j} A_{i,j}(\ttau) = (-1)^{s-1} A_{i,j}(\ttau)\\
 \Longrightarrow \qquad &A_{i,j}(\ttau(-t)) = A_{i,j}(\bar{\ttau}(t)) = (-1)^{s-1} A_{i,j}(\ttau(t))\,.
\end{align}
Using the expansion \eqref{eq:Aijexp} this implies
\begin{equation}
c(\alpha,\beta,\gamma) = 0 \, ,  \qquad {\rm for}\qquad \alpha+\gamma \not\equiv (s-1)\, {\rm mod}\,2\,. \label{eq:condmod2}
\end{equation}

For example we can derive formulae for infinite families of $A_{i,j}(\rho)$ at fixed low value of $i$ and arbitrary $j$.
From \eqref{eq:AijDef} we can then prove by induction over $j$ that
\begin{align}
\label{eq:A0j} \sigma_2^{\frac{j}{2}} A_{0,j}(\rho(t))  = {\sigma_1}^j\, _2F_1\left(\tfrac{1}{2}-\tfrac{j}{2},-\tfrac{j}{2};\tfrac{1}{2}-j \vert \tfrac{3 {\sigma_2}}{{\sigma_1}^2}\right)\,,
\end{align}
and
\begin{align}
&\nn \sigma_2^{\frac{j+3}{2}} A_{1,j}(\rho(t))=\\
&\nn \frac{j (j-1)(j-2) {\sigma_1}^{j-3} {\sigma_2^3} \, _2F_1\left(\tfrac{3-j}{2},2-\tfrac{j}{2};\tfrac{1}{2}-j\vert \tfrac{3 {\sigma_2}}{{\sigma_1}^2}\right)}{(2 j+1) (2 j+3) (2 j+5)} +\frac{j (j+4) {\sigma_1}^{j-1} {\sigma_2}^{2} \, _2F_1\left(\tfrac{1-j}{2},1-\tfrac{j}{2};-j-\tfrac{1}{2}\vert \tfrac{3 {\sigma_2}}{{\sigma_1}^2}\right)}{(2 j+3) (2 j+5)}\\
&\label{eq:A1j}+\frac{{\sigma_1}^{j+1} {\sigma_2} \, _2F_1\left(-\tfrac{j}{2}-\tfrac{1}{2},-\tfrac{j}{2};-j-\tfrac{3}{2}\vert\tfrac{3 {\sigma_2}}{{\sigma_1}^2}\right)}{(2 j+5)}-{\sigma_3} {\sigma_1}^j  \, _2F_1\left(\tfrac{1-j}{2},-\tfrac{j}{2};-j-\tfrac{5}{2}\vert\tfrac{3 {\sigma_2}}{{\sigma_1}^2}\right)\, .
\end{align}
We note that all the hypergeometric functions $_2F_1(a,b;c\vert z)$ that arise are actually polynomials in their argument $z$.

We can then focus on the integrand \eqref{eq:BwijDef} and expand it as
\begin{equation}
B^{w}_{i,j}(t_1,t_2,t_3) = \sigma_2^{\frac{w-s-2}{2}}\!\!\! \sum_{\substack{  \alpha,\beta,\gamma \geq 0\\ \alpha+2\beta+3\gamma = s-1 }}\!\! c(\alpha,\beta,\gamma)\,  \sigma_1^\alpha  \sigma_2^{\beta}  \sigma_3^\gamma\,.\label{eq:Bijwexp}
\end{equation}
From this equation it is rather easy to compute the boundary contribution \eqref{eq:BT} where only the terms with $\gamma=0$ and $\gamma=1$ contribute.
Therefore, for the purpose of extracting the source terms in the differential equation we can restrict our analysis of \eqref{eq:Bijwexp} to terms constant or linear in $\sigma_3$.
In general, these terms take the form
\begin{align}
&\notag B^{w}_{i,j}(t_1,t_2,t_3)= \\
&\label{eq:Bsmalls3}  \sigma_2^{\frac{w-s-2}{2}} \left[ \left( \sum_{n=0}^{\lfloor \frac{i+j}{2}\rfloor} c^{(0)}_{i,j}(n)\sigma_1^{i+j-2n} \sigma_2^{i+n}\right) +\sigma_3 \left( \sum_{n=0}^{\lfloor \frac{i+j-1}{2}\rfloor} c^{(1)}_{i,j}(n)\sigma_1^{i+j-2n-1} \sigma_2^{i+n-1}\right)+O(\sigma_3^2) \right]\,,
\end{align}
where we compactly denoted the coefficients of \eqref{eq:Bijwexp} as $c^{(\ell)}_{i,j}(n) \coloneqq c(i+j-2n-\ell,i+n-\ell, \ell)$.

Although the coefficients $c^{(\ell)}_{i,j}(n)$ are not known in closed form for arbitrary $i,j$ we have found expressions for infinite families at fixed $i$ or at fixed $j$.
For example for the infinite family $B^w_{0,j}(t)$ from \eqref{eq:A0j} we find
\begin{align}
c^{(0)}_{0,j}(n) &\label{eq:c0j0} = \frac{\Gamma (j+1) }{\Gamma \left(j+\frac{1}{2}\right) }\times  \frac{ (- \frac{3}{4})^n  \Gamma \left(j-n+\frac{1}{2}\right)}{\Gamma(n+1) \Gamma (j-2 n+1)}\,,
\end{align}
with $c^{(\ell >0)}_{0,j}(n)=0$. Similarly from \eqref{eq:A1j} we deduce that for the family $B^w_{1,j}(t)$,
\begin{align}
c^{(0)}_{1,j}(n) &\label{eq:c1j0} = \frac{\Gamma (j+1) }{\Gamma \left(j+\sevenh\right) }\times \frac{(- \frac{3}{4})^n \Gamma \left(j-n+\fiveh\right)}{18\Gamma (n+1)  \Gamma (j-2 n+2)} [\,j (9-6 n)+4 (n-7) n+9]\,,\\
c^{(1)}_{1,j}(n) &\label{eq:c1j1} = -\frac{\Gamma (j+1) }{\Gamma \left(j+\sevenh\right) }\times  \frac{(- \frac{3}{4})^n \Gamma \left(j-n+\sevenh\right)}{ \Gamma (n+1)\Gamma (j-2 n+1)}\,,
\end{align}
with $c^{(\ell >1)}_{1,j}(n)=0$. While for $B^w_{i,0}(t)$ we have
\begin{align}
c^{(0)}_{i,0}(n) &\label{eq:ci00}= \frac{\sqrt{\pi }  \,\Gamma (i+1)^2}{3^{3i} \Gamma \left(i+\frac{1}{6}\right)   \Gamma \left(i+\frac{5}{6}\right)} \times\frac{2^{2 n+1}  \Gamma \left(2 i-n+\frac{1}{2}\right)}{  \Gamma (2 n+1) \Gamma (i-2 n+1) \Gamma (i-n+1)}\,, \\
c^{(1)}_{i,0}(n) &\label{eq:ci01} = -\frac{\sqrt{\pi }  \,\Gamma (i+1)^2}{3^{3i} \Gamma \left(i+\frac{1}{6}\right)   \Gamma \left(i+\frac{5}{6}\right)} \times \frac{2^{2n+2}\Gamma \left(2 i-n+\threeh\right)}{ \Gamma (2 n+2) \Gamma (i-2 n) \Gamma (i-n+1)}\,,
\end{align}
where now in general $c^{(\ell >1)}_{i,j}(n)\neq0$ for $i\geq 2$.

We can then focus on the boundary terms produced by a single monomial contribution to $B_{i,j}^w(t_1,t_2,t_3)$, i.e. rather than the sum over $n$ in \eqref{eq:Bsmalls3} we consider the individual term,
\begin{equation}
 B_{i,j}^w(t_1,t_2,t_3)\rightarrow  c^{(0)}_{i,j}(n)\sigma_1^{i+j-2n} \sigma_2^{\frac{w-s-2}{2}+i+n} + c^{(1)}_{i,j}(n)\sigma_1^{i+j-2n-1} \sigma_2^{\frac{w-s-2}{2}+i+n-1} \sigma_3 \,,
\end{equation}
and substitute it into \eqref{eq:BT} to find 
\begin{align}
{\rm{\bt}}  &= \nn    \sum_{\substack{ p_1,p_2,p_3 \in\Lambda' \\ p_1+p_2 +p_3=0}}  3 (2n-i-j) c_{i,j}^{(0)}(n)\int_0^\infty  \,(t_1 t_2)^{\frac{w-s}{2}+i+n}(t_1+t_2)^{i+j-2n-1}  e^{ -\frac{ \pi t_1 |p_1|^2}{\tau_2}-\frac{ \pi t_2 |p_2|^2}{\tau_2}}{\rm d}^2t \\
&+\nn \sum_{\substack{ p_1,p_2,p_3 \in\Lambda' \\ p_1+p_2 +p_3=0}} 3  c_{i,j}^{(1)}(n) \int_0^\infty  (t_1 t_2)^{\frac{w-s}{2}+i+n}(t_1+t_2)^{i+j-2n-1}  e^{-\frac{ \pi t_1 |p_1|^2}{\tau_2}-\frac{ \pi t_2 |p_2|^2}{\tau_2}}{\rm d}^2t\\
&+\sum_{p\in \Lambda' } \frac{6\pi |p|^2}{\tau_2}    c_{i,j}^{(0)}(n)\int_0^\infty (t_1 t_2)^{\frac{w-s}{2}+i+n}(t_1+t_2)^{i+j-2n} e^{-\frac{\pi |p|^2 (t_1+t_2)}{\tau_2}}{\rm d}^2t  \, ,\label{eq:BT1}
\end{align}
once again in the above we have only focused a single monomial contribution to the boundary term.

We see from \eqref{eq:Bsmalls3} that in the above expression the exponent $\alpha$ of each term $(t_1+t_2)^\alpha$ is always a positive integer, hence we can use binomial expansion and integrate term by term, arriving at the sum of three boundary contributions, 
\begin{align}
\nn{\rm{\bt}}   &=    3 \big[ (2n {-} i {-} j) c_{i,j}^{(0)}(n)+c_{i,j}^{(1)}(n)\big]  \sum_{m=0}^{i+j-2n-1} {i{+}j{-}2n{-}1 \choose m} \Gamma\big(\tfrac{w+(i+j-2 m-2 n-1)}{2} \big)\Gamma\big(\tfrac{w-(i+j-2 m-2 n-1)}{2} \big) \\
&\notag  \times \Big[E(\tfrac{w+(i+j-2 m-2 n-1)}{2} ;\tau) E(\tfrac{w-(i+j-2 m-2 n-1)}{2} ;\tau){-}E(w;\tau)\Big]\\
&  +6 c_{i,j}^{(0)}(n) E(w;\tau)\Big[ \sum_{m=0}^{i+j-2n}   {i{+}j{-}2n \choose m}\Gamma\left(\tfrac{w+1+(i+j-2 m-2n)}{2} \right)
\Gamma\left(\tfrac{w+1-(i+j-2 m-2n)}{2}\right) \Big]   \,,\label{eq:sources}
\end{align}
 where we have used the identity
\begin{align}
\sum_{\substack{ p_1,p_2,p_3 \in\Lambda' \\ p_1+p_2 +p_3=0}}  \frac{(\tau_2/\pi)^{s_1+s_2}}{ |p_1|^{2s_1} |p_2|^{2s_2}} &= \sum_{p_1,p_2\in \Lambda' }  \frac{(\tau_2/\pi)^{s_1+s_2}}{ |p_1|^{2s_1} |p_2|^{2s_2}} - \sum_{p\in \Lambda'}  \frac{(\tau_2/\pi)^{s_1+s_2}}{ |p|^{2(s_1+s_2)}} \cr
&\label{eq:latticeId}= E(s_1;\tau)E(s_2;\tau)- E(s_1+s_2;\tau)\,.
\end{align}
Note that we have substituted for the eigenvalue $s=3i+j+1$ in  \eqref{eq:sources} and we can easily see that the indices of the bilinear in non-holomorphic Eisenstein series  are such that $w\pm (i+j -1)\equiv  w - s\, ({\rm mod}\,2)$. Hence, we deduce that when $w$ and $s$ have the same parity the bilinear source terms is the product of two non-holomorphic Eisenstein series with integer indices. Similarly, when $w$ and $s$ have opposite parity the bilinear non-holomorphic Eisenstein series have half-integer indices. Finally, the third line gives terms proportional to the non-holomorphic Eisenstein series $E(w;\tau)$. Thus we have reproduce the claim \eqref{eq:LapcEijw}.

To obtain the complete boundary term we simply need to sum over the boundary terms in \eqref{eq:sources} arising from \eqref{eq:Bsmalls3}. After changing summation variables $(n,m)\to (n,r{\coloneqq} n{+}m - \tfrac{1}{2}(i+j-1))$ and inverting the Laplace equation \eqref{eq:LapIn} (with the caveat explained in footnote \ref{footnote}) we arrive at 
\begin{align}
& \cE_{i,j}^w(\tau) =  \sum_{r=-\frac{(i+j-1)}{2}}^{\frac{(i+j-1)}{2}} b_{i,j}(w,r) \cE(s;\tfrac{w}{2}+r  ,\tfrac{w}{2}-r ;\tau)  + d_{i,j}(w)E(w;\tau)\,,
\label{eq:Eijsolved}
\end{align}
where the coefficients $b_{i,j}(w,r)$ and $d_{i,j}(w)$ are given by 
\begin{align}
\notag b_{i,j}(w,r) =& \Gamma \left(\frac{w}{2}+r \right)\Gamma \left(\frac{w}{2}-r \right)\times\\
&\label{eq:bij}\left[  \sum_{n=0}^{{\rm min}(\tfrac{i+j-1}{2}-r,\tfrac{i+j-1}{2}+r)}3 \big[ (2n-i-j) c_{i,j}^{(0)}(n)+c_{i,j}^{(1)}(n)\big]  {i{+}j{-}2n{-}1 \choose\frac{1}{2} (i+j-1)+ r -n}\right]\,,\\
 d_{i,j}(w)=&\notag - \sum_{r=-\frac{(i+j-1)}{2}}^{\frac{(i+j-1)}{2}} \frac{b_{i,j}(w,r)}{[w(w-1)-s(s-1)]}+   \sum_{r=-\frac{(i+j)}{2}}^{\frac{(i+j)}{2}}\Big\{ \frac{6  \Gamma \left(\frac{w+1}{2}+r \right)\Gamma \left(\frac{w+1}{2}-r \right)}{[w(w-1)-s(s-1)]}\times\\
 &\label{eq:dij} \Big[ \sum_{n=0}^{{\rm min}(\tfrac{i+j}{2}-r,\tfrac{i+j}{2}+r)}  c_{i,j}^{(0)}(n)  {i{+}j{-}2n \choose\frac{1}{2} (i+j)+ r -n}\Big] \Big\}\,.
\end{align}

Note that the GES terms in  \eqref{eq:Eijsolved} are invariant  under the exchange $w\to 2-w$ as one can check from the coefficients \eqref{eq:bij} combined with the reflection property for the GESs
\begin{equation}
\Gamma(s_1)\Gamma(s_2)\cE(s;s_1,s_2;\tau) = \Gamma(1-s_1)\Gamma(1-s_2)\cE(s;1-s_1,1-s_2;\tau)\,, \label{eq:GenEisFunct}
\end{equation}
which can be inferred from its differential equation \eqref{eq:geneisen} combined with the functional identity  \eqref{eq:FuncRelGen}.
On the other hand, the single Eisenstein series term in  \eqref{eq:Eijsolved} is not invariant as one can see from the coefficient  \eqref{eq:dij} and the result is the modified functional equation \eqref{eq:FuncEijw} for $\cE^w_{i,j}(\tau)$.

In the next sections \ref{sec:Alt} and \ref{sec:GenEij} we provide an alternative way of determining the coefficients $b_{i,j}(w,r)$ of the GES part of \eqref{eq:Eijsolved} for general values of $i,j\in \mathbb{N}$. Importantly, we will see that this second method does not rely on knowing the general form for the coefficients $c^{(0)}_{i,j}(n)$ and $c^{(1)}_{i,j}(n)$ in \eqref{eq:Bsmalls3}.

\subsection{Alternative approach}
\label{sec:Alt}

In this section we discuss an alternative method, originally presented in \cite{Green:2008bf}, for obtaining the source terms in the Laplace equation \eqref{eq:LapIn}.
Although this method does not fully capture the single Eisenstein series contribution in \eqref{eq:sources}, it leads to a particularly efficient procedure for determining the explicit coefficients $b_{i,j}(w,r)$ for generic values of parameters, $i$, $j$, $w$, $r$.

For this calculation we make use of the formulation for the theta lift of modular local polynomials in terms of the original variables $\rho$ and $V$.
As discussed just below equation  {\eqref{eq:intlatt}},
 the Maass functions $A_{i,j}(\rho)$ are invariant under the  permutation group of three elements $\mathfrak{S}_3$ as is manifest in terms of the $t_i$ variables.
In particular, the $A_{i,j}(\rho)$ are invariant under $\rho \to 1-\bar{\rho}$. It is important to note that  the symmetry $\ttau_1\mapsto 1-\ttau_1$, implies that the derivative $\partial_{\ttau_1} A_{i,j}(\ttau)$  satisfies the following conditions on the boundary of the  shaded domain $\cF_{\rho} = \Gamma_0(2) \backslash \mathfrak{H}$ in figure~\ref{fig:fundomain},
\bea
\partial_{\ttau_1} A_{i,j}(\ttau) |_{\ttau_1=0}= - \partial_{\ttau_1} A_{i,j}(\ttau)  |_{\ttau_1=1}\,.
\label{eq:t1deriv}
\eea

It is convenient to use the symmetries of $A_{i,j}(\ttau)$ to map the original domain $\mathcal{F}_\rho$ into a three-fold covering of the fundamental domain of ${\rm SL}(2,\mathbb{Z})$ given by $\hat{\cF}_\rho = {\rm SL}(2,\mathbb{Z}) \backslash \mathfrak{H}$. In that case the two vertical boundary lines in $ \cF_\rho$,  $\ttau_1=0$ and $\ttau_1=1$, are respectively mapped to the {lines $\ttau_1=0-\epsilon$ and  $\ttau_1=0+\epsilon$ in $\hat \cF_\rho$, with $\epsilon\to0^+$.} If we now define $ \hat A_{i,j}(\ttau)$ as the mapping of the local Maass function $A_{i,j}(\rho)$ to $\hat{\cF}_\rho$, we see that  the relation \eqref{eq:t1deriv}                                                                                                                                                                                                                                                                                                                                                                  
implies the existence of a discontinuity in $\partial_{\ttau_1}\hat A(\ttau)$ at $\ttau_1=0$.  {It therefore follows from $\partial^2_{\ttau_1} |\ttau_1| = 2 \delta(\ttau_1)$} that inside the domain  $\hat{\cF}_\rho$ we have
\bea
\ttau_2^2\, \partial^2_{\ttau_1} \hat A_{i,j} (\ttau) =2 \ttau_2\,  c_{i,j}(\ttau_2) \, \delta(\ttau_1)\, , 
\label{eq:newA}
\eea
where $c_{i,j}(\ttau_2) = c_{i,j}(1/\ttau_2)$ is a Laurent polynomial in $\ttau_2$ that will play a key role in the following. 

A crucial consequence of this fact is that on the fundamental domain $\hat{\cF}_\rho$ the Laplace equation \eqref{DeltaA} is modified to  
\bea
\label{DeltaAhat}
\left[\Delta_\rho   - s(s-1) \right]  \hat A_{i,j}  (\ttau)=  2  \ttau_2\,  c_{i,j}(\ttau_2) \, \delta(\ttau_1)\,.
\eea
The contribution of the boundary terms to the Laplace equation  may then be determined by an argument similar to \eqref{eq:defEwij2}.   After changing the integration domain in the theta lift \eqref{eq:Ewij2}  from $\cF_{\rho}$ to its triple cover $\hat{\cF}_\rho$, acting with the $\tau$ Laplacian gives
\begin{align}
\Delta_\tau \mathcal{E}^w_{i,j}(\tau)  &= 6 \int_{ \mathbb{R}^+ \times \hat{\cF}_\rho } V^{w} \hat{A}_{i,j}(\rho)  \,\Delta_\rho\Gamma_{2,2}'(\tau;\rho,i V)\,2  \frac{{\rm d}V}{V^2} \frac{{\rm d}^2 \rho}{\rho_2^2}\,,
\end{align}
where  \eqref{eq:LapVar} has been used.
At this point integrating by parts and using \eqref{DeltaAhat} leads to 
\begin{align}
[\Delta_\tau-s(s-1)] \mathcal{E}^w_{i,j}(\tau)  &\sim 6 \int_{ \mathbb{R}^+ \times [1,\infty) } V^{w} c_{i,j}(\ttau_2) \Gamma_{2,2}'(\tau;i\rho_2,i V)\,2  \frac{{\rm d}V}{V^2} \frac{{\rm d} \rho_2}{\rho_2}\,,
\label{eq:lapeq1}
\end{align}
where we have discarded contributions  on the right-hand side, which are linear combinations of non-holomorphic Eisenstein series.

By performing the change of variables  $\ttau_2 = \sqrt{y/x}$, $V=\sqrt{x y}$  equation \eqref{eq:lapeq1} becomes (ignoring a subtlety in the limits of integration to be described below),
 \begin{equation}
[\Delta_\tau-s(s-1)] \mathcal{E}^w_{i,j}(\tau)  \sim 3\sum_{ p_1,p_2 \in\Lambda' } \int_0^\infty  x^{\frac{w-2}{2}} y^{\frac{w-2}{2}}\,c_{i,j}(\sqrt{y/x}) \,e^{- \pi x \frac{|p_1|^2}{\tau_2}- \pi y \frac{|p_2|^2}{\tau_2}} \,{\rm d} x\,{\rm d} y \,.\label{eq:sourcecij}
\end{equation}
In writing this expression we have assumed that the integration over $\ttau_2$ can be extended from the range $[1,\infty)$ to the semi-infinite line $[0,\infty)$ by exploiting symmetry of the integrand under $\ttau_2\to 1/\ttau_2$.
However, this ignores a subtlety that arises from the fact that  the manifest $\mathfrak{S}_3$ symmetry of  \eqref{eq:G22tred} has been lost in mapping the domain $\cF_\rho$ to $\hat\cF_\rho$ and the symmetry $\ttau_2\to 1/\ttau_2$ only exists if we remove the restriction $p_1+p_2\neq 0$.  In other words, the expression \eqref{eq:sourcecij} includes the $p_1+p_2=0$ term but the $p_1= 0$ and $p_2=0$ terms must be omitted   in order to avoid obvious divergences in the expression for the source term.  So in order to maintain $\mathfrak{S}_3$ symmetry  we must subtract the $p_1+p_2=0$ term from \eqref{eq:sourcecij} even though its presence does not cause a divergence.
From \eqref{eq:latticeId} it is easy to see that subtracting the $p_1+p_2=0$ term adds a term proportional to a single non-holomorphic Eisenstein series, while leaving the bilinear source term unaffected.

From   \eqref{eq:AijDef}  the Laurent polynomial $c_{i,j}(\rho_2)$ takes the form
\begin{equation}
c_{i,j}(\rho_2) = \sum_{r=-\frac{(i+j-1)}{2}}^{\frac{(i+j-1)}{2}}s_{i,j}(r) \,\rho_2^{2r} \,.\label{eq:cijrho}
\end{equation}
Importantly, we see that each monomial in the above expression leads to a particular bilinear in Eisenstein series in \eqref{eq:sourcecij},
\begin{equation}
\left[\Delta_\tau-s(s-1) \right] \mathcal{E}^w_{i,j}(\tau)  \sim  \sum_{r=-\frac{(i+j-1)}{2}}^{\frac{(i+j-1)}{2}} 3s_{i,j}(r)  \Gamma \left(\frac{w}{2}+r \right)\Gamma \left(\frac{w}{2}-r \right) E(\frac{w}{2} +r;\tau) E(\frac{w}{2} -r;\tau)\,,\label{eq:source2}
\end{equation}
{where we have made use of the integral representation for the non-holomorphic Eisenstein series \eqref{eq:Eisenint}. }
Comparing this equation to \eqref{eq:Eijsolved} we deduce that the coefficients $b_{i,j}(w,r)$ given in \eqref{eq:bij} can be written in  the simpler form
\begin{equation}
b_{i,j}(w,r) = 3s_{i,j}(r)  \Gamma \left(\frac{w}{2}+r \right)\Gamma \left(\frac{w}{2}-r \right)\,.\label{eq:bijcij}
\end{equation} 

In the next section this alternative computation of the source term \eqref{eq:source2} will be used to derive a general expression valid for all $i,j\in \mathbb{N}$ of the coefficients $b_{i,j}(w,r)$ of the GES part of \eqref{eq:Eijsolved}.

\subsection{General expression for $\cE^w_{i,j}(\tau)$}
\label{sec:GenEij}

We now exploit the analysis of section \ref{sec:Alt} to deduce a general expression for the source term \eqref{DeltaAhat} generated by the modular local Maass function $A_{i,j}(\rho)$ defined in \eqref{eq:AijDef} and hence derive an explicit formula for the particular linear combinations of GESs produced by $\cE^w_{i,j}(\tau)$.
Thanks to \eqref{eq:cijrho} and \eqref{eq:bijcij}, this will in turn allow us to determines the coefficients $b_{i,j}(w,r)$ of the GES part of \eqref{eq:Eijsolved}.

The coefficient $s_{i,j}(r)$ appearing in \eqref{eq:cijrho} for a particular $A_{i,j}(\rho)$, can be obtained by expanding the corresponding modular local polynomial, $u^i v^j $, as a polynomial in $\rho$, 
\ie
u^i v^j = \rho^{2i}(1-\rho)^{2i} ( \rho^2-\rho+1)^j =  \sum _{k=0}^{2 i}  \sum _{\ell =0}^j \sum _{m=0}^\ell  (-1)^{k+m+\ell } C_{i,j; k, \ell, m} \, \rho^{2i +k+\ell+m  }\, ,\label{eq:uvExp}
\fe
where $C_{i,j; k, \ell, m}$ are simply the binomial coefficients, given by 
\ie \label{eq:Cijklm}
C_{i,j; k, \ell, m} \coloneqq \binom{2 i}{k}
   \binom{j}{\ell} \binom{\ell}{m}  \, . 
   \fe
The task of computing the source term \eqref{eq:newA} for a fixed $A_{i,j}(\rho)$ is  then simplified to determining the source contribution that originates from a single monomial term $\rho^{2i +k+ m+\ell }$.
Firstly, we need to use the definition \eqref{eq:Dn2} to perform the Maass raising \eqref{eq:AijDef}
and compute 
\begin{equation}
D^{(n)}_{-2n} [\rho^{q}]  = {(-2i)^n n! \over (2n)!} 
\sum_{m=0}^n \begin{pmatrix} n \cr m \end{pmatrix}\frac{(-n-m)_m}{(2i \rho_2 )^m}\, \frac{\Gamma(q+1)}{\Gamma(q{+}m{-}n{+}1)}\,\ttau^{q+m-n}\,,
\end{equation} 
where the integer $n$ is related to the eigenvalue $s$ via $n= s-1 = 3i+j$.
 We then use \eqref{eq:newA} to derive
 \begin{align}
 \ttau_2^2\, \partial^2_{\ttau_1} \Big( D^{(n)}_{-2n} [\rho^{q}]\Big)&\nn=2 \ttau_2\,  \frac{ (-1)^{n+\frac{q+3}{2}} \Gamma (n{+}1) \Gamma (q{+}1) \rho_2^{q-n}}{\Gamma (2 n{+}1)} \Big[\sum_{m=0}^n \frac{(-1)^m 2^{n-m} \Gamma (m{+}n{+}1)}{\Gamma (m {+} 1) \Gamma (n{-}m{+}1) \Gamma (m{-}n{+}q)} \Big] \, \delta(\ttau_1)\\
 &= 2\ttau_2  \left[ \left(1-(-1)^q \right)  \frac{\Gamma \left(\frac{q}{2} {+} 1\right) \Gamma \left(n {-} \frac{q}{2} {+} 1\right)}{\sqrt{\pi } \, \Gamma \left(n {+} \frac{1}{2}\right)} \ttau_2^{q-n} \right]\delta(\ttau_1)\,.
 \end{align}
From the above equation we use \eqref{eq:sourcecij} to derive the Laplace eigenvalue equation satisfied by the theta lift of the single Maass lifted monomial $D^{(n)}_{-2n} [\rho^{q}]$,  which we denote as $ \mathcal{E}_{s;q}(\tau)$,
\ie
\left[\Delta_\tau-s(s-1) \right] \mathcal{E}_{s;q}(\tau)  = 3 \big[ 1 - (-1)^{q}  \big]  \widetilde{C}_{s; s_1, s_2}  E(s_1; \tau) E(s_2; \tau)  \, , \label{eq:Esq}
\fe
where $s_1, s_2$ have been defined to be
\ie
s_1 = {w + s-q-1  \over 2} \, , \qquad  s_2 = { w- s+q+1 \over 2} \, , 
\fe
 and the coefficient $\widetilde{C}_{s; s_1, s_2}$ is given by
\ie
 \widetilde{C}_{s; s_1, s_2}  \coloneqq  \frac{\Gamma (s_1) \Gamma (s_2)  }{\sqrt{\pi } \Gamma \left(s-\frac{1}{2}\right)}   \Gamma \left(\frac{s+s_1-s_2+1}{2} \right) \Gamma \left(\frac{s-s_1+s_2+1}{2} \right)  \, . \label{eq:Ctilde}
\fe
We see that $\mathcal{E}_{s;q}(\tau)$ is simply a multiple of a single GES with a non-trivial normalisation factor related to $\widetilde{C}_{s; s_1, s_2}$. We anticipate that these coefficients $\widetilde{C}_{s; s_1, s_2}$ will play a key role in section \ref{sec:GenEis}.

Finally, we combine the result \eqref{eq:Esq} for the single monomial with the general expansion \eqref{eq:uvExp} to arrive at
\ie \label{eq:general-GES}
\left[\Delta_\tau-s(s-1) \right]  \cE_{i,j}^{w}(\stau)  = 3\sum _{k=0}^{2 i}  \sum _{\ell= 0}^j \sum _{m=0}^l \left( 1 - (-1)^{k+\ell+m}  \right)  C_{i,j; k, \ell, m} \, \widetilde{C}_{s; s_1, s_2} \,  E(s_1; \tau) E(s_2; \tau) \, , 
\fe
with $s=3i+j+1$, and 
\ie
s_1 = {w \over 2} +  { k + \ell + m -i-j \over 2 }\, , \qquad s_2 = {w \over 2} -  { k + \ell + m -i-j \over 2 }  \, . \label{eq:s1s2}
\fe
We therefore conclude that the theta lift \eqref{eq:Ewij2} of the modular local Maass form $A_{i,j}(\rho)$ is given by
\ie \label{eq:general-GES2}
 \cE_{i,j}^{w}(\stau)  = 3\sum _{k=0}^{2 i}  \sum _{\ell=0}^j \sum _{m=0}^\ell \big[ 1 - (-1)^{k+\ell+m}  \big]  C_{i,j; k, \ell, m} \, \widetilde{C}_{s; s_1, s_2} \,  \cE(s; s_1, s_2; \tau) + d_{i,j}(w) E(w;\tau) \, .
\fe
In the above equation we have also re-instated the boundary term contribution producing a multiple of the non-holomorphic Eisenstein series $ E(w;\tau)$ and evaluated earlier in \eqref{eq:dij}, although  this simpler contribution will not play any role for the following discussion. 

Using 
\eqref{eq:s1s2} it is easy to see  the linear combination of GESs $\cE(s; s_1, s_2; \tau)$ in  \eqref{eq:general-GES2}  appears with a range of indices $s_1, s_2$, which precisely matches the earlier statement \eqref{eq:cijw}.
We conclude that for arbitrary $i,j\in \mathbb{N}$, the coefficients $b_{i,j}(w,r)$ of the generalised Eisenstein contribution to $\cE_{i,j}^{w}(\stau) $ as presented in \eqref{eq:cijw} are given by
\begin{equation}
b_{i,j}(w,r) = 3\sum _{k=0}^{2 i}  \sum _{\ell=0}^j \sum _{m=0}^\ell  \delta(2r-( k + \ell + m -i-j )) \big[ 1 - (-1)^{k+\ell+m}  \big]  C_{i,j; k, \ell, m} \, \widetilde{C}_{s; s_1, s_2}  \,,\label{eq:bijGen}
\end{equation}
with $\delta(r)$ denoting Kronecker delta and $s_1,s_2$ given in \eqref{eq:s1s2}.

Although the general coefficients \eqref{eq:bijGen} are not quite expressed  in a fully explicit form, in section \ref{sec:GenEis} we will see that they do hide rather beautiful mathematical structures. 
%

\section{Structure of generalised Eisenstein series}
\label{sec:GenEis}

This section emphasises some interesting number theoretic aspects of GESs.
In particular, we discuss the occurrence of the $L$-values associated with holomorphic cusp forms in the Fourier mode expansion of GESs. Surprisingly, due to some intriguing cancellations such $L$-values are absent from the modular functions $\cE^w_{i,j}(\tau)$, which are constructed as theta lifts of local Maass functions \eqref{eq:Ewij2} and correspond to MGFs and SMFs that are relevant to superstring scattering amplitudes.

\subsection{Fourier mode expansion}

It is useful to consider the structure of the Fourier mode expansion of GESs with respect to ${\rm Re}\,\tau=\tau_1$.
As in \eqref{eq:geneisen}  we  define a GES as the modular invariant solution to the inhomogeneous Laplace eigenvalue equation
   \bea
(\Delta_\stau -s(s-1) )\, \cE(s;s_1,s_2; \stau) =  E (s_1;\stau)  
E(s_2;\stau) \,,
 \label{eq:LapEq}
 \eea
 subject to the condition that $ \cE(s;s_1,s_2; \stau)$ has moderate growth at the cusp $\tau_2\to \infty$ and fixing the boundary condition that the coefficient of the solution of the homogeneous equation  $\tau_2^s$ vanishes, thus implying the growth condition $ \cE(s;s_1,s_2; \stau) = O(\tau_2^{s_1+s_2})$ as $\tau \to i\, \infty$. For $s\in \mathbb{N}$, which is the case of present interest, $\cE(s;s_1,s_2; \stau) $ is the unique modular invariant solution to \eqref{eq:LapEq} subject to the above boundary conditions.
 
 It follows from the Fourier mode decomposition of the non-holomorphic Eisenstein series \eqref{eq:EisenExpansion}, that the GESs $\cE(s;s_1,s_2;\tau)$ can be decomposed in Fourier modes as
\begin{equation}
\cE(s;s_1,s_2;\tau) = \sum_{n,m= 0 }^\infty \cE^{(n,m)}(s;s_1,s_2;\tau_2) q^n \bar{q}^m\,,\label{eq:qqbar}
\end{equation}
 where $q=e^{2 \pi i \stau}$ and $\bar q = e^{-2\pi i\bar \stau}$.  The mode number is given by $k=n-m$ and modes with $k>0$ represent the contributions of instantons while modes with $k<0$ are contributions of anti-instantons.  It is useful to separate the terms with $k=0$ into purely perturbative contributions, coming from the $m=n=0$ term, and instanton/anti-instanton contributions where $m=n >0$.
 
 The $\cE^{(0,0)}$  term is a  Laurent polynomial that is  given by\footnote{\label{footnoteHalf}There are some special cases for the parameters $s_1,s_2$ and $s$ for which either the differential equation \eqref{eq:LapEq} does not admit a modular invariant solution, or the Laurent polynomial has to be slightly modified, see e.g. \cite{Klinger-Logan:2018sjt}. In particular, we note that for the case where either $s_1=\frac{1}{2}$ or $s_2 = \frac{1}{2}$, the Laurent polynomial requires the addition of a logarithmic term $\log(\tau_2)$. None of these cases seem to be important for either MGFs or SMFs.}
\begin{align}
\label{genEisPert}
&\cE^{(0,0)}(s;s_1,s_2;\tau_2) = \nn \\ 
& \frac{4 \pi^{-2s_1-2s_2} \zeta(2s_1) \zeta(2s_2)}{\pi^{s_1+s_2}(s_1+s_2-s)(s_1+s_2+s-1)} (\pi \tau_2)^{s_1+s_2} + \frac{4 {\pi^{-{1\over 2}-2s_2}}\Gamma(s_1-\half) \zeta(2s_1-1)\zeta(2s_2)}{(s_2-s_1+s)(s_2-s_1-s+1) \Gamma(s_1)} (\pi \tau_2)^{1-s_1+s_2} \nn \\
 +&  \frac{4 \pi^{-{1\over 2}-2s_1}\Gamma(s_2-\half) \zeta(2s_2-1)\zeta(2s_1)}{(s_1-s_2+s)(s_1-s_2-s+1)
 \Gamma(s_2)} (\pi \tau_2)^{1-s_2+s_1}
\\
+ & \frac{4 \Gamma(s_1-\half)\Gamma(s_2-\half) \zeta(2s_1-1)\zeta(2s_2-1)}{\pi (s_1+s_2-s-1)(s_1+s_2+s-2) \Gamma(s_1)\Gamma(s_2)} (\pi \tau_2)^{2-s_1-s_2}
+ \beta(s;s_1,s_2) \,   (\pi \tau_2)^{1-s} \,, \nn
\end{align}
where the first four terms can be obtained by matching powers of $\tau_2$ on the left-hand and right-hand sides of \eqref{eq:LapEq} and the last term satisfies the homogeneous equation and  its coefficient is given by \cite{Green:2005ba,Green:2008bf}\footnote{In these references, this coefficient is determined by projecting the Laplace equation \eqref{eq:LapEq} on $E(s;\stau)$. Alternatively, it may be obtained from a Poincar\'e series representation as in \cite{Dorigoni:2021jfr}.}

 \begin{equation}\label{eq:beta}
 \beta(s;s_1,s_2)  \coloneqq {4 \pi^{s-1}\over
  \Gamma(s_{1})\Gamma(s_{2})}\,
{\zeta^*(s-s_{1}-s_{2}+1)\zeta^*(s+s_{1}-s_{2})
\zeta^*(s-s_{1}+s_{2})\zeta^*(s+s_{1}+s_{2}-1)\over (1-2s)\,\zeta^*(2s)}\,,
\end{equation}
 with $\zeta^*(s)\coloneqq \zeta(s)\Gamma(s/2)/\pi^{s/2}$.  
 
 In addition to the purely perturbative terms, the zero-mode sector contains an infinite tower of instanton/anti-instanton contributions,  (i.e. all $\cE^{(n,n)}$  terms with $n>0$), which have the form
\begin{equation}
\cE^{(n,n)}(s;s_1,s_2;\tau_2)  =\frac{ n^{s_1 + s_2 - 2} \sigma_{1-2s_1}(n)\sigma_{1-2s_2}(n) }{\Gamma(s_1)\Gamma(s_2)}\Phi_{s;s_1,s_2}( 4\pi n \tau_2)\, ,
\label{eq:Fqq}
\end{equation}
where {$\sigma_s(n) \coloneqq \sum_{d\vert n}d^s$ denotes the standard divisor sigma function.} When $s_1,s_2\in \NN$ the expression $\Phi_{s; s_1,s_2}( 4\pi n\tau_2)$ is a polynomial  in inverse powers of $\tau_2$ of degree $s_2+s_1-2$. The first two perturbative orders in the $(n,n)$ sector do not depend on the eigenvalue $s$ and have the form
\begin{align}
\Phi_{s;s_1,s_2}( y)  =  \frac{8}{y^2}+ \frac{8[s_1(s_1-1)+s_2(s_2-1) -4]}{y^3}+O(y^{-4})\, .
\end{align}
 Higher-order coefficients do depend on $s$ and can be computed from the differential equation \eqref{eq:LapEq} as performed in \cite{Chester:2020vyz, Fedosova:2022zrb} or via resurgence analysis methods \cite{Dorigoni:2019yoq,Dorigoni:2022bcx,Dorigoni:2023nhc}. Note that similar instanton/anti-instanton contributions contribute to the general $k^{th}$ Fourier mode associated with terms in \eqref{eq:qqbar},  
\begin{equation}
\cE^{(n,m)}(s;s_1,s_2;\tau_2)= \frac{ n^{s_1 -1}m^{s_2-1} \sigma_{1-2s_1}(n)\sigma_{1-2s_2}(m) }{\Gamma(s_1)\Gamma(s_2)}\Big(
\frac{1}{4 nm\, (\pi \tau_2)^2} +O(\tau_2^{-3})\Big) + (n\leftrightarrow m) \, , \label{eq:Fqqbn1n2}
\end{equation}
where $k=n-m\ne 0$ and both $m, n$ are non-vanishing.  

Analysis of the  $(n,0)$ and $(0,m)$ terms requires more care since in these sectors we have the freedom of adding solutions of the homogeneous equation  to the differential equation \eqref{eq:LapEq} without spoiling either the boundary condition or the moderate growth condition at the cusp. 
From the differential equation \eqref{eq:LapEq} it is easy to see that the solutions of the homogeneous equation  in the zero-mode sector are simply $\tau_2^s$ and $\tau_2^{1-s}$. 
While the coefficient of the $\tau_2^s$ term has been set to zero thanks to our choice of boundary condition, the coefficient of $\tau_2^{1-s}$  must hence take the value \eqref{eq:beta} as a consequence of modular invariance.
For the non-zero Fourier mode the story is more interesting since the solution of the homogeneous equation  in the $n^{th}$ Fourier mode sector (with $n\neq0$) is  {proportional to $K_{s-\half}(2\pi |n| \tau_2)$, which satisfies} \footnote{{Note that there is a second, linearly independent, solution of the homogeneous equation  that is proportional to the modified Bessel function of the first kind $I_{s-\half}(2\pi |n| \tau_2)$. This solution to the homogeneous equation is discarded since it grows exponentially at the cusp $\tau_2\gg1$.}}
\begin{equation}
(\Delta_\tau-s(s-1)) \Big[ e^{2\pi i n \tau_1}\sqrt{|n| \tau_2} K_{s-\half}(2\pi |n| \tau_2)\Big] = 0\, . 
\label{eq:homosoln}
\end{equation}
The modified Bessel function is exponentially suppressed at the cusp, i.e.
\begin{equation}
\sqrt{|n| \tau_2} K_{s-\half}(2\pi |n| \tau_2) = e^{-2 \pi  |n| \tau_2 } \left(\frac{1}{2}+ \frac{s(s-1) }{8 \pi  |n| \tau_2 }+ O(\tau_2^{-2})\right)\,,\qquad \tau_2\gg1\,.\label{eq:Klarge}
\end{equation}
Furthermore we note for future convenience that at the origin {$\tau_2 \to 0$} this solution of the homogeneous equation behaves as
\begin{equation}
\sqrt{|n| \tau_2} K_{s-\half}(2\pi |n| \tau_2) =  \frac{  (\pi |n| \tau_2)^{1-s}\Gamma (s -\half )}{2 \sqrt{\pi} } \left( 1 - \frac{2 (\pi |n|\tau_2)^2}{2s-3} + O(\tau_2^4)\right)\,,\qquad \tau_2\to 0\,.
\label{eq:Ksmall}
\end{equation}

A key point is that a particular solution of the differential equation \eqref{eq:LapEq} is not necessarily  modular invariant.
To ensure that we have indeed constructed the unique modular invariant solution to \eqref{eq:LapEq} (given the boundary condition above) it may be necessary to add to the particular solution a suitable solution to the homogeneous equation 
\bea
\left(\Delta_\tau-s(s-1) \right) H(\tau)=0\,,
\label{eq:homogen}
\eea
which is given by the linear combination\footnote{Here we are interested in constructing modular invariant solutions to \eqref{eq:LapEq} that are even under the upper-half plane involution $\tau\to-\bar{\tau}$. This condition forces the solution of the homogeneous equation  to take the form of  \eqref{eq:HomoGen} and the coefficients $\alpha_n$ to be real. In section \ref{sec:Discussion} we comment on a different family of modular invariant functions introduced in \cite{Dorigoni:2021jfr} that are odd under this involution and are related to cuspidal MGFs. }
\begin{equation}\label{eq:HomoGen}
H(\tau) = \sum_{n=1}^\infty \alpha_n \sqrt{n \tau_2} K_{s-\half}(2\pi n \tau_2) \big( e^{2\pi i n \tau_1}+ e^{-2\pi i n \tau_1})\,,
\end{equation}
with particular real coefficients $\alpha_n$.
We now show that this is precisely the case when constructing GESs relevant for MGFs and SMFs, and the homogeneous coefficients $\alpha_n$ are crucially related to the Fourier coefficients of certain holomorphic cusp forms.

\subsection{$L$-values and generalised Eisenstein series}
\label{sec:Lvalues}

In this subsection we clarify how to determine the coefficients in the solution of the  homogeneous equation \eqref{eq:HomoGen} so as to obtain a modular invariant solution to \eqref{eq:LapEq}.
We first discuss how to construct a particular solution to the differential equation \eqref{eq:LapEq} and how to check whether or not  this is modular invariant. Since the discussion is rather different for the case of GESs relevant to MGFs (where $s_1$, $s_2\in \NN^+$)  and the case relevant to SMFs (where $s_1$, $s_2\in \NN+{1\over 2}$)  we will discuss these two cases separately.
However, the basic arguments that determine  the solution of the homogeneous equation are the same for both families of GESs.  We will  present a conjectured closed-form expression for the coefficient of the solution of the homogeneous equation for the GESs relevant for MGFs \eqref{eq:lambdaMGF2t} that reproduces all previously known results. Strikingly, we prove that the analytic continuation \eqref{eq:SolSp2} of this conjectural expression to the case of GESs relevant for SMFs also produces the correct solution for the coefficient of the solution of the homogeneous equation, which will be determined thanks to some recent results on convolution identities for divisor sums \cite{Fedosova:2023cab}. We will highlight these important parallels amongst the two families when they arise in the following discussion.

\subsubsection*{Two-loop Modular Graph Functions}

In \cite{Dorigoni:2021jfr}, two related methods for constructing solutions to the differential equation \eqref{eq:LapEq} were presented for the case 
\begin{equation}
\label{eq:spec1}
s_1,s_2\in \mathbb{N}\,, \quad {\rm with}\quad s_1,s_2\geq 2\,,\qquad s\in\{|s_1-s_2|+2,|s_1-s_2|+4, \ldots ,s_1+s_2-4,s_1+s_2-2\}\,.
\end{equation}
A first approach is via Poincar\'e series where the key idea is to write one of the non-holomorphic Eisenstein series on the right hand side of  \eqref{eq:LapEq}, say $E(s_1;\tau)$, as a sum over images under the action of $B(\mathbb{Z})\backslash {\rm SL}(2,\mathbb{Z})$ of the much simpler function $\tau_2^{s_1}$ where $B(\mathbb{Z})\coloneqq \{ \left( \begin{smallmatrix} 1 & n \\ 0 & 1 \end{smallmatrix}\right)\,\vert\,n\in \mathbb{Z}\}$ is the Borel stabiliser of the cusp $\tau= i \infty$.
It is then possible to find a solution to this modified differential equation via standard methods. Upon summing this solution over all the images under $B(\mathbb{Z})\backslash {\rm SL}(2,\mathbb{Z})$ we retrieve the modular invariant GES solution.

One of the advantages of this approach is that it immediately generates a modular invariant solution to \eqref{eq:LapEq}. However, the drawback is that while it is possible \cite{Dorigoni:2019yoq,Dorigoni:2022bcx} to extract the zero-mode sector $\mathcal{E}^{(n,n)}$ \eqref{eq:Fqq} from the Poincar\'e series, it becomes much harder, if not impossible, to say anything about the instanton or anti-instanton sectors  $\mathcal{E}^{(n,0)}$ and $\mathcal{E}^{(0,n)}$.

Interestingly, this Poincar\'e series approach suggests a different way of tackling the problem of obtaining a particular solution to  \eqref{eq:LapEq} for the spectrum of sources and eigenvalues of present interest. In \cite{Dorigoni:2021jfr} it was realised that passing to the Poincar\'e seeds for the GESs motivates a certain notion of \textit{depth} for the functional space under consideration. Building on {\cite{Brown:I, Brown:II, Broedel:2018izr},}  in \cite{Dorigoni:2021jfr} it was shown that a possible Poincar\'e representation of the GESs with spectrum \eqref{eq:spec1} can be given in terms of depth-one iterated integrals of holomorphic Eisenstein series, and their complex conjugates, thus suggesting that GESs themselves are given by depth-two iterated integrals of holomorphic Eisenstein series and their complex conjugates plus possibly lower depth terms.

While invariance under a $T$-transformation $\tau\to \tau+1$ is manifest in the space of iterated Eisenstein integrals, invariance under a $S$-transformation $\tau\to -1/\tau$ is not and has to be checked case by case. Given a particular solution constructed in terms of iterated integrals of two holomorphic Eisenstein series, its $S$-transform can be evaluated and it can be checked whether or not the particular solution is modular invariant. 
The difference between an iterated integral of holomorphic Eisenstein series and its $S$-dual is controlled by a family of periods called \textit{multiple modular values}. While at depth-one the only multiple modular values we find are standard Riemann zeta values, the same is not true at higher depth.
In particular, as discussed in \cite{Dorigoni:2021jfr,Dorigoni:2021ngn}  multiple modular values at depth-two display (amongst other things) $L$-values of holomorphic cusp forms. As a consequence,  we see that for the spectrum \eqref{eq:spec1} the particular solution to \eqref{eq:LapEq} constructed via depth-two iterated Eisenstein integrals fails to be modular invariant whenever the eigenvalue $s$ is such that $ \dim \mathcal{S}_{2s} \neq 0$
where $\mathcal{S}_{2s} $ is the vector space of holomorphic cusp forms with weight $2s$.
In appendix \ref{app:Cusps} we review some important properties of holomorphic cusp forms and their associated $L$-values.

As argued above, if the particular solution is not modular invariant we must add a suitable solution of the homogeneous equation  of the form  \eqref{eq:HomoGen} to obtain the unique GES modular invariant solution to \eqref{eq:LapEq}.
In \cite{Dorigoni:2021jfr,Dorigoni:2021ngn} it was shown that the failure of modularity for the particular depth-two iterated Eisenstein integral solution can be cancelled by a suitable multiple of the non-modular invariant solution of the homogeneous equation
\begin{align}
H_{\Delta}(\tau)&\label{eq:Hdelta}\coloneqq \sum_{n=1}^\infty  \frac{a_\Delta(n)}{n^s} \sqrt{n\tau_2 }  K_{s-\frac{1}{2}}(2 \pi n \tau_2)\big( e^{2\pi i n \tau_1}+e^{-2\pi i n \tau_1}\big)\cr
& =   (-1)^s \frac{\pi^{s} i }{\Gamma(s)} \tau_2^{1-s}\int_\tau^{i\infty} (\tau-\tau')^{s-1} (\bar\tau-\tau')^{s-1}\,\Delta(\tau') \,{\rm d}\tau' + {\rm c.c.}\,,
\end{align}
which can be thought of as a depth-one iterated integral of the Hecke-normalised, holomorphic cusp form $\Delta(\tau)= \sum_{n=1}^\infty a_\Delta(n)q^n \in \mathcal{S}_{2s}$. Hecke normalisation implies $a_\Delta(1)=1$ and we refer again to appendix \ref{app:Cusps} for important aspects on the theory of holomorphic cusp forms.

It should be  stressed that the Eichler integral $H_{\Delta} (\tau)$ is not modular invariant on its own.  However, the addition of a suitable multiple of $ H_{\Delta}(\tau)$ to the particular solution $\cE_{{\rm p}}(s;s_1,s_2;\tau)$  is crucial for obtaining a modular invariant solution to \eqref{eq:LapEq}. As a result the GESs can be expressed as
\begin{equation}\label{eq:PartplusHomo}
 \cE(s;s_1,s_2;\tau)=  \cE_{{\rm p}}(s;s_1,s_2;\tau) + \sum_{\Delta\in \cS_{2s}} \lambda_{\Delta}(s;s_1,s_2) H_{\Delta}(\tau)\,,
\end{equation}
where the sum is over a basis of Hecke-normalised cusp forms for $\mathcal{S}_{2s}$. 
In this expression  $\cE_{{\rm p}}(s;s_1,s_2;\tau)$ denotes the particular solution constructed in \cite{Dorigoni:2021jfr}, which contains only iterated integrals of holomorphic Eisenstein series with depth less than or equal to two and their complex conjugates. 

Recalling the formula for the dimension of the vector space of holomorphic cusp forms:
\begin{equation}\label{eq:DimS}
 \dim \mathcal{S}_{2s} = \left\lbrace \begin{array}{lc}
\left\lfloor \frac{2s}{12}\right\rfloor -1 & 2s\equiv 2\, {\rm mod }\,12\,,\\[2mm]
 \left\lfloor \frac{2s}{12}\right\rfloor \phantom{-1}& \mbox{otherwise}\,,
\end{array}\right.
\end{equation}
we see that the first instance where the particular solution fails to coincide with the GESs happens at eigenvalue $s=6$ for which we have $ \mathcal{S}_{12} = {\rm span}\{\Delta_{12}\}$ where $\Delta_{12}= \sum_{n=1}^\infty \tau(n) q^n$ is the discriminant modular form whose $q$-series coefficients are given by the Ramanujan tau-function $\tau(n)$ (not to be confused with the modular parameter $\tau$).
For the spectrum \eqref{eq:spec1} the eigenvalue $s=6$ is attained by considering sources with indices $(s_1,s_2)\in\{(2,6),(3,5),(4,4)\}$ and we find that the corresponding GESs are given by \cite{Dorigoni:2021jfr,Dorigoni:2021ngn}
\begin{align}
\mathcal{E}(6;2,6;\tau) &\notag =  \cE_{{\rm p}}(6;2,6;\tau) + \frac{2}{17275} \frac{\Lambda(\Delta_{12},13)}{\Lambda(\Delta_{12},11)} H_{\Delta_{12}}(\tau)\,,\\
\mathcal{E}(6;3,5;\tau) &\label{eq:E6MGFs} =  \cE_{{\rm p}}(6;3,5;\tau) - \frac{5}{11056} \frac{\Lambda(\Delta_{12},13)}{\Lambda(\Delta_{12},11)} H_{\Delta_{12}}(\tau)\,,\\
\mathcal{E}(6;4,4;\tau) &\notag =  \cE_{{\rm p}}(6;4,4;\tau) + \frac{7}{10365} \frac{\Lambda(\Delta_{12},13)}{\Lambda(\Delta_{12},11)} H_{\Delta_{12}}(\tau)\,.
\end{align}
 Here $\Lambda(\Delta,t)$ denotes the standard completed $L$-function, defined in \eqref{eq:Lambda}, of the cusp form $\Delta\in \mathcal{S}_{2s}$.

While in \cite{Dorigoni:2021jfr} the coefficients $\lambda_{\Delta}(s;s_1,s_2)$ have not been determined in closed form for general parameters $s,s_1,s_2$ in \eqref{eq:spec1}, they appear to take the form
\begin{equation}
\lambda_{\Delta}(s;s_1,s_2) = \kappa_{\Delta,s,s_1,s_2} \frac{\Lambda(\Delta,w+s-1)}{\Lambda(\Delta,2s-1)}\,,\label{eq:lambdaMGF}
\end{equation}
where $w=s_1+s_2$.  
{From the spectrum \eqref{eq:spec1} it is easy to see that since $w=s_1+s_2\geq s+2$ it follows that the ratio of $L$-values in the coefficients \eqref{eq:lambdaMGF} always consists of an odd critical $L$-value in the {denominator} and an odd non-critical $L$-value in the {numerator} where, for a cusp form $\Delta\in \mathcal{S}_{2s}$, the critical $L$-values are the numbers $\Lambda(\Delta,t)$ with $t \in \{1, \ldots  ,2s-1\}$.}
{The general expression for the coefficients $\kappa_{\Delta,s,s_1,s_2}$ is not determined at this stage  but they can be evaluated on a case by case basis by evaluating the multiple modular values associated with the particular solution part of \eqref{eq:PartplusHomo} and they are generically in the number field associated with $\mathcal{S}_{2s}$.\footnote{Whenever $\dim \mathcal{S}_{2s} =1$ this number field actually coincides with the rational numbers. However, when $\dim \mathcal{S}_{2s} >1$ that is no longer true.  For example,  with $2s=24$ we have  $\dim  \mathcal{S}_{24} =2$ and the Fourier coefficients of the two Hecke eigenforms basis elements of $\mathcal{S}_{24}$ belong to the number field $\mathbb{Q}[\sqrt{144169}]$. by the Fourier coefficients $a(n)$ of $\Delta(\tau) = \sum_{n=1}^\infty a(n)q^n$.} However, we will shortly propose a strongly motivated conjectural expression for the coefficients in  \eqref{eq:lambdaMGF} based on the expression \eqref{eq:lambdaIC} for these coefficients in the context of SMFs, which will  be discussed in the next sub-section.}

 {In order to make contact with the expression for $\lambda_\Delta$ that we will find for SMFs in the next sub-section  it is convenient to re-express \eqref{eq:lambdaMGF} so that the critical $L$-value in the  denominator is converted to the Petersson norm $\langle \Delta,\Delta \rangle$ defined in  \eqref{eq:PeterNorm}.
 This makes use of a  classic result by Rankin \cite{Rankin} that states that the Petersson norm of a Hecke eigenform $\Delta$ can be written in terms of the product of two critical $L$-values with opposite parity.  For example the Petersson norm of the Ramanujan cusp form $\Delta_{12}$,~ i.e. the unique Hecke normalised element of $\mathcal{S}_{12}$, can be written as {(see (9.1) of \cite{Rankin})}
\begin{equation}
{\langle \Delta_{12},\Delta_{12} \rangle}=\frac{691 }{7680}  \Lambda (\Delta_{12},8) \Lambda (\Delta_{12},11)\,.\label{eq:norm12}
\end{equation}
The general expression for the norms of arbitrary cusp forms follows from Rankin’s theorem stated explicitly in \eqref{eq:Rankin}. }

{Furthermore, a beautiful result by Manin \cite{Manin} known as the periods theorem and briefly reviewed in appendix \ref{app:Cusps}, proves that the ratio of any two even critical $L$-values or any two odd critical $L$-values are rational over the algebraic number field generated by the Fourier coefficients of the cusp forms.
Hence by combining \eqref{eq:Rankin} and Manin's theorem we can rewrite \eqref{eq:lambdaMGF}  as}
\begin{equation}
\lambda_{\Delta}(s;s_1,s_2)  = \tilde{\kappa}_{\Delta,s,s_1,s_2}  \frac{ \Lambda (\Delta, s+s_1-s_2)\Lambda (\Delta, s+w-1)}{\langle \Delta, \Delta\rangle}\,.\label{eq:lambdaMGF2t} 
\end{equation}

{According to \ref{Conj1} presented in section \ref{sec:Results}  the expression for the coefficients of the solution of the homogeneous equation,  $\lambda_{\Delta}(s;s_1,s_2)$, in the MGF case (with integer $s_1$ and $s_2$)    are the same as those in the SMF case with (half-integer $s_1$ and $s_2$).  An explicit expression for the latter will be derived in the next sub-section and is given in \eqref{eq:SolSp2}-\eqref{eq:lambdaIC}. As a result the coefficient $\tilde{\kappa}_{\Delta,s,s_1,s_2}$ which appears in  \eqref{eq:lambdaMGF2t} is given by
\begin{equation}
\tilde{\kappa}_{\Delta,s,s_1,s_2}  = (-1)^{\frac{s+s_1-s_2+2}{2} }   \frac{  \Gamma(s) \Gamma\left(\frac{w-s}{2}\right)}{2^{2s-3}\Gamma\left(\frac{w+s}{2}\right)\widetilde{C}_{s;s_1,s_2}} \,, \label{eq:ktilde}
\end{equation}
with $w=s_1+s_2$ and where $\widetilde{C}_{s;s_1,s_2}$ is the rational number defined in \eqref{eq:Ctilde}.}
{ Our choice of normalisation ensures that the overall coefficient $\tilde{\kappa}_{\Delta,s,s_1,s_2}$ in \eqref{eq:ktilde} is manifestly rational for the spectrum \eqref{eq:spec1}.  Furthermore, for this spectrum $\lambda_\Delta(s;s_1,s_2)$ always contains the product of an even critical $L$-value and an odd non-critical $L$-value.
Lastly, while the particular results provided in \cite{Dorigoni:2021jfr,Dorigoni:2021ngn} have the completed $L$-values multiplied by an overall coefficient which explicitly belongs to the number field associated with $\mathcal{S}_{2s}$, we see that in our expression \eqref{eq:lambdaMGF2t} the dependence on  the number field is implicit and it is entirely captured by the Petersson norm $\langle \Delta, \Delta\rangle$ in the denominator, as we see in Rankin's expression \eqref{eq:Rankin} for the Petersson norm.}

{We have checked that our conjectural expression \eqref{eq:lambdaMGF2t} does indeed reproduce all particular examples presented in the ancillary file of \cite{Dorigoni:2021jfr}. }
If we rewrite the particular examples \eqref{eq:E6MGFs} using \eqref{eq:norm12} in order to implement the above representation we can check that \eqref{eq:lambdaMGF2t} is correct for $s=6$ and we find 
\begin{align}
\mathcal{E}(6;2,6;\tau) &\notag =  \cE_{{\rm p}}(6;2,6;\tau) + \frac{1}{184320} \frac{\Lambda(\Delta_{12},10)\Lambda(\Delta_{12},13)}{\langle \Delta_{12},\Delta_{12}\rangle} H_{\Delta_{12}}(\tau)\,,\\
\mathcal{E}(6;3,5;\tau) &\label{eq:E6MGFs2} =  \cE_{{\rm p}}(6;3,5;\tau) - \frac{125}{32 } \cdot \frac{1}{184320} \frac{\Lambda(\Delta_{12},10)\Lambda(\Delta_{12},13)}{\langle \Delta_{12},\Delta_{12}\rangle} H_{\Delta_{12}}(\tau)\,,\\
\mathcal{E}(6;4,4;\tau) &\notag =  \cE_{{\rm p}}(6;4,4;\tau) +\frac{35}{6 } \cdot \frac{1}{184320} \frac{\Lambda(\Delta_{12},10)\Lambda(\Delta_{12},13)}{\langle \Delta_{12},\Delta_{12}\rangle} H_{\Delta_{12}}(\tau)\, , 
\end{align}
{where we have expressed all the critical $L$-values in terms of $\Lambda(\Delta_{12},10)$ by using Manin's relations \cite{Manin}.}
We will shortly see that the coefficients in front of the solutions of the homogeneous equation  appearing in \eqref{eq:E6MGFs2} are actually deeply connected with the particular combinations of GESs obtained from the theta-lifted local Maass functions discussed in section \ref{sec:LapEq}.
However, before doing that we will now  review how a similar interplay between the particular solution and the solution of the homogeneous equation  related to holomorphic cusp forms is also present in the context of GESs associated with SMFs.

\subsubsection*{S-dual Modular Functions}

We now consider the GES solution to \eqref{eq:LapEq} with spectrum\footnote{As mentioned in footnote \ref{footnoteHalf}, the case with either $s_1$ or $s_2$  equal to $\frac{1}{2}$ is slightly different from the other half-integer cases since the Fourier zero-mode sector does contain a term proportional to $\log(\tau_2)$. Although our present discussion encompasses this case as well, these particular modular invariant functions do not seem to be relevant when discussing SMFs.}
\begin{equation}
\label{eq:spec2}
s_1,s_2\in {\mathbb{N}}+\frac{1}{2}\,,\qquad s\in \{s_1+s_2+1,s_1+s_2+3, \ldots \}\,.
\end{equation}
An algorithm for constructing a particular solution, $\cE_{{\rm p}}(s;s_1,s_2;\tau)$, to \eqref{eq:LapEq} for the special case  $\cE_{{\rm p}}(4;\threeh,\threeh;\tau)$ in \cite{Green:2014yxa} (see also \cite{Chester:2020vyz}), and later generalised in \cite{Fedosova:2022zrb} to the spectrum \eqref{eq:spec2}.
Particularly important for the present discussion is the form of the particular solution for $\cE^{(n,0)}_{{\rm p}}$ and $\cE^{(0,n)}_{{\rm p}}$ with $n\neq 0$ as defined in \eqref{eq:qqbar}.

Given the differential equation \eqref{eq:LapEq} and the Fourier mode expansion for the non-holomorphic Eisenstein series \eqref{eq:EisenExpansion} we see that $\cE^{(n_1,n_2)}_{{\rm p}}$  with $n_1+n_2\neq0$ can be represented as
\begin{equation}
e^{-2\pi (n_1+n_2) \tau_2}\cE^{(n_1,n_2)}_{{\rm p}}(s;s_1,s_2;\tau_2)  =  \hat{g}_{n_1,n_2}(\tau_2)\,,
\end{equation}
where the functions $ \hat{g}_{n_1,n_2}(\tau_2)$ are solutions to the second order differential equations
\begin{align}
& \left[\tau_2^2\partial_2 -(2\pi n  \tau_2)^2 -s(s-1) \right] ( \hat{g}_{n,0}(\tau_2) +  \hat{g}_{0,n}(\tau_2)) =
\nn \\
&\Big( \frac{2 \zeta(2 s_1)}{\pi^{s_1}}{\tau_2^{s_1}} +   \frac{2\Gamma(s_1 - \frac 12)}{\pi^{s_1-\frac12}\Gamma(s_1)} \zeta(2s_1-1)\tau_2^{1-{s_1}} \Big) \frac{4 \sqrt{\tau_2}}{\Gamma(s_2)}  n^{s_2-\half}
    \sigma_{1-2s_2}(n) \, 
      K_{s_2 - \frac 12} (2 \pi  n\tau_2) +(s_1\leftrightarrow s_2) \, , 
      \label{eq:gn0}
      \end{align}
      and 
 \begin{align}
& \left[\tau_2^2\partial_2 -(2\pi (n_1+n_2)  \tau_2)^2 -s(s-1) \right] \hat{g}_{n_1,n_2}(\tau_2) = \nn \\
& \frac{16\tau_2}{\Gamma(s_1)\Gamma(s_2)}  n^{s_1+s_2-1}
    \sigma_{1-2s_1}(n_1) \sigma_{1-2s_2}(n_2) \, K_{s_1 - \frac 12} (2 \pi  n_1\tau_2) \,
      K_{s_2 - \frac 12} (2 \pi  n_2\tau_2) +(s_1\leftrightarrow s_2) \,.
      \label{eq:gn1n2}
\end{align} 
{The terms $\hat{g}_{n,0}(\tau_2) $ and $\hat{g}_{0,n}(\tau_2)$ in \eqref{eq:gn0} originate from contributions to the source term coming from the product of the $n$-instanton sector of $E(s_1;\tau)$  and the perturbative expansion of $E(s_2;\tau)$ (together with the terms with $s_1\leftrightarrow s_2$) while  $\hat{g}_{n_1,n_2}(\tau_2)$ in \eqref{eq:gn1n2}  comes from contributions from  the product of the $n_1$-instanton sector of $E(s_1;\tau)$  with the $n_2$-instanton sector of $E(s_2;\tau)$. }

Equation \eqref{eq:gn0} can be solved by standard methods, while \eqref{eq:gn1n2} can be brought into a nicer form by writing it as 
\begin{equation}
\hat{g}_{n_1,n_2}(\tau_2)  = \frac{16 }{\Gamma(s_1)\Gamma(s_2)} n_1^{-s_1+\frac{1}{2}} n_2^{-s_2+\frac{1}{2}}\,\sigma_{2s_1-1}(n_1)\sigma_{2s_2-1}(n_2) G(2\pi n_1,2\pi n_2,\tau_2)\,,\label{eq:gG}
\end{equation}
where we have used $\sigma_{-\alpha}(n) =n^{-\alpha}\sigma_{\alpha}(n)$. The auxiliary function $G(n_1,n_2,y)$ solves:
\begin{equation}\label{eq:Gode}
\left[y^2 \partial_y^2 - ( (n_1+n_2)  y)^2 -s(s-1) \right] G(n_1,n_2,y)=y K_{s_1-\frac{1}{2}}( n_1 y)K_{s_2-\frac{1}{2}}( n_2 y)\,.
\end{equation}

The results in \cite{Fedosova:2022zrb} lead to a particular solution of the form:
\begin{equation}\label{eq:Gsol}
 G(n_1,n_2,y)  = \sum_{i,j=0}^1 \eta_{ij}(n_1,n_2,y) K_i(n_1 y) K_j (n_2 y)\,,
\end{equation}
with $ \eta_{ij}(n_1,n_2,y)$ a rational function depending on the parameters $s_1,s_2$ and $s$.
The key question now is whether this particular solution of \eqref{eq:LapEq} is modular invariant and is therefore a GES.
In order to answer this question  we recall an important lemma proved in \cite{Green:2014yxa}.

\vskip 0.2cm
\textbf{\setword{Lemma}{LA}.}  \textit{If ${\rm{F}}(\tau)$ is an ${\rm SL}(2, \mathbb{Z})$ invariant function on the upper half-plane such that at the cusp $\tau_2\to\infty$, with $\tau_2= \rm{Im}\,\tau$, it satisfies the growth condition ${\rm{F}}(\tau) = O(\tau_2^w)$ with $w>1$, then each of its Fourier modes ${\rm{F}}_n(\tau_2) = \int_0^1 {\rm{F}}(\tau) e^{-2\pi i n \tau_1} {\rm d}\tau_1$ satisfies the bound ${\rm{F}}_n(\tau_2) = O(\tau_2^{1-w})$ in the limit $\tau_2\to 0$.}
{\textit{It is important to note that in the special case where the modular invariant function  ${\rm{F}}(\tau)$ satisfies the growth condition ${\rm{F}}(\tau) = O(\tau_2)$, we can obtain the slightly weaker bound ${\rm{F}}_n(\tau_2) = O(\tau_2^{-\epsilon})$ in the limit $\tau_2\to 0$ for all $\epsilon>0$.}}
\vspace{0.2cm}

 The proof of this lemma is based on the fact that a cuspidal growth of order $\tau_2^w$ implies that the modular invariant function ${\rm F}(\tau)$ must be bounded by {$E(w; \tau)$} on the whole upper half-plane and since for small $\tau_2$ we have {$E(w; \tau)  = O(\tau_2^{1-w})$}, then the same bound must hold for ${\rm F}_n (\tau)$.

We will now argue that thanks to the result of \cite{Fedosova:2023cab}, the particular solution constructed in \cite{Fedosova:2022zrb} and given by the solution of \eqref{eq:gG}-\eqref{eq:Gsol} violates the \ref{LA} whenever the eigenvalue $s$ is such that $\dim \mathcal{S}_{2s}\neq 0$ and hence it is not a modular invariant solution to \eqref{eq:LapEq}. However, by adding a suitable multiple of exactly the same solution of the homogeneous equation  \eqref{eq:Hdelta} previously considered, we will be able to construct a modular  invariant solution to \eqref{eq:LapEq} for the spectrum \eqref{eq:spec2}.

From the asymptotic expansion as $\tau_2\to \infty$ presented in \eqref{genEisPert}, we see that the GES $\cE(s;s_1,s_2;\tau)$ grows at the cusp as $\tau_2^{s_1+s_2} = \tau_2^{w}$. Hence from the above \ref{LA} we know that the behaviour of any of its Fourier modes at the origin is bounded by $\tau_2^{1-w}$.
In the zero-mode sector we immediately see that the perturbative solution of the homogeneous equation  $\tau_2^{1-s}$ violates the growth condition at the origin for the spectrum \eqref{eq:spec2} since $s\geq w+1$. However, the zero-mode sector also contains the infinite tower \eqref{eq:Fqq} of instanton/anti-instanton terms. While these terms are are exponentially suppressed at the cusp $\tau_2\gg1$, it is possible to show either by direct calculation 
\cite{Green:2014yxa,Dorigoni:2022bcx} or via ${\rm SL}(2,\mathbb{Z})$ spectral theory \cite{Dorigoni:2023nhc} that near the origin $\tau_2\to0$ the sum of instanton/anti-instanton terms behaves as $\tau_2^{1-s}$ and cancels the unwanted power of $\tau_2^{1-s}$ coming from the perturbative part, thus proving that in the zero-mode sector the growth condition implied by the \ref{LA} is indeed respected.

We now turn to consider the small-$\tau_2$ behaviour of the non-zero mode part, which is given by
\begin{equation}
 \int_0^1\cE_{{\rm p}}(s;s_1,s_2;\tau) e^{-2\pi i n \tau_1} {\rm d}\tau_1 =  \hat{g}_{n,0}(\tau_2)+\hat{g}_{0,n}(\tau_2) + \sum_{\substack{n_1+n_2=n\\ n_1,n_2\neq0}} \hat{g}_{n_1,n_2}(\tau_2)\,.\label{eq:FourN}
\end{equation} 
The small-$\tau_2$ limits of $\hat{g}_{n,0}(\tau_2)$ and $\hat{g}_{0,n}(\tau_2)$ are easy to obtain from the differential equation \eqref{eq:gn0}, but the small-$\tau_2$ limit of the infinite sum of the $\hat{g}_{n_1,n_2}(\tau_2)$ contributions is more complicated.
Firstly, we notice from \eqref{eq:Gsol} that since the modified Bessel functions $K_0(x),K_1(x)$ decrease exponentially fast for large and positive $x$ \eqref{eq:Klarge}, the sum over $n_1+n_2=n$ appearing in \eqref{eq:FourN} is absolutely convergent for all $n\neq 0$.
Hence we can first expand $\hat{g}_{n_1,n_2}(\tau_2)$ for small-$\tau_2$ before  performing the sum over $n_1+n_2=n$.
Using the results of  \cite{Fedosova:2022zrb} the expansion of the particular solution $\hat{g}_{n_1,n_2}(\tau_2)$ given by  \eqref{eq:gG}-\eqref{eq:Gsol}  in the limit $\tau_2\to 0$  with $n_1+n_2=n$ and $n_1,n_2\neq0$  is given by 
\begin{equation}\label{eq:gsmall}
\hat{g}_{n_1,n_2}(\tau_2)\stackrel{\tau_2\to0}{\sim} R_{s;s_1,s_2}  \tau_2^{1-s} \frac{1}{n^{s+s_1+s_2-1}} \Big[ \sigma_{2s_1-1}(n_1)\sigma_{2s_2-1}(n_2) Q^{2s_1-1,2s_2-1}_{s-s_1-s_2}\Big(\frac{n_2-n_1}{n_2+n_1}\Big)  \Big] +O(\tau_2^{3-s})\,,
\end{equation}
where $Q_d^{r_1,r_2}$ is a Jacobi function of the second kind and the coefficients $R_{s;s_1,s_2}$ are given by: 
\begin{equation}\label{eq:Rss1s2}
R_{s;s_1,s_2} \coloneqq 4 \, \frac{ \Gamma \left( \frac{s-s_1 -s_2 +1}{2}\right) \Gamma \left(\frac{s+s_1 +s_2 -1}{2}\right)}{\pi^s \, \widetilde{C}_{s;s_1,s_2}}\,,
\end{equation}
and the coefficients $\widetilde{C}_{s;s_1,s_2}$ have been defined in \eqref{eq:Ctilde}. Although we do not have a proof of this expression, we have checked its validity by computing the particular solution \eqref{eq:Gsol} using the methods of \cite{Fedosova:2022zrb}  for a multitude of values for the parameters $s,s_1,s_2$ in the spectrum \eqref{eq:spec2}.

Furthermore, although \eqref{eq:gsmall} only displays the leading power of  the small-$\tau_2$ expansion of $\hat{g}_{n_1,n_2}(\tau_2)$ we have checked that, for the range of parameters $s,s_1,s_2$  in the spectrum \eqref{eq:spec2}, the particular solution $\hat{g}_{n_1,n_2}$ does produce all power-like terms $\tau_2^{-\ell}$ in the range $\ell\in \{w,w+2, \ldots ,s-3,s-1\}$ with $w=s_1+s_2$ that violate the boundary condition at small-$\tau_2$. Importantly, these ``unwanted'' terms can be all written in the form  \eqref{eq:gsmall}  with exactly the same prefactor function of $n_1,n_2$ multiplying a polynomial in inverse powers of $(n\tau_2)$ that  is identical to the small-$\tau_2$ expansion of $\sqrt{ n \tau_2}\,K_{s-1/2}(2\pi n \tau_2)$ given in \eqref{eq:Ksmall}.

To check whether the complete $n^{th}$ Fourier mode \eqref{eq:FourN} does in fact violate the \ref{LA} we still need to perform the sum over $n_1+n_2=n$ in \eqref{eq:gsmall}. To this end we utilise the following crucial theorem.
\vspace{0.1cm}

\vskip 0.2 cm
{\textbf{\setword{Theorem FKR}{TH1}.}} \textit{(Fedosova, Klinger-Logan, Radchenko \cite{Fedosova:2023cab}) For any $d\geq 1$ and $r_1,r_2 \in 2\mathbb{Z}_{>0}$ we have:
\begin{equation}
\sum_{\substack{ n_1+n_2=n \\ n_1,n_2\neq 0} }\sigma_{r_1}(n_1)\sigma_{r_2}(n_2) Q^{r_1,r_2}_{d}\Big(\frac{n_2-n_1}{n_2+n_1}\Big) = (-1)^d C_d^{r_1,r_2}(n)\sigma_{r_1}(n)-C_d^{r_2,r_1}(n)\sigma_{r_2}(n) + \frac{\tilde{a}(n)}{n^d}\,,
\end{equation}
where 
\begin{equation}
C_d^{r_1,r_2} (n) = \frac{(r_2-1)!(r_1+d)!}{2(r_1+r_2+d)!} \zeta(r_2) n^{r_2}+ { d+r_2 \choose d}\frac{\zeta'(-r_2)}{2}\,,
\end{equation}
while $\tilde{\Delta}= \sum_{n=1}^\infty \tilde{a}(n)q^n \in \mathcal{S}_{k}$ is a holomorphic cusp form of weight $k$:
\begin{equation}
k=2d+r_1+r_2+2\,,
\end{equation}
given by $\tilde{\Delta}(\tau)=\sum_{\Delta \in  \mathcal{S}_{k}} \mu'_\Delta \Delta$ where $\Delta$ runs over normalised Hecke eigenforms in $\mathcal{S}_{k}$ and
\begin{equation}
\mu'_\Delta = \frac{\pi(-1)^{d+\frac{r_2}{2}+1}}{2^k} {k-2 \choose d} \frac{\Lambda(\Delta,d+1)\Lambda(\Delta, r_1+d+1)}{\langle \Delta,\Delta \rangle}\,.\label{eq:lambdadelta}
\end{equation}}
\vskip 0.1 cm

If we specialise this theorem to the present case \eqref{eq:gsmall} we have $r_1= 2s_1-1$, $r_2 =2s_2-1$ while $d=s-s_1-s_2$. Hence in our case $L$-values of holomorphic cusp forms will appear whenever
\begin{equation}
k= 2d+r_1+r_2+2 = 2(s-s_1-s_2) +(2s_1-1) +(2s_2-1) + 2 = 2 s\,,
\end{equation}
with ${\rm dim}\,\mathcal{S}_{k} \neq 0$.
We then apply this theorem to the infinite sum over $n_1+n_2=n$ of the small-$\tau_2$ expansion \eqref{eq:gsmall} for the particular solution $\hat{g}_{n_1,n_2}$ to deduce the schematic form:
\begin{equation}
\sum_{\substack{ n_1+n_2= n \\ n_1,n_2\neq0}}\hat{g}_{n_1,n_2}(\tau_2) \stackrel{\tau_2\to0}{\sim} \Big[\frac{\zeta (2 s_1 ) +\zeta(2s_2)}{n^{s+s_1+ s_2 -1}}+\frac{\zeta (2 s_1 -1) }{n^{ s+s_2 -s_1}}+\frac{\zeta (2 s_2 -1) }{n^{s+s_1 -s_2 }}+\frac{ \tilde{a}(n)}{ n^{2 s-1}}\Big] \tau_2^{1-s}+ O(\tau_2^{3-s})\,.\label{eq:gn1n2sum}
\end{equation}
Substituting this behaviour in the full $n$-instanton sector \eqref{eq:FourN} we see that the terms proportional to Riemann zetas cancel against the same terms coming from the small-$\tau_2$ expansion of $\hat{g}_{n,0}+ \hat{g}_{0,n}$, in exactly the same fashion as for the zero-mode sector. However, when ${\rm dim}\,\mathcal{S}_{2s} \neq 0$ we see that the power behaved term proportional to $ \tilde{a}(n)\tau_2^{1-s}$ does not vanish and therefore violates the growth condition coming from the \ref{LA}.
We deduce that whenever the eigenvalue $s$ is such that ${\rm dim}\,\mathcal{S}_{2s} \neq 0$ the particular solution constructed using the methods of \cite{Fedosova:2022zrb} cannot possibly be a GES.

However,  although the small-$\tau_2$ limit of the $n^{th}$ Fourier mode contains power-like terms $\tau_2^{-\ell}$ in the range $\ell\in \{w,w+2, \ldots ,s-3,s-1\}$ with $w=s_1+s_2$,  which all violate the \ref{LA}, as mentioned below equation \eqref{eq:gsmall} they all multiply the same overall factor $\tilde{a}(n)/n^s$ and their relative coefficients coincide with the small-$\tau_2$ expansion of $\sqrt{ n \tau_2}K_{s-1/2}(2\pi n \tau_2)$.
This tells us that if we add to the particular solution a suitable multiple of the solution of the homogeneous equation  $H_{\tilde{\Delta}}(\tau)$ given in \eqref{eq:Hdelta} for the holomorphic cusp form $\tilde{\Delta}$ defined in \ref{TH1}, we can cancel all of the unwanted powers of $\tau_2$ in \eqref{eq:gn1n2sum} that do not satisfy the \ref{LA} without spoiling the differential equation \eqref{eq:LapEq}, i.e. for the spectrum \eqref{eq:spec2} we are led to the solution
\begin{align}\label{eq:SolSp2}
\!\!\! \mathcal{E}(s;s_1,s_2;\tau) =  \cE_{{\rm p}}(s;s_1,s_2;\tau) + [\widetilde{C}_{s;s_1,s_2}]^{-1}\! \sum_{\Delta \in \mathcal{S}_{2s}} \!\lambda'_\Delta(s;s_1,s_2) H_{\Delta} (\tau)\,,
\end{align}
where $\widetilde{C}_{s;s_1,s_2}$ is given in \eqref{eq:Ctilde} and the coefficients $\lambda'_\Delta$ are obtained from \eqref{eq:lambdadelta}
\begin{equation}\label{eq:lambdaIC} 
\lambda'_\Delta(s;s_1,s_2)  =    (-1)^{\frac{s+s_1-s_2+2}{2} }  \frac{ \Gamma(s) \Gamma\left(\frac{w-s}{2}\right)}{2^{2s-3}\Gamma\left(\frac{w+s}{2}\right)} \frac{ \Lambda (\Delta, s+s_1-s_2)\Lambda (\Delta, s+w-1)}{\langle \Delta, \Delta\rangle}\,,
\end{equation}
with $w=s_1+s_2$. Using the functional equation \eqref{eq:Lambda} for the completed $L$-function it is easy to show that $\lambda'_\Delta$ is symmetric under $s_1\leftrightarrow s_2$. Importantly, when compared with the identical conjectural coefficients \eqref{eq:lambdaMGF} for the case \eqref{eq:spec1} relevant for MGFs, we see that $\lambda'_\Delta$  for the spectrum \eqref{eq:spec2}  is now always given by the product of two critical even $L$-values.

Crucially, the solution \eqref{eq:SolSp2} produces the GES solution to \eqref{eq:LapEq} with spectrum  \eqref{eq:spec2}. The proof of this statement follows again from the \ref{LA}. Firstly, recall that the particular solution defined in \eqref{eq:SolSp2} was constructed so as not to violate the growth  condition at the origin. While this growth condition is a necessary condition for the modular invariance of $\mathcal{E}(s;s_1,s_2;\tau) $ in this case it turns out to be sufficient as well. This can be seen as follows.  

 Firstly, we can use ${\rm SL}(2,\mathbb{Z})$ spectral methods to show  that the modular invariant GES solution to \eqref{eq:LapEq} for the spectrum of parameters \eqref{eq:spec2} does in fact exist \cite{Klinger-Logan:2018sjt}. Suppose now that the solution \eqref{eq:SolSp2} is not modular invariant and therefore differs from the GES solution, even though it respects the growth condition at the origin for any of its Fourier modes. Since both \eqref{eq:SolSp2} and the GES solution solve the same differential equation  \eqref{eq:LapEq} they must differ by a solution to the homogeneous equation.  However, it is not possible to add any such solution to \eqref{eq:SolSp2} without spoiling the growth condition at the origin.  In the zero-mode sector this would amount to either introducing a $\tau_2^s$ term or changing the coefficient \eqref{eq:beta} of the $\tau_2^{1-s}$ term, neither of which is allowed. In the non-zero mode sector, this would amount to adding a new term  of the form \eqref{eq:HomoGen}. However, since we know that the solution \eqref{eq:SolSp2} grows like $O(\tau_2^{1-w})$ at the origin, adding a new  solution to the homogeneous equation will re-introduce an unwanted  $\tau_2^{1-s}$ term in the limit $\tau_2\to 0$, thus spoiling the necessary growth condition coming from the \ref{LA}.
We conclude that~\eqref{eq:SolSp2} must be the modular invariant GES solution to~\eqref{eq:LapEq} for the SMF spectrum of parameters~\eqref{eq:spec2}.

A similar argument regarding the necessity of having to add a solution of the homogeneous equation  can be made for the GES relevant for the MGFs, with spectrum given in \eqref{eq:spec1}.
Firstly, note that for a GES $\cE(s;s_1,s_2; \tau)$ with parameters satisfying \eqref{eq:spec1} we have that a term proportional to $\tau_2^{1-s}$, originating from the particular solution constructed, does not violate the growth condition at the origin of the form $\tau_2^{1-s_1-s_2}$ dictated by the \ref{LA}. However, we simply need to consider a more refined version of the \ref{LA} where rather than $\cE(s;s_1,s_2; \tau)$, we analyse $F = \cE(s;s_1,s_2; \tau) + \alpha E(s_1{+}s_2; \tau)$ where $\alpha$ is chosen in such a way as to cancel the leading $\tau_2^{s_1+s_2}$ cuspidal growth. The auxiliary function $F$ has now the tamer growth at the cusp  $\tau_2^{s_1+1-s_2}$ (without loss of generality we may assume $s_1\geq s_2$), thus
 implying a behaviour at the origin bounded by $\tau_2^{s_2-s_1}$.
If $s_1> s_2$ we can furthermore construct yet another ``improved'' modular invariant function by considering $F' = F+\beta E(s_1{-}s_2{+}1; \tau)$ with $\beta$ chosen as to cancel the cuspidal behaviour $\tau_2^{s_1+1-s_2}$ so that $F'= O(\tau_2)$ at the cusp.   We can then use the \ref{LA} and bound  $F'$  by the Eisenstein  series  $E(1{+}\epsilon; \tau)$ for all $\epsilon>0$.  For the diagonal case, $s_1=s_2$, we have $F= O(\tau_2)$ at the cusp and  this is directly bounded by $E(1{+}\epsilon; \tau)$ for all $\epsilon>0$.    As a consequence, we deduce that every Fourier modes must diverge at the origin at most as $\tau_2^{-\epsilon}$. The particular solutions constructed in \cite{Dorigoni:2021jfr} via iterated integrals of holomorphic Eisenstein series do not always satisfy this bound, hence they are not in general modular invariant functions and a suitable multiple of the homogenous solution \eqref{eq:PartplusHomo} must be added so as to obtain the GES solution to \eqref{eq:geneisen} with spectrum \eqref{eq:spec1}. 

It should be stressed that there is no analogue of \ref{TH1} for the GESs with {integer $s_1$ and $s_2$,~i.e., with the spectrum \eqref{eq:spec1}}, and so far the only way to fix the coefficient of the solution of the homogeneous equation so as to construct a modular invariant solution to the inhomogeneous Laplace equation has relied on the analysis of \cite{Dorigoni:2021jfr,Dorigoni:2021ngn}. 
In addition to the asymptotic argument just presented, we believe that there should be a more general version of \ref{TH1}  responsible for the validity of \ref{Conj1}. 

Returning to the case of SMFs, we see that the first instance where $L$-values of holomorphic cusp forms appear in the GESs \eqref{eq:SolSp2} is again for eigenvalue $s=6$ where we know that $ \cS_{12}$ is a one-dimensional vector spaced spanned by Ramanujan cusp form $\Delta_{12}$. From the spectrum \eqref{eq:spec2} we see that an eigenvalue $s=6$ necessitates weight $w=s_1+s_2=3$ or $w=5$. In these cases the GESs are given by 
\allowdisplaybreaks{
\begin{align}
\mathcal{E}(6;\half, \fiveh;\tau) &\notag =  \cE_{{\rm p}}(6;\half, \fiveh; \tau) - \frac{625 }{ 6048} \cdot \frac{\Lambda(\Delta_{12},10)^2}{\pi \langle \Delta_{12},\Delta_{12}\rangle}H_{\Delta_{12}}(\tau)\,,\\
\mathcal{E}(6;\threeh,\threeh;\tau) &\notag =  \cE_{{\rm p}}(6;\threeh,\threeh;\tau) + \frac{25 }{ 72} \cdot \frac{\Lambda(\Delta_{12},10)^2}{\pi \langle \Delta_{12},\Delta_{12}\rangle}H_{\Delta_{12}}(\tau)\,,\\
\mathcal{E}(6;\half, \nineh;\tau) &\label{eq:E6IC} =  \cE_{{\rm p}}(6;\half, \nineh; \tau) -\frac{32}{6615}     \cdot \frac{ \Lambda \left(\Delta _{12},10\right)^2}{  \pi \langle \Delta_{12},\Delta_{12}\rangle}H_{\Delta_{12}}(\tau)\,,\\
\mathcal{E}(6;\threeh,\sevenh;\tau) &\notag =  \cE_{{\rm p}}(6;\threeh,\sevenh;\tau) +\frac{10}{189}     \cdot \frac{ \Lambda \left(\Delta _{12},10\right)^2}{  \pi \langle \Delta_{12},\Delta_{12}\rangle}H_{\Delta_{12}}(\tau)\,,\\
\mathcal{E}(6;\fiveh,\fiveh;\tau) &\nn  =  \cE_{{\rm p}}(6;\fiveh,\fiveh;\tau)- \frac{8}{81}   \cdot \frac{ \Lambda \left(\Delta _{12},10\right)^2}{  \pi \langle \Delta_{12},\Delta_{12}\rangle}H_{\Delta_{12}}(\tau)\,, 
\end{align}}
\!\!\!\! where, again, we have used Manin's relations to express the results in terms of $\Lambda(\Delta_{12},10)$ only. 
As was the case with \eqref{eq:E6MGFs2} the seemingly random rational numbers appearing in front of the solutions of the homogeneous equation will shortly prove to be rather special.

To summarise, in this section we have shown that for both spectra \eqref{eq:spec1} and \eqref{eq:spec2} (relevant to MGFs and SMFs, respectively) it is possible to find GES solutions of the form \eqref{eq:PartplusHomo} and \eqref{eq:SolSp2}.  In either case the modular invariant solution is obtained by first constructing a particular solution whose Fourier modes have asymptotic expansions at the cusp with coefficients that are quadratic forms in Riemann zeta values.  Furthermore, in each case it is necessary to add a suitable multiple of the solution of the homogeneous equation  that contributes only to the $(0,n)$ and $(n,0)$ instanton sector and has an asymptotic expansion at the cusp with coefficients controlled by the $q$-series coefficients of holomorphic cusp forms and their associated critical and non-critical $L$-values.

\subsection{Absence of $L$-values in theta-lifted   local Maass functions}
\label{sec:Absence}
 Although there is no known lattice representation for an individual GES we now use the results of section \ref{sec:LapEq} to show that the lattice sum constructed as a theta lift of a local Maass function \eqref{eq:Ewij2}  produces special linear combinations of GESs for both  MGFs and SMFs.  Furthermore, these four-dimensional lattice sum representations produce precisely the rational linear combinations of GESs \eqref{eq:PartplusHomo} and \eqref{eq:SolSp2} for which  the cuspidal contribution $H_\Delta$ cancels.
Once again, we will present the cases of MGFs and SMFs separately.

\subsubsection*{Two-loop Modular Graph Functions}

Using the results of section \ref{sec:LapEq}, we see that if we want to obtain linear combinations of GESs with parameters in the spectrum \eqref{eq:spec1} relevant for MGFs we have to consider the theta lift \eqref{eq:Ewij2}, or equivalently \eqref{eq:Ewijt}, for $w\geq 4$ and $1\leq s\leq w-2$ with the same parity.
As a consequence we see from \eqref{eq:Aijexp} that the integrand \eqref{eq:BwijDef} reduces to a symmetric polynomial in $t_i$ of degree $w-3$.

It is then straightforward to perform the integration \eqref{eq:Ewijt} using the basic integral representation of a MGF given in \eqref{eq:cabcint}.   We deduce that the theta lift \eqref{eq:Ewij2} of a local Maass function with $w\geq 4$ and $1\leq s\leq w-2$ produces a rational linear combination of MGFs  ($C_{a,b,c}(\tau)$ with $w=a+b+c$), that solves an inhomogeneous Laplace equation with eigenvalue $s$ and source terms given by bilinears, $E(s_1; \tau)E(s_2; \tau)$, in non-holomorphic Eisenstein series with $w=s_1+s_2$ and possibly a multiple of the single non-holomorphic Eisenstein series $E(w; \tau)$.

Moreover, the dimension of the vector eigenspace $\cV_C(w,s)$ of MGFs of a given weight $w=a+b+c$ and eigenvalue $s$ was determined in \cite{DHoker:2015gmr} to be
\begin{equation}
\label{eq:dimC}
\dim \mathcal{V}_C(w,s) = \left\{\begin{array}{cl} \left\lfloor \frac{s+2}{3}\right\rfloor &\text{for $1\leq s\leq w{-}2$ and $s,w$ of same parity,} \\[2mm]
0 &\text{otherwise.}
\end{array} \right. 
\end{equation}
Note that this matches the total number of functions $B^{w}_{i,j}$ defined in \eqref{eq:BwijDef} consistent with the spectrum \eqref{eq:spec1} given that $s=3i+j+1$.  Hence we see that the theta lift off local Maass functions with $w\geq 3$ and $1\leq s\leq w-2$ results in the vector space of $C_{a,b,c}(\tau)$ conveniently arranged in terms of eigenvectors of the Laplace operator.

However, a similar calculation of the number of independent GESs $\cE(s;s_1,s_2; \tau)$ with spectrum \eqref{eq:spec1} gives 
\begin{equation}
\label{eq:dimF}
\dim \mathcal{V}_{\cE}(w,s) = \left\{\begin{array}{cl} \left\lfloor \frac{s}{2}\right\rfloor &\text{for $2\leq s\leq w{-}2$ and $s,w$ of the same parity,}\\[2mm]
0 &\text{otherwise,}\end{array}\right.
\end{equation}
where $w=s_1 + s_2$. Note that  $\cV_C(w,s)$ includes the eigenvalue $s=1$, which occurs when $w$ is odd.
In this case, the Laplace equation reduces in complexity and the corresponding eigenvector becomes a non-holomorphic Eisenstein series plus a constant.  This special case with $s=1$ corresponds to the theta lifts of $B^w_{0,0}$ for $w\geq 3$.  For example, when $w=3$ we have  $B^3_{0,0}(t)=1$ and 
\begin{align}
 \cE^3_{0,0}(\tau) &= \sum_{\substack{ p_1,p_2,p_3 \in\Lambda' \\ p_1+p_2 +p_3=0}} \int_{( \mathbb{R}^+)^3 }    \exp\Big( -\frac{\pi}{\tau_2} \sum_{j=1}^3 t_j |p_j|^2 \Big)\,  {\rm d}^3 t =  C_{1,1,1}(\tau) = E(3;\tau)+\zeta(3)\,.
\end{align}

Comparing the dimensions~\eqref{eq:dimC} and~\eqref{eq:dimF} we see that there are in general more GESs $\cE(s;s_1,s_2; \tau)$ with spectrum \eqref{eq:spec1} than the $C_{a,b,c}(\tau)$. As noticed in \cite{Dorigoni:2021jfr} for $s\geq2$ the difference in dimensions gives strikingly,
\begin{equation}
\dim \mathcal{V}_{\cE}(w,s) - \dim \mathcal{V}_C(w,s)  = \dim \mathcal{S}_{2s}\,.\label{eq:dimES}
\end{equation}
Since $\dim \mathcal{S}_{2s}=0$ for $s<6$ and since $s<w-2$ it follows that  at weight $w=8$ and at any weight $w\geq 10$, the number of independent GESs is 
strictly larger than that of $C_{a,b,c}(\tau)$. More precisely, the number of GESs at weight $w$ and eigenvalue $s$ in the MGF spectrum \eqref{eq:spec1} that cannot be written in terms of MGFs equals the number of holomorphic cusp forms at modular weight $2s$ (which is given in  \eqref{eq:DimS}).

The first case for which this mismatch starts playing a role is $s=6$, for which $\cS_{12}$ is the one-dimensional vector space $ \cS_{12}= {\rm span}\{\Delta_{12}\}$,  and necessarily $w\ge 8$.
From the spectrum \eqref{eq:spec1} we see that for $(s,w)=(6,8)$ there are three GESs, $\cE(6;2,6; \tau), \cE(6;3,5; \tau)$ and $\cE(6;4,4; \tau)$,  each of which can be written in terms of iterated integrals of holomorphic Eisenstein series of depth at most two and crucially a non-zero homogeneous contribution originating from an iterated integral of the Ramanujan cusp form $\Delta_{12}$ as seen in \eqref{eq:E6MGFs}-\eqref{eq:E6MGFs2}.

Conversely, there are only two theta-lifted  local Maass functions \eqref{eq:Ewijt} corresponding to MGFs with weight $8$ and eigenvalue $6$ given by $\cE_{1,2}^8(\tau)$ and $\cE_{0,5}^8(\tau)$. Thanks to the results of section \ref{sec:LapEq}, we have 
\begin{align}
\cE_{1,2}^8(\tau) &\notag = \sum_{\underset {p_1+p_2+p_3=0} {p_1,p_2, p_3\ne 0}} \int_0^\infty \frac{\left(7 {\sigma_1}^3 {\sigma_2}-63 {\sigma_1}^2 {\sigma_3}+3 {\sigma_1} {\sigma_2}^2+21 {\sigma_2} {\sigma_3}\right)}{63} \exp\Big( - \frac{\pi}{\tau_2}\sum_{i=1}^3 t_i |p_i|^2  \Big) \, {\rm d}^3 t \\
&\label{eq:cE128}= -\frac{24}{7} \big[112\, \cE(6;3,5;\tau)+75 \,\cE(6;4,4;\tau) \big] - \frac{436}{7} \, E(8;\tau) \\
&\notag =\frac{8}{7} \big[ 14 \,C_{1,2,5}(\tau)+24\, C_{1,3,4}(\tau)-2 \,C_{2,2,4}(\tau)-C_{2,3,3}(\tau) \big]\,,
\end{align}
where the notation in the first line utilises  the basis of symmetric polynomials in three variables $\sigma_1=t_1+t_2+t_3$, $\sigma_2 = t_1 t_2 +t_1 t_3+ t_2 t_3$ and $\sigma_3 =t_1t_2t_3$. Similarly we find,
\begin{align}
\cE_{0,5}^8(\tau) &\notag = \sum_{\underset {p_1+p_2+p_3=0} {p_1,p_2, p_3\ne 0}} \int_0^\infty \frac{ {\sigma_1} \left(21 {\sigma_1}^4-70 {\sigma_1}^2 {\sigma_2}+45 {\sigma_2}^2\right)}{21} \exp\Big( - \frac{\pi}{\tau_2}\sum_{i=1}^3 t_i |p_i|^2  \Big)  {\rm d}^3 t \\
&\label{eq:cE058} = - \frac{180}{7}\big[ 140 \, \cE(6;2,6;\tau) + 112 \,\cE(6;3,5;\tau) + 51\, \cE(6;4,4;\tau) \big] + \frac{87270}{91} \, E(8;\tau) \\
&\notag = \frac{60}{7} \big[ 42\, C_{1,1,6}(\tau)+28\, C_{1,2,5}(\tau)+18\, C_{1,3,4}(\tau)+2 \, C_{2,2,4}(\tau)+C_{2,3,3}(\tau) \big]\,.
\end{align}
At this point it is straightforward to substitute {\eqref{eq:E6MGFs2}}  in the above relations and check that indeed the solution of the homogeneous equation  involving $L$-values of holomorphic cusp forms do drop out and both $\cE_{1,2}^8(\tau)$ and $\cE_{0,5}^8(\tau)$ can be written as linear combinations of iterated integrals involving solely holomorphic Eisenstein series (and their complex conjugate).

This phenomenon whereby the holomorphic cusp form contributions  to GESs conspire to cancel out in linear combinations ultimately given by lattice sums corresponding to MGFs had been studied in \cite{Dorigoni:2021jfr,Dorigoni:2021ngn}. These magical cancellations provide the first hint of a beautiful algebraic structure at play behind the scenes. 
To better understand this story we need to consider a wider class of non-holomorphic modular covariant forms introduced by Francis Brown.
The intriguing mathematical properties of MGFs stimulated Brown's construction of an infinite family of non-holomorphic modular forms dubbed equivariant iterated Eisenstein integrals \cite{Brown:mmv,Brown:I,Brown:II} but the connection between MGFs and equivariant iterated Eisenstein integrals is indirect, see e.g. \cite{Dorigoni:2022npe,Dorigoni:2024oft}.

In \cite{Brown:I,Brown:II} Brown has shown how to construct non-holomorphic forms as combinations of non-modular invariant iterated integrals of holomorphic Eisenstein series and their complex conjugates. However, not every equivariant iterated integral  corresponds to a higher-loop MGF.
Notably, in the Fourier expansion of modular graph forms we encounter only a very small subset of the vast number of multiple modular values given by single-valued multiple zeta values and, in particular, we do not encounter completed $L$-values of holomorphic cusp forms, which can instead be present in a given equivariant iterated integral.
However, if we consider the generating series of all modular graph forms constructed in \cite{Gerken:2019cxz,Gerken:2020yii} we see that only special linear combinations of Brown's equivariant iterated integrals correspond to modular graph forms.

To extract a particular MGF from the generating series we simply need to extract the coefficient of certain non-commutative sign-post variables, which {conjecturally} furnish a representation of an important mathematical object called Tsunogai's derivation algebra~\cite{Tsunogai}.
As a consequence of this fact, we find that MGFs are {conjecturally} related only to specific linear combinations of equivariant iterated Eisenstein integrals, namely those that follow from the so-called Tsunogai's relations~\cite{Tsunogai,Pollack}, see also~\cite{Brown:II, Gerken:2020yii, Dorigoni:2021ngn, Dorigoni:2022npe,Dorigoni:2024iyt}. 
While the modular completion of iterated integrals of holomorphic Eisenstein series necessitates the addition of iterated integrals of holomorphic cusp forms~\cite{Brown:mmv, Brown:II}, such as for the depth-two examples \eqref{eq:E6MGFs}, the linear combinations of equivariant iterated integrals selected by Tsunogai's relations and hence {conjecturally} corresponding to MGFs are expressible in terms of iterated integrals involving solely holomorphic Eisenstein series (and their complex conjugates) and in addition the periods arising in this construction are conjecturally restricted solely to single-valued MZVs. 

As a concrete consequence of this structure, we find that the theta lift of modular local polynomials corresponding to the spectrum \eqref{eq:spec1} produces precisely the linear combinations of GESs in \eqref{eq:cE128}-\eqref{eq:cE058}, or equivalently of depth-two equivariant iterated Eisenstein integrals for which the cusp form contributions have dropped out. We expect that a generalisation of \eqref{eq:Ewij2} to a higher-dimensional lattice sum will produce higher-loop MGFs and hence the corresponding linear combinations of higher-depth iterated Eisenstein integrals selected by Tsunogai's relations.

\subsubsection*{S-dual Modular Functions}

We now turn to consider the theta lift of local Maass functions giving rise to linear combinations of GESs with the spectrum \eqref{eq:spec2}, in which case $s_1$ and $s_2$ are half-integral.
From the general structure \eqref{eq:Eijsolved}, we see that in order to  obtain linear combinations of GESs with this spectrum we need to consider the case where $w$ and $s$ have opposite parity.
Furthermore, since in general $i+j\leq s-1$ where $s=3i+j+1$, we see that the indices $s_1,s_2$ of the GESs $\cE(s;s_1,s_2; \tau)$, which appear in  \eqref{eq:Eijsolved} always lie between $(w-s+2)/2\leq s_1,s_2 \leq (w+s-2)/2$. Hence in order to obtain the spectrum \eqref{eq:spec2} where $s\geq w+1$, some of these GESs might have negative indices.
This is however not a problem since if $s_1$ is a negative half-integer the functional relation \eqref{eq:GenEisFunct} can be used to consider instead $\cE(s;1-s_1,s_2; \tau)$. As previously mentioned, this comes at the price of spoiling uniform transcendentality  in \eqref{eq:Eijsolved}.

We can now analyse the vector space of theta-lifted Maass functions at fixed weight $w$ and eigenvalue $s$, as we did earlier for MGFs.
From the analysis of the previous section we expect that the lowest eigenvalue for which some cusp form might appear to be $s=6$, which from \eqref{eq:spec2} requires either $w=3$ or $w=5$. 
This leads us to consider the theta lifts $\cE^w_{1,2}(\tau), \cE^w_{0,5}(\tau)$ with $w\in\{3,5\}$.
It is now a matter of exploiting \eqref{eq:Eijsolved} combined with \eqref{eq:bij}-\eqref{eq:dij} to obtain
\allowdisplaybreaks{
\begin{align}
\cE_{1,2}^3(\tau) &\notag = \sum_{\underset {p_1+p_2+p_3=0} {p_1,p_2, p_3\ne 0}} \int_0^\infty  \, \frac{\left(7 {\sigma_1}^3 {\sigma_2}-63 {\sigma_1}^2 {\sigma_3}+3 {\sigma_1} {\sigma_2}^2+21 {\sigma_2} {\sigma_3}\right)}{63\, \sigma_2^{\fiveh}} \exp\Big( - \frac{\pi}{\tau_2}\sum_{i=1}^3 t_i |p_i|^2  \Big) {\rm d}^3 t\\*
&\label{eq:E123}=-\frac{ \pi }{14}  \left[84 \,\cE(6;\half, \fiveh;\tau)+25\, \cE(6;\threeh,\threeh;\tau)\right]-\frac{\pi}{2} \, E(3;\tau) \,,\\
\cE_{0,5}^3(\tau) &\notag = \sum_{\underset {p_1+p_2+p_3=0} {p_1,p_2, p_3\ne 0}} \int_0^\infty d^3 t \, \frac{ {\sigma_1} \left(21 {\sigma_1}^4-70 {\sigma_1}^2 {\sigma_2}+45 {\sigma_2}^2\right)}{21\,\sigma_2^{\fiveh}} \exp\Big( - \frac{\pi}{\tau_2}\sum_{i=1}^3 t_i |p_i|^2  \Big)\\*
&\label{eq:E053}=-\frac{15  \pi }{56}  \left[168 \,\cE(6;\half, \fiveh;\tau)+34 \,\cE(6;\threeh,\threeh;\tau)+105\, \cE(6;\threeh,\sevenh;\tau)\right]+\frac{225\pi}{32}\, E(3;\tau)\,,
\end{align}}
with similar equations for $\cE^5_{1,2}(\tau), \cE^5_{0,5}(\tau)$, which can be expressed as 
\begin{align}
\cE_{1,2}^5(\tau)  &\label{eq:E125} =-\frac{15\pi}{56} \left[ 28\, \cE(6;\threeh,\sevenh;\tau) +15\, \cE(6;\fiveh,\fiveh;\tau)\right] -\frac{125\pi }{56}\,{E(5;\tau)}  \,,\\
\cE_{0,5}^5(\tau)  &\label{eq:E055}= -\frac{45 \pi}{112} \left[490\, \cE(6;\half,  \nineh;\tau) +140\,\cE(6;\threeh,\sevenh;\tau) +51\,\cE(6;\fiveh,\fiveh;\tau)\right]-\frac{16347\pi }{224} \,E(5;\tau)\,.
\end{align}
 Note that the GESs on the right-hand side of  \eqref{eq:E053} do not have uniform weight $3$ but  also contain $\cE(6;\threeh,\sevenh; \tau)$, which arises from inverting the negative index of   $\cE(6;-\half, \sevenh; \tau)$ using the functional equation.

Crucially, if we substitute \eqref{eq:E6IC} into the linear combinations \eqref{eq:E123}-\eqref{eq:E055} we can easily check that the cuspidal contributions  to the solution of the  homogeneous equations do in fact cancel out.
In other words we see that the linear combinations of GESs belonging to the spectrum \eqref{eq:spec2} that are selected by the theta lift of local Maass functions are precisely the same as those for which the $L$-values of holomorphic cusp forms are absent.
In contrast to the MGF case, in the present case there is no Tsunogai's derivation algebra, which 
could be responsible for such cancellations. However, it is entirely possible that upon studying higher-dimensional lattice versions of these particular theta lifts, which play the analogue of higher-loop MGFs in the previous section, such a structure might emerge thus justifying this seemingly magical cancellation of holomorphic cusp forms.

{The results described in this section provide substantial evidence for the validity of \ref{Conj2} presented in section \ref{sec:Results}, which implies that  the modular invariant function ${\cal E}_{i,j}^w(\tau)$, given by the theta lift of 
{$B_{i,j}^w(t) $} defined in \eqref{eq:Ewijt}, is a rational linear combination of GESs, which admits a four-dimensional lattice sum representation and receives no contributions from $L$-values of holomorphic cusp forms. This applies both to MGFs  and to SMFs. }   
It is worth stressing that for the case of SMFs, thanks to the result of \cite{Fedosova:2023cab} we have derived a general expression \eqref{eq:SolSp2}-\eqref{eq:lambdaIC} for the $L$-value contribution to an arbitrary GES in the spectrum \eqref{eq:spec2}.  Hence what we will now show amounts to a proof of \ref{Conj2} specialised to the case of SMFs.
However, for the MGFs the expression for the $L$-value contribution \eqref{eq:lambdaMGF2t} to an arbitrary GES in the spectrum \eqref{eq:spec1} is only conjectural.

Let us consider the theta lift of {$B_{i,j}^w(t) $} as defined in \eqref{eq:Ewijt}. From the general expression \eqref{eq:general-GES2} we know how to write $\cE_{i,j}^{w}(\stau) $ as a rational linear combination of GESs. Thanks to \eqref{eq:general-GES2} we then deduce that the $L$-value contribution to $\cE_{i,j}^{w} ( \tau)$ coming from a single Hecke eigenform $\Delta \in \mathcal{S}_{2s}$, which we denote by $ \cE_{i,j}^{w} ( \tau )\vert_{\Delta} $, must take the form
\ie 
 \cE_{i,j}^{w}(\stau)\Big\vert_{\Delta}  = 3\sum _{k=0}^{2 i}  \sum _{\ell=0}^j \sum _{m=0}^\ell \big[ 1 - (-1)^{k+\ell+m}  \big]  C_{i,j; k, \ell, m} \, \widetilde{C}_{s; s_1, s_2} \,  \cE(s; s_1, s_2; \tau) \Big\vert_\Delta \, , \label{eq:Eijcusp}
\fe 
with $ s_1 = (w+k + \ell + m -i-j)/2$ and  $s_2 = [w-(k + \ell + m -i-j)]/2$.
For the MGF case, the GESs appearing in the linear combination on the right-hand side of this equation belong to the spectrum of parameters \eqref{eq:spec1}, while for the SMFs they belong to the spectrum \eqref{eq:spec2}. 
Hence, if we want to extract the $L$-value contribution to \eqref{eq:Eijcusp} specialised to the MGFs we must use the conjectural expressions  \eqref{eq:lambdaMGF2t}, while for SMFs we have proved that the $L$-value contribution is given by \eqref{eq:SolSp2}-\eqref{eq:lambdaIC}.

Importantly, our \ref{Conj1} states that the $L$-value contribution for GESs with spectrum \eqref{eq:spec1} has the same functional dependence on the parameters $s,s_1,s_2$ as \eqref{eq:SolSp2}-\eqref{eq:lambdaIC}.
Substituting \eqref{eq:SolSp2}-\eqref{eq:lambdaIC}, we deduce that for both MGFs and SMFs the $L$-value contribution \eqref{eq:Eijcusp} takes the form
\begin{equation}
 \cE_{i,j}^{w}(\stau)\Big\vert_{\Delta}  =  3   \frac{ \Gamma(s) \Gamma\left(\frac{w-s}{2}\right)}{2^{2s-4}\Gamma\left(\frac{w+s}{2}\right)} \frac{\Lambda (\Delta, s+w-1)}{\langle \Delta, \Delta\rangle} \mathcal{K}_{i,j}(\Delta)\,,\label{eq:EijH}
\end{equation}
where we have defined
\ie \label{eq:general-L}
\mathcal{K}_{i,j}(\Delta) \coloneqq  {1 \over 2} \sum _{k=0}^{2 i}  \sum _{\ell=0}^j \sum _{m=0}^\ell \big[  (-1)^{k+\ell+m}  -1 \big]  C_{i,j; k, \ell, m} \, (-1)^{\frac{2 i+k+\ell+m+1}{2}}  \Lambda (\Delta_{2s}, 2 i+k+\ell+m+1)  \, ,
   \fe
and again $s=3i+j+1$.
It is important to note that the coefficients $\widetilde{C}_{s; s_1, s_2}$ defined in \eqref{eq:Ctilde} and appearing in \eqref{eq:Eijcusp} cancel out when we consider the cuspidal contributions \eqref{eq:lambdaMGF2t} and \eqref{eq:SolSp2}.
   
We shall prove that $\mathcal{K}_{i,j} (\Delta)$ vanishes identically for all $i,j\in \mathbb{N}$ and for every Hecke eigenforms $\Delta\in \mathcal{S}_{2s}$ with $s=3i+j+1$, i.e.
\begin{equation}
\mathcal{K}_{i,j} (\Delta) = 0 \,,\qquad \qquad \forall \,i,j\in \mathbb{N}\,, \quad \forall \,\Delta\in \mathcal{S}_{6i+2j+2}\,.
\end{equation}
To prove this statement we proceed as follows. Firstly, we rewrite the $L$-values in \eqref{eq:general-L} using the integral representation discussed in appendix \ref{app:Cusps},
\begin{equation}
\Lambda(\Delta,\ell+1) = i^{-\ell-1} \int_0^{i\infty} \Delta(\tau) \tau^\ell {\rm d}\tau\,.\label{eq:LambdaInt}
\end{equation}
The sum of completed $L$-values \eqref{eq:general-L} can then be written simply as
\begin{align}
\mathcal{K}_{i,j}(\Delta) &\notag = {1 \over 2} \int_0^{i\infty} \Delta(\tau)  \Big[ \sum _{k=0}^{2 i}  \sum _{\ell=0}^j \sum _{m=0}^\ell \big[  (-1)^{k+\ell+m}  -1  \big]  C_{i,j; k, \ell, m} \tau^{2 i+k+\ell+m+1}\Big]{\rm d}\tau \\
&\label{eq:general} = {1 \over 2} \int_0^{i \infty} \Delta(\tau) \,  \tau ^{2i} \left[   (\tau -1)^{2i} (\tau^2 - \tau +1)^j - (\tau +1)^{2i}  \left(\tau ^2+\tau +1\right)^j \right] {\rm d} \tau\, , 
\end{align}
where we have used the definition \eqref{eq:Cijklm} for the coefficients $C_{i,j; k, \ell, m}$ originating from having expanded \eqref{eq:uvExp}.
The two terms in \eqref{eq:general} are actually identical as one can easily check by changing the integration variable of the first term to $\tau = S\cdot \tau' = -1/\tau'$ and using the fact that $\Delta$ is a modular form of weight $2s=2(3i+j+1)$.
We then deduce that \eqref{eq:general} is identical to
\begin{equation}
\mathcal{K}_{i,j}(\Delta)  = \int_0^{i \infty} \Delta(\tau) \, u^i v^j {\rm d} \tau\,,\label{eq:cuspij}
\end{equation}
where $u^i v^j$ is precisely the modular local polynomial defined in \eqref{eq:uvdef} (here a function of the variable $\tau$) whose theta lift $\mathcal{E}^w_{i,j}(\tau)$ we are presently studying.

To prove that \eqref{eq:cuspij} vanishes identically for all $i,j\in \mathbb{N}$ and for every Hecke eigenforms $\Delta\in \mathcal{S}_{2s}$ with $s=3i+j$ we simply need to consider the action of the order-three generator of ${\rm SL}(2,\mathbb{Z})$ given by $U= \left(\begin{smallmatrix} 1&-1\\ 1 & 0 \end{smallmatrix}\right)$, which maps 
\begin{align*}
&U\cdot 0 = i \infty \,,\qquad\qquad U\cdot 1 = 0 \,,\qquad \qquad U \cdot i\infty = 1\,.
\end{align*}
Since the modular local polynomial $u^iv^j$  transforms  under $U$ with modular weight $-2(3i+j)=-2(s-1)$ and given that $\Delta\in \mathcal{S}_{2s}$ we see that $\Delta(\tau) \, u^i v^j {\rm d} \tau$ is invariant under $U$.  We deduce that under the change of integration variables $\tau = U\cdot \tau'$ equation  \eqref{eq:cuspij}  becomes  
\begin{align}
\mathcal{K}_{i,j}(\Delta)  &\notag =   \big( \int_0^{1} + \int_1^{i\infty} \big) \Delta(\tau) \, u^i v^j {\rm d} \tau \\
& =    \big( \int_1^{i\infty} + \int_{i\infty}^0 \big) \Delta(\tau') \, u'^i v'^j {\rm d} \tau'\,,\label{eq:U}
\end{align}
where with $u'^i v'^j$ simply means the same modular local polynomial that was defined in \eqref{eq:uvdef} in the variable $\tau'$. Similarly, we rewrite \eqref{eq:cuspij} using the change of integration variables $\tau = U^2\cdot \tau''$, 
\begin{align}
\mathcal{K}_{i,j}(\Delta)  &\notag =   \big( \int_0^{1} + \int_1^{i\infty} \big) \Delta(\tau) \, u^i v^j {\rm d} \tau \\
& =   \big( \int_{i\infty}^{0} + \int_{0}^1 \big) \Delta(\tau'') \, u''^i v''^j {\rm d} \tau''\,,\label{eq:U2}
\end{align}
Combining \eqref{eq:U} with \eqref{eq:U2} we deduce that $\mathcal{K}_{i,j}(\Delta)=0$ for all $i,j\in \mathbb{N}$ and for every Hecke eigenforms $\Delta\in \mathcal{S}_{2s}$ with~$s=3i+j+1$.  
Finally, thanks to \eqref{eq:EijH} we conclude that ${\cal E}_{i,j}^w(\tau)$ receives no contributions from $L$-values of holomorphic cusp forms contained in each individual GES. 

Furthermore, a quick counting argument shows that \textit{any} rational linear combination of GESs, $\mathcal{E}(s;s_1,s_2;\tau)$, at fixed weight $w=s_1+s_2$ and eigenvalue $s$ with parameters $(s,s_1,s_2)$ in either  \eqref{eq:spec1} or \eqref{eq:spec2} and for which the cusp form contribution cancels must be a linear combination of theta lifts ${\cal E}_{i,j}^w(\tau)$ with $s=3i+j+1$.
For the spectrum of parameters \eqref{eq:spec1} it is easy to see that at fixed eigenvalue $s$ we have $\lfloor \frac{s}{2}\rfloor$ linearly independent GESs at fixed weight $w=s_1+s_2$.  Similarly, for the spectrum of parameters \eqref{eq:spec2} at fixed eigenvalue $s$ the largest number of GESs is found at weight $w=s-1$ where there are  again  $\lfloor \frac{s}{2}\rfloor$ linearly independent GESs.
In both cases we can write a general rational linear combination of GESs and impose cancellation of the $L$-values, using either the conjectural expression \eqref{eq:lambdaMGF2t} or \eqref{eq:lambdaIC}, and we obtain a system of ${\rm dim}\,\mathcal{S}_{2s}$ equations\footnote{Since we are working with rational linear combinations of GESs and the cusp form coefficients \eqref{eq:lambdaMGF2t} and \eqref{eq:lambdaIC} take values in the number field associated with $\mathcal{S}_{2s}$ we deduce that a single linear equation with number field coefficients becomes a system of ${\rm dim}\,\mathcal{S}_{2s}$ equations with rational coefficients.}  in 
$\lfloor\frac{s}{2}\rfloor$ variables.
From the dimension formula \eqref{eq:DimS} we see that this system of equations has exactly $\lfloor \frac{s-1}{3} \rfloor$ linearly independent solutions, which is precisely the number of ${\cal E}_{i,j}^w(\tau)$ with $s=3i+j+1$.

 We conclude this section by noting that the cuspidal contribution defined in \eqref{eq:general-L} for arbitrary $i,j\in \mathbb{N}$ as a finite sum over completed $L$-values can be simplified dramatically for the particular infinite families of $\cE^w_{i,j}(\tau)$ considered in section \ref{sec:bc}.
For example, if we consider the infinite family $\cE^w_{i,0}$ with eigenvalue $s=3i+1$ and integer weight $w\leq s-1$ constructed using \eqref{eq:ci00}-\eqref{eq:ci01}, we find that the vanishing of the corresponding cuspidal contribution $\mathcal{K}_{i,0}(\Delta_{6i+2})$ with $\Delta_{6i+2}\in \mathcal{S}_{6i+2}$ reduces to
\begin{equation}
\mathcal{K}_{i,0}(\Delta_{6i+2}) \propto \sum_{r=-\frac{i-1}{2}}^{\frac{i-1}{2}} (-1)^{r+\frac{i-1}{2}}  \frac{\Lambda (\Delta_{6 i+2},(3i+1)+2 r)}{ \Gamma (i+1+2r) \Gamma (i+1-2r)}\,.\label{eq:cusp1}
\end{equation} 
The first value of $i$ for which the identity $\mathcal{K}_{i,0}(\Delta_{6i+2})=0$ becomes non-trivial is $i=3$ for which we must consider the unique Hecke normalised cusp form of weight $6i+2 = 20$ and, after using the $L$-function reflection identity \eqref{eq:Lambda}, equation \eqref{eq:cusp1} for $i=3$ reduces  to
\begin{equation}
-\frac{\Lambda(\Delta_{20},10)}{36} + \frac{\Lambda(\Delta_{20},12)}{60} =0\,,
\label{eq:L20}
\end{equation}
which is a consequence of Manin's periods theorem \cite{Manin}. 
It is easy to show that using the integral representation \eqref{eq:LambdaInt}, equation \eqref{eq:cusp1} coincides with the general vanishing expression \eqref{eq:general} specialised to $j=0$. 

If we repeat a similar argument for the infinite family $\cE^w_{0,j}(\tau)$ with eigenvalue $s=j+1$ and weight $w\leq s-1$ and use \eqref{eq:c0j0} to compute the 
cuspidal contribution $\mathcal{K}_{0,j}(\Delta_{2j+2})$ we find
\begin{equation}
\mathcal{K}_{0,j}(\Delta_{2j+2}) \propto \sum _{r =-\frac{j-1}{2} }^{\frac{j-1}{2}} \frac{(-1)^{r } \, _2F_1\left(-\frac{j}{2}-r +\half-\frac{j}{2}+r +\frac{1}{2};\frac{1}{2}-j\vert \frac{3}{4}\right) \Lambda (\Delta_{2 j+2},2 r +j+1)}{\Gamma (j-2 r +1) \Gamma (j+2 r +1)}\,.\label{eq:cusp4}
\end{equation}
Surprisingly, for this infinite family of linear combination of GESs, the cuspidal contribution \eqref{eq:cusp4} becomes a rather complicated linear combination of completed $L$-values, although it is the result of  the integral representation \eqref{eq:LambdaInt}  by specialising to $i=0$. We have not performed an analytic derivation of this starting from  \eqref{eq:LambdaInt}, however, with the aid of Pari/GP \cite{parigp} we have numerically verified that \eqref{eq:cusp4} is satisfied for all cusp forms of weight $2s\leq 100$ within the numerical precision of $10^{-50}$ utilised.

\section{Discussion}
\label{sec:Discussion}

We conclude with comments on some possible extension of the current work.

\subsection{Cuspidal generalised Eisenstein series}

Within the context of the low-energy expansion for genus-one string scattering amplitudes, it is also necessary to consider a different class of odd MGFs \cite{DHoker:2019txf,Gerken:2020yii}. These odd MGFs are again modular invariant functions that can be represented in terms of lattice sums similar to \eqref{eq:Cabc}.
However, they are odd under the involution of the upper-half plane $\tau \to -\bar\tau$, thus necessarily leading to a cuspidal expansion at $\tau_2\gg1$, i.e. rather than having polynomial growth as for the GESs \eqref{genEisPert} their asymptotic expansion at the cusp must necessarily be $O(e^{2\pi i \tau})$.
In \cite{Dorigoni:2021jfr,Dorigoni:2021ngn} it was realised that equation \eqref{eq:geneisen} can be modified so as to obtain a second vector space of \textit{cuspidal generalised Eisenstein series},
\begin{equation}\label{eq:cuspGES}
(\Delta_\tau-s(s-1)) \mathcal{E}^-(s;s_1,s_2;\tau) = \frac{\nabla E(s_1;\tau) \, \overline\nabla E(s_2;\tau) - \nabla E(s_2;\tau) \, \overline\nabla E(s_1;\tau) }{2 \tau_2^2}\,,
\end{equation}
with $\nabla \coloneqq 2i \tau_2^2 \partial_\tau$.
It is easy to see that the modular invariant solutions to the above equation must be odd under the involution $\mathcal{E}^-(s;s_1,s_2;-\bar\tau) = - \mathcal{E}^-(s;s_1,s_2;\tau)$.

In the same way as the GESs with spectrum of parameters \eqref{eq:spec1} are related to MGFs,  rational combinations of 
cuspidal GESs with spectrum of parameters 
\begin{equation}
s_1,s_2\in \mathbb{N}\,, \quad {\rm with}\quad s_1,s_2\geq 2\,,\qquad s\in\{|s_1-s_2|+1,|s_1-s_2|+3, \ldots , s_1+s_2-3,s_1+s_2-1\}\,,
\end{equation}
produce all odd MGFs.
Importantly,  it was shown in \cite{Dorigoni:2021jfr,Dorigoni:2021ngn} that whenever the eigenvalue $s$, which now has the opposite parity to the weight $w=s_1+s_2$, is such that $\dim \mathcal{S}_{2s}\neq0$, the function $\mathcal{E}^-(s;s_1,s_2;\tau)$ receives a non-vanishing contribution from an odd variant of the solution of the homogeneous equation  \eqref{eq:Hdelta}. However, compared to \eqref{eq:lambdaMGF2t} the  coefficient  of the solution to the homogeneous equation is proportional to the product of an even non-critical $L$-value times an odd critical one.
Cancellation of $L$-values in the rational linear combinations of cuspidal GESs that produce odd MGFs is again a consequence of the same Tsunogai's derivation algebra as discussed in section \ref{sec:Absence} for MGFs.

Although these modular functions have not yet appeared in the physics literature, it is rather natural to ask whether  in the odd SMFs case a generalisation of \eqref{eq:cuspGES} to the case of half-integer indices $s_1,s_2\in \mathbb{N}+1/2$ and eigenvalue with the same parity as the weight $w=s_1+s_2$ plays a role analogous to that of \eqref{eq:spec2}. Firstly, we believe that a suitable modification of \ref{TH1} \cite{Fedosova:2023cab} ought to produce an $L$-value contribution which, in parallel to \eqref{eq:lambdaIC}, should now result in the product of two critical odd values.
We are not aware of an expression similar to \eqref{eq:cuspij} that is responsible for the cancellation of the odd critical $L$-values in the rational linear combinations of cuspidal GESs corresponding to odd MGFs or odd SMFs. Furthermore, even in the case of odd MGFs where a lattice sum representation is available, we do not know whether these objects can be constructed in terms of some modified theta lift akin to \eqref{eq:thetlift0}.

\subsection{Finite-$N$ integrated correlators}

As mentioned in the introduction, it was conjectured in \cite{Alday:2023pet} that the correct vector space of modular invariant functions containing the coefficients of the large-$N$ expansion of a certain integrated correlator, $\mathcal{H}_N(\tau)$, of four superconformal primary operators in the stress-tensor multiplet in $\mathcal{N}=4$ SYM is given by the direct sum of non-holomorphic Eisenstein series with half-integer index and SMFs  $\mathcal{E}^w_{i,j}(\tau)$ where $w$ and $s=3i+j+1$ have opposite parity.
An interesting open question is how can this integrated correlator be represented at generic finite value of $N$. 

Inspired by a simpler case~\cite{Dorigoni:2021bvj, Dorigoni:2021guq} where the same four-point correlator is integrated with respect to a different measure, we may expect that in passing from large-$N$ to finite $N$, the vector space of modular invariant functions changes to the direct sum of non-holomorphic Eisenstein series with integer index and MGFs $\mathcal{E}^w_{i,j}(\tau)$ where $w$ and $s=3i+j+1$ have the same parity.
At finite $N$ we expect the integrated correlator to be given by a formal sum over infinitely many elements of this vector space.

For computational purposes, it might be useful to try and enlarge this vector space and use as a basis all GESs with spectrum \eqref{eq:spec1}. However, at any given $N$  this physical observable can be computed from a well-known matrix model originating from supersymmetric localisation, which is believed to not produce $L$-values of holomorphic cusp forms. It is of interest to find explicitly the expression for the integrated correlator in terms of MGFs, thus immediately proving the absence of $L$-values, or as a sum of GESs with spectrum \eqref{eq:spec1} and subsequently verify whether the cusp forms do indeed cancel in the infinite sum.

Guided by~\cite{Dorigoni:2021bvj, Dorigoni:2021guq}, we conjecture that at finite $N$ the representation of the integrated correlator $\mathcal{H}_N(\tau)$ as an infinite sum over  MGFs and integer-index non holomorphic Eisenstein series should be equivalent to
\begin{equation}
\mathcal{H}_N(\tau) = \int_{(\mathbb{R}^+)^3} B_N(t) \, \Gamma_{2,2}(\tau;t)\,{\rm d}^3 t\,.\label{eq:HN}
\end{equation}
Two additional key properties  motivate this conjecture.
Firstly, the finite-$N$ expression \eqref{eq:HN} when expanded at large-$N$ ought to reproduce the results derived in \cite{Alday:2023pet}. In particular, the integrand function $B_N(t)$ originates from the resummation of the infinite series of  MGFs with coefficients depending from $N$. Order by order at large-$N$, the expansion of $B_N(t)$ should produce a linear combination of integrands $B^w_{i,j}(t)$ of the form \eqref{eq:BwijDef}, which upon integrating over $t$ produces the linear combination of SMFs found in \cite{Alday:2023pet}. This is exactly what happens in the simpler case~\cite{Dorigoni:2021bvj, Dorigoni:2021guq}. 

Secondly, in \eqref{eq:HN} we have used the complete theta-series \eqref{eq:G22t} which  includes terms in the lattice sum  with vanishing momenta $p_i=0$.  While for a single $\mathcal{E}^w_{i,j}(\tau)$ the inclusion of these terms leads to an ill-defined singular integral in \eqref{eq:Ewijt}, for the resummed expression in \eqref{eq:HN} we expect the integral to converge even in the absence of the exponential suppression coming from \eqref{eq:G22tred}.
Furthermore, if we consider $\mathcal{E}^w_{i,j}(\tau)$ by extending the theta-series to the complete lattice sum as in \eqref{eq:intlatt}, we can regularise the integral by introducing a cutoff, i.e. by integrating over $(t_1,t_2,t_3) \in [0,L]^3$ rather than over $(\mathbb{R}^+)^3$. One can then easily check that terms where one or three momenta vanish\footnote{Due to momentum conservation $p_1+p_2+p_3=0$ in the lattice sum \eqref{eq:G22t}, if two momenta are zero then the third one is necessarily zero.} produce divergent terms as $L\gg1$, which are either constant or proportional to integer index non-holomorphic Eisenstein series when $w$ and $s=3i+j+1$ have the same parity, or half-integer index when $w$ and $s$ have opposite parity.
We believe that this phenomenon is a signal of the fact that the large-$N$ expansion of \eqref{eq:HN}  produces both integer powers in $1/N$ terms, which are linear combinations of SMFs, as well as the half-integer powers in $1/N$, which are rational linear combinations of non-holomorphic Eisenstein series with half-integer indices.

\subsection{S-dual modular forms}

The generic type IIB superstring amplitudes and correlators in $\mathcal{N}=4$ SYM are not necessarily invariant under ${\rm SL}(2, \mathbb{Z})$. In particular, these include $U(1)$-violating interactions, which transform covariantly  with non-zero modular weight under ${\rm SL}(2, \mathbb{Z})$. A natural extension of the cases studied in this paper involves the maximal $U(1)$-violating amplitudes~\cite{Green:2019rhz} and their holographic dual correlators in $\mathcal{N}=4$ SYM~\cite{Green:2020eyj, Dorigoni:2021rdo}. 

Explicit examples discussed in~\cite{Green:2019rhz} include the maximal $U(1)$-violating scattering amplitudes of four gravitons and $h$ axion-dilatons excitations in type IIB superstring theory, which are described by modular forms with holomorphic and anti-holomorphic weights $(h, -h)$, here denoted by $\cE_h(s; s_1, h_1; s_2, h_2; \tau)$. In general,  such modular forms  transform as
\ie
f_h( \gamma \cdot \tau) =  \left( {c\tau+d \over c \bar{\tau} +d } \right)^h  f_h( \tau) \, , \quad {\rm for } \quad \gamma = \left(\begin{matrix} a & b \\ c & d \end{matrix}\right) \in {\rm SL}(2,\ZZ)\, . 
\fe
It was argued in~\cite{Green:2019rhz}, the coefficients in the low-energy expansion of such amplitudes obey a Laplace eigenvalue equation of the form,
\ie \label{eq:modform}
\left[ \Delta^{(h)}_{(+)}  -(s+h)(s-h-1) \right] \cE_h(s; s_1, h_1; s_2, h_2; \tau) = E_{h_1}(s_1; \tau) E_{h_2}(s_2; \tau) \, , 
\fe
where $h_1+h_2=h$, and 
\ie
\Delta^{(h)}_{(+)}   =4\overline{\cD}_{-h-1}   \cD_{h}  \, ,   \qquad {\rm with} \quad \cD_{h} = i \left( \tau_2 {\partial \over \partial_{\tau}} -i {h \over 2} \right)  \, , 
\fe
The modular forms $ E_{h}(s; \tau)$ in the source term of \eqref{eq:modform} are defined by
\ie
 E_{h}(s; \tau)  = {2^h \Gamma(s) \over \Gamma(s{+}h)} \cD_{h-1} \cdots \cD_0 E(s; \tau) = \sum_{(m,n)\neq (0,0)} \left( {m+n \bar{\tau} \over m+n  \tau } \right)^h {(\tau_2/\pi)^s \over |m+n \tau|^{2s}} \, .
 \fe
When $h=0$, $E_{0}(s; \tau)$ reduces to the non-holomorphic Eisenstein series, $E(s; \tau)$. 

The Laplace eigenvalue equation \eqref{eq:modform} is a very natural generalisation of \eqref{eq:geneisen} that was studied in this paper.  An interesting case that is relevant to superstring amplitudes was considered in~\cite{Green:2019rhz} corresponds to $s=4, s_1=s_2=\threeh$ and $h>0$.  In this case solutions to \eqref{eq:modform} can be related to the action of  the covariant derivatives $ \cD_{w}$ on $\cE(4; \threeh,\threeh; \tau)$ with appropriate values of $w$.   It is of interest to study  the solution to equation \eqref{eq:modform} with more general parameters  (with $s_1,s_2\in \mathbb{N} $ and $s_1,s_2\in \mathbb{N} + {1 \over 2}$, respectively), and in particular examine the roles played by the $L$-values of holomorphic cusp forms and their potential cancellations.

\section*{Acknowledgements}
We thank  Kathrin Bringmann, Guillaume Bossard, Ksenia Fedosova, Jens Funke, Abhiram Kidambi, Axel Kleinschmidt, Kim Klinger-Logan, Caner Nazaroglu, Oliver Schlotterer for helpful discussions.  We would also like to thank Axel Kleinschmidt and Oliver Schlotterer for providing extremely helpful comments on the draft.   
DD is indebted to the Albert Einstein Institute and in particular to Axel
Kleinschmidt, Hermann Nicolai and Stefan Theisen for the hospitality and financial support.
CW is supported by a Royal Society University Research Fellowship,  URF$\backslash$R$\backslash$221015 and partly a STFC Consolidated Grant, ST$\backslash$T000686$\backslash$1 ``Amplitudes, strings \& duality".

\appendix

\section{Non-holomorphic Eisenstein series}
\label{sec:nonhol}

Non-holomorphic Eisenstein series constitute an important class of modular invariant functions that arise both in the study of one-loop MGFs as well as in the analysis of  low-order coefficients in the low energy effective action of  type IIB superstring theory.  

They are special examples of Maass functions for the group ${\rm SL}(2,\mathbb{Z})$ which we recall are modular invariant functions $f(\tau)$, i.e.
\begin{equation}
f(\gamma \cdot \tau) = f(\tau)\,,\qquad \qquad \forall\, \gamma = \left(\begin{matrix} a & b \\ c & d \end{matrix}\right) \in {\rm SL}(2,\ZZ)\,,
\end{equation}
with $\gamma\cdot \tau \coloneqq \frac{a\tau+b}{c\tau+d}$, of moderate growth for large $\tau_2$, i.e. there exists $N\in \mathbb{N}$ such that $f(\tau) = O(\tau_2^N)$ as $\tau_2\gg1$, and importantly $f(\tau)$ satisfies the Laplace eigenvalue equation 
\begin{equation}
[\Delta_\tau -s(s-1)] f(\tau) = 0\, ,
\label{eq:EisenLap}
\end{equation}
for some $s\in \mathbb{C}$.
The non-holomorphic Eisenstein series, $E(s;\tau)$, is perhaps the simplest example of a Maass function  and can be defined by the two-dimensional lattice sum, 
\begin{align}
  E(s; \tau)
   \label{eq:EisenExpansion}=&\sum_{(m,n)\ne(0,0)} \frac{(\tau_2/\pi)^s}{|m+n\tau|^{2s}}\,.
   \end{align}
   The function $E(s; \tau)$ satisfies the Laplace equation \eqref{eq:EisenLap} and has a Fourier expansion given by
   \begin{align}
   E(s; \tau) = & \, \label{eq:EisenFourier} \frac{2 \zeta(2 s)}{\pi^s}{\tau_2^s} +   \frac{2\Gamma(s - \frac 12)}{\pi^{s-\frac12}\Gamma(s)} \zeta(2s-1)\tau_2^{1-s} \\
    &\notag+ \frac{4 \sqrt{\tau_2}}{\Gamma(s)} \sum_{k\ne 0} \abs{k}^{s-\half}
    \sigma_{1-2s}(k) \, 
      K_{s - \frac 12} (2 \pi  \abs{k}\tau_2) \, e^{2 \pi i k \tau_1} \, ,
\end{align}
where $K_{s - \frac 12} (2 \pi  \abs{k}\tau_2) $ is a modified Bessel function of the second kind and $\sigma_s(n) \coloneqq \sum_{d\vert n} d^s$ is the divisor sigma function. Although the lattice sum definition \eqref{eq:EisenExpansion} only converges for $\mbox{Re}(s)>1$,  it is possible to analytically continue $ E(s; \tau) $ to a meromorphic function of $s\in \mathbb{C}$, with a simple {pole located at $s=1$}. For later convenience we note that with the normalisation convention chosen the non-holomorphic Eisenstein series satisfies the functional equation
\begin{equation}
\Gamma(s) E(s;\tau) = \Gamma(1-s)E(1-s;\tau)\,,
\label{eq:functional}
\end{equation}
and we also recall the  integral representation
\begin{align}
  E(s; \tau) =  \sum_{(m,n)\neq(0,0)}\frac{1}{\Gamma(s)} \int_0^\infty  t^{s-1} e^{-\pi t \frac{ |m + n \tau|^2}{\tau_2}}   {\rm d}t \,,
\label{eq:Eisenint}
\end{align}
valid for $\mbox{Re}(s)>1$.

\section{Modular local polynomials}
\label{app:ModPol}

In this appendix we provide more details regarding the modular local polynomials presented in section \ref{sec:ModLoc}.
Following \cite{BringmannMaas} we are interested in analysing the exceptional set $E_D$, as defined in \eqref{eq:ExSet}, where the modular local polynomials become non-differentiable. In particular, we now prove the claim made in the main text that for the case of discriminant $D=1$, modular local polynomials with exceptional set $E_1$ are uniquely specified by their values on the fundamental domain of $\Gamma_0(2)$.  

We start by noting that the exceptional set $E_D$ defined in \eqref{eq:ExSet} is $ {\rm SL}(2, \mathbb{Z})$ invariant since for all $\gamma\in {\rm SL}(2, \mathbb{Z})$ and for any $Q=[a,b,c] \in \mathcal{Q}_D$ (where $D=b^2-4 a c$ is the discriminant)  we have
\begin{align}
S_Q \vert \gamma  =  \left\lbrace \rho \in \mathfrak{H}\,\Big\vert\,a|\gamma\cdot \rho|^2 + b\,{ \mbox Re}(\gamma\cdot  \rho)+c =0\right\rbrace= S_{Q'}\,,
\end{align}
with $Q'=[a',b',c'] \in \mathcal{Q}_D$ given by\begin{align}
a' &\notag= a I^2+ b I  K+c K^2\,,\\
b'&= 2 a I J+ b I  L+b J K+2 c K L \,,\\
c' &\notag= a J^2+b J L+c L^2\,,
\end{align}
and 
\begin{equation}
\gamma = \left(\begin{matrix} I & J \\ K & L \end{matrix}\right)\in  {\rm SL}(2, \mathbb{Z})\,.
\end{equation}

From now on, we focus our attention to the case where the quadratic forms have discriminant $D=1$ and note that the class $[0,1,c]$ with $c\in \mathbb{Z}$   belongs to $ \mathcal{Q}_1$ with corresponding set
\begin{equation}
S_{[0,1,c]} =  \left\lbrace \rho \in \mathfrak{H}\,\Big\vert\, \mbox{Re}(\rho) +c=0\right\rbrace\,.
\end{equation}
We also see that  $[-1,1,0]\in \mathcal{Q}_1$ with set
\begin{equation}
S_{[-1,1,0]} =  \left\lbrace \rho \in \mathfrak{H}\,\Big\vert\,  |\rho-\frac{1}{2}|^2 = \frac{1}{4} \right\rbrace\,.
\end{equation}

The subsets of $E_1$ given by $S_{[0,1,0]}$, $S_{[0,1,1]}$ and $S_{[-1,1,0]}$ coincide precisely with the boundary of the fundamental domain of $\Gamma_0(2)$, hence we claim that the connected component $\mathcal{C}_{\rho^*} \subset \mathfrak{H}\setminus E_1$ which contains $\rho^* \coloneqq e^{\frac{\pi i }{3}}$ coincides with the fundamental domain of $\Gamma_0(2)$, i.e. 
\begin{equation}
 \mathcal{C}_{\rho^*}  = \mathcal{F}_{\Gamma_0(2)} \coloneqq \left\lbrace \rho \in \mathfrak{H}\,\Big\vert\, 0<{\mbox Re}(\rho) <1 \,,\,  |\rho-\frac{1}{2}|^2 >  \frac{1}{4}\right\rbrace\,.
\end{equation}

\textit{Proof:} Firstly we show that $\rho^*$ cannot possibly belong to the exceptional set, i.e. $\rho^* \neq E_1$. By contradiction, if $\rho^*  \in E_1$ then there must exist  $Q=[a,b,c] \in \mathcal{Q}_1$ such that 
\begin{equation}
a |\rho^*|^2 + b\,  \mbox{Re}(\rho^*) +c =0 \,\Rightarrow b = - 2(a+c)\,,
\end{equation}
which would imply $D = b^2 -4 ac$ even thus contradicting the fact that $D=1$.

Now we {will show} that the sets $S_{[0,1,0]}$, $S_{[0,1,-1]}$ and $S_{[-1,1,0]}$ form  the boundaries of $\mathcal{F}_{\Gamma_0(2)} $, $\mbox{Re}(\rho)=0$, $\mbox{Re}(\rho) = 1$ and $  |\rho-\frac{1}{2}|^2 = \frac{1}{4}$,  respectively.
Suppose there exists $Q=[a,b,c]\in \mathcal{Q}_1$ such that $S_Q \cap \mathcal{F}_{\Gamma_0(2)} \neq \emptyset$, without loss of generality we can assume that $a\geq 0$, since 
\begin{equation}
[a,b,c]\in \mathcal{Q}_1 \iff [-a,-b,-c]\in \mathcal{Q}_1\,,
\end{equation} 
and furthermore 
\begin{equation} 
S_{[a,b,c]} \cap \mathcal{F}_{\Gamma_0(2)} \neq \emptyset \iff S_{[-a,-b,-c]} \cap \mathcal{F}_{\Gamma_0(2)} \neq \emptyset \,.
\end{equation}
We can furthermore discard the case $a=0$ since we already know that $S_{[0,1,c]}$ either does not intersect $\mathcal{F}_{\Gamma_0(2)} $ or produces the two vertical boundaries.

Let us suppose there exists $\tilde{\rho} = x+ i y \in S_{[a,b,c]} \cap \mathcal{F}_{\Gamma_0(2)}$, i.e. 
\begin{equation}
a(x^2+y^2 )+b x +c =0\,,\qquad \rho^\star \in \mathcal{F}_{\Gamma_0(2)}\,.\label{eq:inF}
\end{equation}
Firstly, we see immediately that if $c<0$ it is impossible to satisfy the discriminant condition $D=b^2 -4 ac =1$ unless $a=0, b= 1$ which we have already considered, hence we can assume $c\geq 0$.

Since $\tilde{\rho}\in \mathcal{F}_{\Gamma_0(2)}$ we know already that $0<x<1$ and $x^2+y^2 >x$ so we deduce
\begin{equation}
a(x^2+y^2 )+b x +c > (a+b)x +c\,.\label{eq:const}
\end{equation}
If we assume that $(a+b)\geq 0$, since as discussed above $c\geq 0 $, we find that the only possibility of satisfying \eqref{eq:inF} given the constraint \eqref{eq:const} and $D=1$ is to consider $S_{[1,-1,0]} = S_{[-1,1,0]}$ which correspond to the lower boundary of $\mathcal{F}_{\Gamma_0(2)}$.
Let us consider then the case $(a+b)<0$. Since $\tilde{\rho} \in \mathcal{F}_{\Gamma_0(2)}$ we must have $x<1$ hence for $(a+b)<0$ \eqref{eq:const} becomes
\begin{equation}
a(x^2+y^2 )+b x +c > a+b+c\,.\label{eq:const2}
\end{equation}
To satisfy \eqref{eq:inF} given this constraint we must have $-b>a+c \geq 0$ hence the discriminant must satisfy
\begin{equation}
1= D = b^2 -4 a c \geq (a-c)^2\,.
\end{equation}
This implies either $a=c\in \mathbb{N}^+$ and hence $b = - \sqrt{4a^2+1}\notin \mathbb{Z}$ or $a=c\pm1$ and hence $b=-2c \mp 1$  so that the constraint \eqref{eq:const2} becomes $a(x^2+y^2 )+b x +c >0$ and again $\tilde{\rho}$ can only live at the boundary of $\mathcal{F}_{\Gamma_0(2)}$.

\section{Holomorphic cusp forms and their $L$-values}
\label{app:Cusps}

In this appendix we review some important facts about holomorphic cusp forms for ${\rm SL}(2,\mathbb{Z})$.
We start by defining $M_k$ as the vector space of modular forms of weight $k\in \mathbb{Z}$ under ${\rm SL}(2,\mathbb{Z})$, which are furthermore analytic in the upper half-plane $\mathfrak{H}$. It is well-known (see for example \cite{ApostolTomM1976MfaD}) that when $k$ odd, or  if $k<0$, or $k=2$  then the only element of $M_k$ is given by the zero-function while for $k=0$ the only element of $M_0$ is the constant function.
For $k>2$ and even, $M_k$ splits naturally as
\begin{equation}
M_k = {\rm span}\{ {\rm G}_k\} \oplus \mathcal{S}_{k}\,,\label{eq:holosplit}
\end{equation}
where $G_k$ denotes the holomorphic Eisenstein series
\begin{equation}
G_k(\tau) = \sum_{n=0}^\infty \alpha_k(n) q^n\,,\label{eq:holoEisen}
\end{equation}
where $q= \exp(2\pi i \tau)$ and the Fourier coefficients $\alpha_k(n)$ are given by 
\begin{equation}
\alpha_k(0)=1\,,\qquad \alpha_k(1) =\alpha_k = -\frac{2k}{B_k}\,,\qquad\alpha_k(n) = \alpha_k \sigma_{k-1}(n)\,,\label{eq:EisenFour}
\end{equation}
with $B_k$ denoting the $k^{th}$ Bernoulli number and $\sigma_s(n) = \sum_{d\vert n}d^s$ the divisor sigma function.

The second term in \eqref{eq:holosplit} is given by the vector space $\mathcal{S}_{k}$ of ${\rm SL}(2,\mathbb{Z})$ holomorphic cusp forms.
A function $\Delta$ analytic in the upper half-plane $\mathfrak{H}$ belongs to $\mathcal{S}_k$ if it is a modular form of weight $k$ and vanishes at the cusp $\tau\to i\infty$, i.e. its $q$-series Fourier expansion must not contain the constant term
\begin{equation}
\Delta(\tau) = \sum_{n=1}^\infty a(n) q^n\,.
\end{equation}
We furthermore say that the cusp form $\Delta\in \mathcal{S}_{k}$ is \textit{Hecke normalised} if $a(1)=1$.
The dimension of the vector space of holomorphic cusp forms is given by
\begin{equation}\label{eq:DimSApp}
 \dim \mathcal{S}_{k} = \left\lbrace \begin{array}{lc}
\left\lfloor \frac{k}{12}\right\rfloor -1 & k\equiv 2\, {\rm mod }\,12\,,\\[2mm]
 \left\lfloor \frac{k}{12}\right\rfloor \phantom{-1}& \mbox{otherwise}\,.
\end{array}\right.
\end{equation}

In \cite{Hecke:1937} Hecke introduced an infinite family of linear operators $T_n:\mathcal{S}_k \to \mathcal{S}_k$ with $n\in \mathbb{N}$, now called Hecke operators, defined by
\begin{equation}
[T_n \Delta](\tau) = n^{k-1} \sum_{d\vert n} d^{-k} \sum_{b=0}^{d-1} \Delta\left(\frac{ n \tau+bd}{d^2}\right) \,.\label{eq:Hecke}
\end{equation} 
A preferred basis for $\mathcal{S}_k$ can be constructed by considering the basis of Hecke eigenforms which simultaneously diagonalise all $T_n$.
Importantly, the eigenvalues of a Hecke eigenform $\Delta(\tau) = \sum_{n=1}^\infty a(n) q^n\ \in \mathcal{S}_k$  are precisely given by its Fourier $q$-series coefficients, i.e. we have $[T_n \Delta](\tau) =  a(n) \Delta(\tau)$ so that from \eqref{eq:Hecke} we see that a Hecke eigenform is automatically Hecke normalised.

Since the Hecke operators are multiplicative, i.e. $T_m T_n = T_{m\cdot n}$ if $(m,n)=1$, it follows that the Fourier coefficients $a(n)$ of every Hecke eigenform must be multiplicative as well, i.e. $a(m)a(n) = a(m\cdot n)$ for $(m,n)=1$. This important property leads to the fact that in general the coefficients of a Hecke eigenform live in a number field extension of the rationals, which is a finite degree field extension of $\mathbb{Q}$ by the roots of an associated defining polynomial of degree $\dim \mathcal{S}_{k}$. 
Whenever $\dim \mathcal{S}_{k} =1$ we have that this number field actually coincides with the rational numbers. However, when $\dim \mathcal{S}_{k} >1$ that is not true any longer. For example if we consider $k=24$, for which $\dim  \mathcal{S}_{24} =2$, we find that the Fourier coefficients of the two unique Hecke normalised basis elements of $\mathcal{S}_{24}$ belong to the number field $\mathbb{Q}[\sqrt{144169}]$ originating from the defining polynomial $x^2 - x - 36042$.

Given a Hecke eigenform  $\Delta(\tau)= \sum_{n=1}^\infty a(n)q^n\in \mathcal{S}_k$, an important quantity to consider is the associated $L$-function
defined via the Dirichlet series
\begin{equation}
 L(\Delta,s) \coloneqq \sum_{n=1}^\infty \frac{a(n)}{n^s}\,. 
\end{equation}
In general this series converges for ${\rm Re}(s)$ large enough but it can be analytically continued to a meromorphic function of $s\in \mathbb{C}$.
From $ L(\Delta,s)$ we construct the completed $L$-function for a Hecke eigenform,
\begin{equation}
\Lambda(\Delta,s) \coloneqq (2\pi)^{-s} \Gamma(s) L(\Delta,s)\,,\label{eq:Lambda}
\end{equation}
which satisfies the important functional equation
\begin{equation}
\Lambda(\Delta,s)=  (-1)^{\frac{k}{2}}\Lambda(\Delta,k-s)\,.\label{eq:LambdaFunctional}
\end{equation}

We remind the reader that the critical strip for the completed $L$-function $\Lambda(\Delta,s)$ of a cusp form $\Delta\in \cS_{k}$ is defined as the strip ${\rm Re}\,s\in (0,k)$ and the critical $L$-values of $\Delta$ are given by $\Lambda(\Delta,s)$ with $s \in  (0,k)\cap \mathbb{N}$ while values outside of this range are called non-critical. A beautiful result by Manin \cite{Manin}, known as the periods theorem, proves that the ratio of any two even critical $L$-values or any two odd critical $L$-values are all rational over the algebraic number field generated by the Fourier coefficient of the cusp forms. This theorem effectively means that there are only two independent critical values for any given cusp forms, usually denoted as $\omega^\pm_\Delta$ and called the periods of the cusp form. On the other hand, all non-critical $L$-values are conjecturally transcendentally independent from one another.

Manin's results rely heavily on the notion of period polynomials. For every Hecke-normalised cusp form $\Delta\in \mathcal{S}_{k}$ we define its associated period polynomial
\begin{equation}
r_\Delta(X) \coloneqq \int_0^{i\infty} \Delta(\tau) (\tau-X)^{k-2} {\rm d}\tau\,,\label{eq:PeriodPoly}
\end{equation}
which is a degree $k-2$ polynomial in the variable $X$. We can use the explicit $q$-series representation $\Delta(\tau) = \sum_{n=1}^\infty a(n) q^n$ with $q=\exp(2\pi i \tau)$ to show that
\begin{equation}\label{eq:LambdaIntApp}
\Lambda(\Delta,\ell+1) = i^{-\ell-1} \int_0^{i\infty} \Delta(\tau) \tau^\ell {\rm d}\tau\,.
\end{equation}
Using \eqref{eq:LambdaIntApp} it is easy to prove that the period polynomial  \eqref{eq:PeriodPoly} takes the form,
\begin{equation}
r_\Delta(X) = \sum_{\ell=0}^{k-2}i^{\ell+1} (-X)^{k-2-\ell}  { k-2\choose \ell} \Lambda(\Delta,\ell+1)\,.
\end{equation}
We immediately note that the coefficients of this polynomial are the completed $L$-values for the cusp form $\Delta$, and  $r_\Delta(X)$ carries important information about the underlying modular group ${\rm SL}(2,\mathbb{Z})$. 
In particular, considering the order-$2$ and $3$ generators of ${\rm SL}(2,\mathbb{Z})$ respectively given by $S= \left(\begin{smallmatrix} 0&-1\\ 1 & 0 \end{smallmatrix}\right)$ and $U= \left(\begin{smallmatrix} 1&-1\\ 1 & 0 \end{smallmatrix}\right)$, the period polynomial must satisfy the two cocycle conditions
\begin{equation}
r_\Delta \vert_{\mathbbm{1}} (X)+r_\Delta \vert_S (X)= r_\Delta \vert_{\mathbbm{1}}(X)+r_\Delta \vert_{U}(X)+r_\Delta \vert_{U^2}(X) = 0\,,\label{eq:SUcocycle}
\end{equation}
where the ${\rm SL}(2,\mathbb{Z})$ slash-action on $r_\Delta$ is given by
\begin{equation}
r_\Delta \vert_\gamma (X) \coloneqq (cX+d)^{k-2} r_\Delta\left(\frac{a X+b}{cX+d}\right)\,,\qquad \gamma = \left(\begin{matrix} a & b \\ c & d \end{matrix}\right)\in {\rm SL}(2,\mathbb{Z}).
\end{equation}
It is easy to check that for the critical $L$-values, the reflection identity $\Lambda(\Delta,s) = (-1)^{k/2} \Lambda(\Delta,k-s)$ is a direct consequence of the $S$-cocycle condition in \eqref{eq:SUcocycle}. 
The aforementioned periods theorem of Manin \cite{Manin} is a consequence of considering both $S$ and $U$ cocycle conditions in \eqref{eq:SUcocycle} as well as the action on $r_\Delta(X)$  induced by the Hecke operators \eqref{eq:Hecke} which are sensitive to the number field generated by the $q$-series coefficients of the Hecke normalised cusp forms in $\mathcal{S}_{k}$ given that their eigenvalues are the precisely Fourier coefficients of the Hecke eigenform considered.

Lastly, it is important that  it is possible to put a norm, called the Petersson norm, on the vector space $\mathcal{S}_k$, which is defined by
\begin{equation}\label{eq:PeterNorm}
\langle \Delta,\Delta \rangle\coloneqq \int_{\mathcal{F}} |\Delta(\tau)|^2 \tau_2^{k-2}\, {\rm d}^2 \tau \,, \qquad \Delta\in \mathcal{S}_k\,,
\end{equation} 
where the integral is over the fundamental domain $\mathcal{F}= {\rm SL}(2,\mathbb{Z})\backslash \mathfrak{H}$.  
In section \ref{sec:Lvalues}, we use a classic result by Rankin \cite{Rankin} which allows us to write the product of two critical $L$-values with opposite parity, i.e. one even and one odd, in terms of the Petersson norm.  
For the $j^{th}$ Hecke normalised eigenform $\Delta^{(j)}\in \mathcal{S}_k$ with $j\in \{1, \ldots , {\rm dim}\,\mathcal{S}_{k}\}$, Rankin's theorem states
\begin{equation}
\langle \Delta^{(j)},\Delta^{(j)} \rangle = - \frac{q \,r \Lambda(\Delta^{(j)},k-1)\Lambda(\Delta^{(j)},q) }{ B_q B_r \,\kappa(j,q,r)}\,,\label{eq:Rankin}
\end{equation}
where $B_n$ is the $n^{th}$ Bernoulli number and $q$ and $r$ are any even integers such that $q+r = k$ and $ 4\leq r \leq \frac{k}{2}-2$. 
The function $\kappa(j,q,r)$ takes value in the number field generated by the Fourier coefficients of the Hecke eigenforms of $\mathcal{S}_k$ and it is given by
\begin{equation}
\kappa(j,q,r) \coloneqq \sum_{n=1}^{{\rm dim}\,\mathcal{S}_{k}} \big(H^{-1}\big)_{jn} \left[ \sum_{m=0}^n \alpha_q(m) \alpha_r(n-m) - \alpha_k(n)\right]\,,
\end{equation}
with $\alpha_k(n)$ given in \eqref{eq:EisenFour}. 
The $({\rm dim}\,\mathcal{S}_{k})\times ({\rm dim}\,\mathcal{S}_{k})$ matrix $H$ is defined by $H_{nj} \coloneqq a^{(j)}(n)$, i.e. the entry in the $n^{th}$ row and $j^{th}$ column of the matrix $H$ is precisely the $n^{th}$ $q$-series coefficient of the $j^{th}$ Hecke eigenform, i.e. $\Delta^{(j)}(\tau) = \sum_{n=1}^\infty a^{(j)}(n) q^n\in \mathcal{S}_k$.
A particularly simple expression can be derived when we specialise \eqref{eq:Rankin} to the cases where ${\rm dim}\,\mathcal{S}_k=1$, i.e. $k\in \{12,16,20,22,26\}$.
In this case the $1\times 1$ matrix $H$ reduces to $1$ from the Hecke normalisation, and if we furthermore choose $r=4,q=k-4$ we find that \eqref{eq:Rankin}
 reduces to 
\begin{equation}
{\langle \Delta,\Delta \rangle} = -\frac{15\ 2^{5-k} (k-4) B_k \,\Lambda (\Delta,k-4) \Lambda (\Delta,k-1)}{\big[(k-4) B_k - (k+120 B_k)B_{k-4}\big]}\,.\label{eq:Petersson}
\end{equation}
For example, for the Ramanujan cusp form $\Delta_{12}$ which is the unique Hecke normalised element of $\mathcal{S}_{12}$, equation \eqref{eq:Petersson} reduces to  {(see (9.1) in \cite{Rankin})}
\begin{equation}
{\langle \Delta_{12},\Delta_{12} \rangle}=\frac{691 \Lambda (\Delta_{12},8) \Lambda (\Delta_{12},11)}{7680} = \frac{10! \times 14511}{2^{24} \pi^{19}} L(\Delta_{12},8) L(\Delta_{12},11)\,.
\end{equation}

\end{document}